\documentclass[12pt,preprint]{aastex}
\slugcomment{{\sc Accepted to ApJS:} March 11, 2009} 

\shorttitle{Magnetohydrosynamics and Adaptive Mesh Refinment}
\shortauthors{Cunningham et al.}

\newcommand{\E}{\mathcal{E}}
\newcommand{\ph}{+1/2}
\newcommand{\mh}{-1/2}
\newcommand{\pmh}{\pm 1/2}
\renewcommand{\vec}[1]{{\bf #1}}

\begin{document}
\title{Simulating Magnetohydrodynamical Flow with Constrained Transport and Adaptive Mesh Refinement; Algorithms \& Tests of the AstroBEAR Code}

\author{Andrew J. Cunningham\altaffilmark{1}, Adam Frank\altaffilmark{1}, Peggy Varni{\`e}re\altaffilmark{1,2}, Sorin Mitran\altaffilmark{3}, Thomas W. Jones\altaffilmark{4}}
\altaffiltext{1}{Department of Physics and Astronomy, University of Rochester, Rochester, NY  14054}
\altaffiltext{2}{LAOG, Universit J. Fourier  UMR5571, France}
\altaffiltext{3}{University of North Carolina, Department of Mathematics, Chapel Hill, NC 27599}
\altaffiltext{4}{Department of Astronomy, University of Minnesota, 116 Church Street SE, Minneapolis, MN 55455}
\email{ajc4@pas.rochester.edu}

\begin{abstract}
A description is given of the algorithms implemented in the AstroBEAR
adaptive mesh refinement code for ideal magnetohydrodynamics.  The
code provides several high resolution, shock capturing schemes which
are constructed to maintain conserved quantities of the flow in a
finite volume sense.  Divergence free magnetic field topologies are
maintained to machine precision by collating the components of the
magnetic field on a cell-interface staggered grid and utilizing the
constrained transport approach for integrating the induction
equations.  The maintenance of magnetic field topologies on adaptive
grids is achieved using prolongation and restriction operators which
preserve the divergence and curl of the magnetic field across
co-located grids of different resolution.  The robustness and
correctness of the code is demonstrated by comparing the numerical
solution of various tests with analytical solutions or previously
published numerical solutions obtained by other codes.
\end{abstract}


\section{Introduction}
The development of efficient and accurate numerical algorithms for
astrophysical flow has become of great interest to the astrophysical
community.  Variable resolution approaches have provided an avenue for
efficient simulation of hydrodynamical flow including multi-physical
effects which involve substantial variation in length scale.  Adaptive
Mesh Refinement (AMR) has been recognized as one of the most versatile
and efficient approaches to enable the simulation of multi-scale
phenomena for which fixed-resolution simulation is either impractical
or impossible.  AMR discretizations employ a hierarchy of grids at
different levels.  High resolution is applied only to those regions of
the flow which would otherwise be subject to unacceptably large
truncation error. The utility of the AMR approach is underscored by
the extensive list of codes that are targeted toward astrophysical
research which utilize AMR.  The list includes AstroBEAR, Enzo
\citep{enzo}, Flash \citep{flash}, Orion
\citep{truelove,klein,crockett}, Nirvana \citep{nirvana-amr}, Ramses
\citep{ramses}, RIEMANN \citep{balsara-amr} and AMRVAC \citep{amrvac}
and the list of codes for which the development AMR capability is in
progress including Athena \citep{gardiner} and Pluto \citep{pluto}.

The simulation of magnetized flow is of particular interest to
astrophysical researchers owing to the utility of numerical
magnetohydrodynamics (MHD) in modeling a wide range of astrophysical
phenomena.  The leading line of recent research in this area has
focused on the application of higher order Godunov methods to
numerical MHD \citep{rj-rs,balsara-rs}.  The conservative formulation
and proper upwinding employed by these methods enable accurate
simulation of strongly supersonic flow.  Because of this unique
capability, such methods are often referred to ``high resolution shock
capturing'' (HRSC) methods.

While HRSC methods have long been recognized as the defacto standard
for the simulation of supersonic hydrodynamical phenomena, their
popularity among researchers interested in magnetized flow has been
slowed because standard HRSC approaches to MHD fail to maintain the
solenoidality constraint on the magnetic field ($\vec{\nabla} \cdot
\vec{B} = 0$).  If not corrected, local divergences in the magnetic
field arising from this short coming usually grow rapidly, causing
anomalous magnetic forces and unphysical plasma transport which
eventually destroys the correct dynamics of the flow
\citep{brackbill}.  Early practitioners of numerical MHD therefore
relied heavily on finite difference methods as employed by codes such
as Zeus \cite{zeus-mhd} which maintain the solenoidality constraint
exactly despite their inferior shock capturing ability \citep{falle}.
Later works focused on improved HRSC which either eliminate the
development of divergences in the magnetic field, or mitigate the
effect of local divergence errors on the dynamics of the flow.  In one
such approach, a projection operator is devised, usually by solving a
Poisson equation, which removes numerical divergences from the grid
after each time-step \citep{balsara98,jiang,kim,zachary,rjf}.  The
primary limitation of this approach is that non-trivial boundary value
problems become indeterminate \citep{rj-ct}.  In the second so called
``8-wave'' approach first explored by \cite{powell}, alternative
formulations of the MHD equations are constructed to prevent the local
build up of magnetic divergence by advecting monopoles to other
regions of the grid where they are of less consequence to the dynamics
of the flow.  The work of \cite{dender} augments this approach by
adding source terms to the system which act to counter the effect of
local divergence in the magnetic field on the dynamics of the flow.  A
third approach known as constrained transport utilizes a
multidimensional, divergence-preserving update procedure for the
magnetic field components which are collated on a staggered grid
centered on the computational volume interfaces.
\citep{evans,rj-ct,balsara-spicer,dai,londrillo,nirvana}.  This
approach has been shown to provide the most accurate results in the
tests of \cite{toth} and \cite{balsara-comp}.

The combination of AMR spatial discretizations with HRSC would seem a
natural choice in order to satisfy the desire for a high accurate,
computationally efficient and versatile strategy for the simulation of
magnetized plasma flow.  The implementation of the aforementioned
adaptations to HRSC methods for MHD in an AMR framework however, poses
several challenges.  Divergence cleaning schemes utilizing a Poisson
projection operator are ill-suited for AMR applications owing to
difficulty in handling the projection step along internal boundaries
on a patchwork of grids at different resolutions. \cite{powell} found
that the application of their 8-wave method on an AMR grid hierarchy,
local divergence errors on one level of the AMR hierarchy caused local
divergence of comparable magnitude on all levels. The main drawback of
this method in an AMR context is that the unphysical effects of local
divergences in the magnetic field are not diminished by the
application of additional refinement.  \cite{crockett}, however, have
constructed an approach suitable for AMR applications which combines
an approximate projection operator with the divergence advection and
dampening the effects of local divergence of \cite{powell} and
\cite{dender}.

Retaining the divergence-free property of the solution obtained
through the application of the constrained transport approach on
hierarchical grids requires application of a divergence-preserving
prolongation operator which interpolates the magnetic field from a
coarse mesh a fine mesh, a divergence-preserving restriction operator
which maps the fine mesh magnetic field onto a coarser mesh and
furthermore requires that the evolution of the magnetic field be
consistent between collocated meshes of different resolution.  Two
approaches to these challenges have emerged.  \cite{balsara-amr} has
generalized the divergence free reconstruction procedure of
\cite{balsara-spicer} to devise a prolongation operator based on a
piece-wise quadratic interpolation procedure that is divergence
preserving in the RIEMANN MHD code.  \cite{li} present an adaptation
of Balsara's procedure that simplifies its implementation for problems
involving arbitrary refinement ratios.  \cite{toth-roe} have devised a
prolongation operator by solving an algebraic system which enforces
the maintenance of the volume average curl and divergence between
grids of different resolution in the AMR hierarchy.

In this paper we provide a concise description of the algorithms and
tests of the AstroBEAR HRSC AMR MHD code.  AstroBEAR is comprised of
several numerical solvers, integration schemes, and radiative cooling
modules for astrophysical fluids.  The code's AMR capability is
derived from the AMR engine of the BEARCLAW boundary embedded adaptive
mesh refinement package for conservation laws.  This code utilizes the
constrained transport approach to adapting HRSC methods to the MHD
system of equations.  To our knowledge, AstroBEAR is the first AMR
code to utilize the prolongation operator of \cite{toth-roe} to
maintain the $\vec{\nabla} \cdot \vec{B} = 0$ constraint.  By
combining multi-physics capabilities relevant to simulation
astrophysical plasma flow, AMR, and a wide selection of HRSC
integration procedures, AstroBEAR will serve as a valuable research
tool.  The authors intend that this paper will serve as a reference
for future works that apply the code and provide a concise recipe for
robust, reliable and accurate HRSC solution strategies for MHD on AMR
grid hierarchies.  In \S \ref{method} we describe the several HRSC
schemes and divergence preservation strategies implemented in the
code.  In \S \ref{AMR} we provide an overview of the AMR algorithm,
highlighting the stages which require special treatment of the
magnetic field.  In \S \ref{prores} we provide a concise description
of the prolongation, restriction and coarse to fine refluxing
procedures required to preserve the divergence and consistency of the
magnetic field across an AMR hierarchy of grids.  In \S \ref{tests} we
comment on the results of several test and example problems with
particular emphasis on the relative strengths and weaknesses of the
various HRSC schemes implemented in the code.  In \S \ref{conclusion}
we provide a synopsis and discussion of the main results of the paper.

\section{Numerical Method} \label{method}
AstroBEAR provides an adaptive mesh refinement framework for the
integration of conservation laws of the form,
\begin{equation}
\frac{\partial}{\partial t} \vec{Q} + 
\frac{\partial}{\partial x} \vec{F}_x(\vec{Q}) + 
\frac{\partial}{\partial y} \vec{F}_y(\vec{Q}) + 
\frac{\partial}{\partial z} \vec{F}_z(\vec{Q}) = \vec{S}(\vec{Q}).
\label{mhd}
\end{equation}
In this work we focus on the equations of ideal MHD which are written
in the conservative form as:
\begin{eqnarray}
& \frac{\partial}{\partial t} \left[ \begin{array}{c}
\rho \\ \rho v_x \\\rho v_y \\ \rho v_z \\ \E \\
B_x \\ B_y \\ B_z
\end{array}\right] + 
\frac{\partial}{\partial x} \left[ \begin{array}{c}
\rho v_x \\ \rho v_x^2+P+B^2/2-B_x^2 \\\rho v_y v_x-B_xB_y \\ \rho v_z v_x-B_xB_z \\ (\E+P+\vec{B}^2/2)v_x - B_x(\vec{B} \cdot \vec{v}) \\
0 \\ -E_z \\ E_y 
\end{array}\right] + \nonumber \\
& \frac{\partial}{\partial y} \left[ \begin{array}{c}
\rho v_y \\ \rho v_x v_y-B_yB_x \\\rho v_y^2+P+B^2/2-B_y^2 \\ \rho v_z v_y-B_yB_z \\ (\E+P+\vec{B}^2/2)v_y - B_y(\vec{B} \cdot \vec{v}) \\
E_z  \\ 0 \\ -E_x
\end{array}\right] +
\frac{\partial}{\partial z} \left[ \begin{array}{c}
\rho v_z \\ \rho v_x v_z-B_zB_x \\\rho v_y v_z-B_zB_y \\ \rho v_z^2+P+B^2/2-B_z^2 \\ (\E+P+\vec{B}^2/2)v_y - B_z(\vec{B} \cdot \vec{v}) \\
-E_y \\ E_x \\ 0
\end{array}\right] = \vec{S}
\label{fullmhd}
\end{eqnarray}
with gas density $\rho$, velocity $\vec{v}$, total volumetric energy
density $\E$, thermal pressure $P$, magnetic field $\vec{B}$, and
electric field $\vec{E}$.  In the proceeding system of equations, we
have chosen units for the electric and magnetic field in the plasma
so that factors of $4\pi$ do not appear in the equations.  The last
three equations in the system follow from Faraday's law,
\begin{equation}\label{faraday}
\frac{\partial}{\partial t} \vec{B} + \vec{\nabla} \times \vec{E} = 0,
\end{equation}
implicit to which is the constraint that initially solenoidal magnetic
field topologies remain solenoidal,
\[ \vec{\nabla} \cdot \vec{B} = 0.\]
The equations are brought to a closed form via Ohm's law for a perfectly
conducting medium,
\begin{equation}\label{ohm}
\vec{E} = -\vec{v} \times \vec{B}
\end{equation}
and the polytropic equation of state for an ideal gas,
\begin{equation}\label{EOS}
P = (\gamma-1) (\E - \rho \vec{v}^2/2 - \vec{B}^2/2).
\end{equation}

In the remainder of this section, we describe the details of the shock
capturing numerical schemes available in our code to integrate the
solution to equations of the form of equation \ref{mhd} and
modifications thereof which ensure the solenoidal constraint on the
magnetic field.  The purpose of this description is to provide a
concise and complete illustration of the steps necessary to build the
code.  The details of any of the particular solution strategies may be
obtained by consulting the original work credited for the particular
strategy.  For a more pedagogically oriented review of high resolution
shock capturing schemes, we refer the reader to the excellent books of
\cite{leveque} and \cite{toro}.  In the remainder of this section, we
consider only the solution method for the homogeneous part of the
conservation law with $\vec{S}=0$.  The effect of non-zero source
terms which may be used to include additional physics are handled
using the operator splitting technique described in \S2.1 of
\cite{sourceI}.  The micro-physical source terms included in the code
to model the effects of radiative cooling, and time dependent,
non-equilibrium ionization and $H_2$ chemistry are cataloged in the
appendix to \cite{sourceI}.  In appendix A we catalog the MHD source
terms and modifications to the numerical scheme that are employed in
applications involving cylindrical axisymmetric flow.  We adopt the
notation that superscripts denote time, the first subscript denotes
the direction of vector components, and the last three subscripts
denote the spatial location on the computational grid.  The
superscript is omitted from temporally varying quantities which are to
be evaluated at time $t$.  Equations which demonstrate operations that
take the same form in each of the $x$, $y$, and $z$ directions are
written only for the $x$-direction sweep. In these cases, the $y$
sweep can can be recovered by replacing $y \to x$ and $j \to i$ and
the $z$ sweep can can be retrieved by replacing $z \to x$ and $k \to
i$.

Numerical integration of the system of conservation laws is achieved
using the finite volume method.  The basis of the finite volume
quadrature can be realized by discretizing the integral form of
equation \ref{mhd}, yielding the following unsplit procedure for
advancing the conserved field forward in time by an increment $\Delta
t$:
\begin{eqnarray}
\vec{Q}_{i,j,k}^{t+\Delta t} &=& \vec{Q}_{i,j,k}^{t} +
   \frac{\Delta t}{\Delta x} \left(\vec{\tilde F}_{x,i\mh,j,k}^n-\vec{\tilde F}_{x,i\ph,j,k}^n\right) + \nonumber \\
&&  \frac{\Delta t}{\Delta y} \left(\vec{\tilde F}_{y,i,j\mh,k}^n-\vec{\tilde F}_{y,i,j\ph,k}^n\right) +
   \frac{\Delta t}{\Delta z} \left(\vec{\tilde F}_{z,i,j,k\mh}^n-\vec{\tilde F}_{z,i,j,k\ph}^n\right). \label{update}
\end{eqnarray}
where $\vec{\tilde F}_{x}^n$, $\vec{\tilde F}_{x}^n$, and $\vec{\tilde
F}_{x}^n$ are suitably accurate numerical approximations to the
inter-cell flux, spatially averaged over the inter-cell area and
temporally averaged from $t$ to $t+\Delta t$.  Construction of higher
order schemes in more than one dimension can, in some cases, be
simplified by utilizing a direction-split approach
\begin{eqnarray}
\vec{Q}_{i,j,k}^{(0)} &=& \vec{Q}_{i,j,k}^{t} \nonumber \\
\vec{Q}_{i,j,k}^{(1)} &=& \vec{Q}_{i,j,k}^{(0)} +
   \frac{\Delta t}{\Delta x_1} \left(\vec{\tilde F}_{x1,i\mh,j,k}^n(\vec{Q}_{i,j,k}^{(0)})-\vec{\tilde F}_{x1,i\ph,j,k}^n(\vec{Q}_{i,j,k}^{(0)})\right) \nonumber \\
\vec{Q}_{i,j,k}^{(2)} &=& \vec{Q}_{i,j,k}^{(1)} +
   \frac{\Delta t}{\Delta x_2} \left(\vec{\tilde F}_{x2,i\mh,j,k}^n(\vec{Q}_{i,j,k}^{(1)})-\vec{\tilde F}_{x2,i\ph,j,k}^n(\vec{Q}_{i,j,k}^{(1)})\right) \nonumber \\
\vec{Q}_{i,j,k}^{t+\Delta t} &=& \vec{Q}_{i,j,k}^{(0)} +
   \frac{\Delta t}{\Delta x_3} \left(\vec{\tilde F}_{x3,i\mh,j,k}^n(\vec{Q}_{i,j,k}^{(2)})-\vec{\tilde F}_{x3,i\ph,j,k}^n(\vec{Q}_{i,j,k}^{(2)})\right). \label{splitupdate}
\end{eqnarray}
In equation (\ref{splitupdate}) we have used the superscript in
parenthesis to denote intermediate states between direction sweeps.
The ordering of the component directions $(x_1,x_2,x_3)$ cycles over
different permutations of the three coordinate directions on
successive time-steps.  In two dimensions the ordering is
$[(x,y), (y,x)]$ and in three dimensions we have found the most
consistent results by cycling over both the cyclic and anti-cyclic
permutations: 
\[ [(x,y,z), (z,x,y), (y,z,x), (z,y,x), (x,z,y), (y,x,z)] \]

The subject of the following subsections is the procedure to compute
the numerical flux which is comprised of three steps 1) reconstruction
(interpolation) to zone edges \S\ref{sr}, 2) upwinding the solution of
the Riemann problem at each zone edge \S\ref{upwind}, and 3) temporal
evolution of the field of conserved quantities \S\ref{te}.  Our code
implements several methods for carrying out each of these steps which
may be utilized in the combination that best tailors the integration
strategy to the requirements of the application at hand.

\subsection{Spatial Reconstruction} \label{sr}
\begin{figure}[!h]
\begin{center}
\includegraphics[clip=true,width=0.75\textwidth]{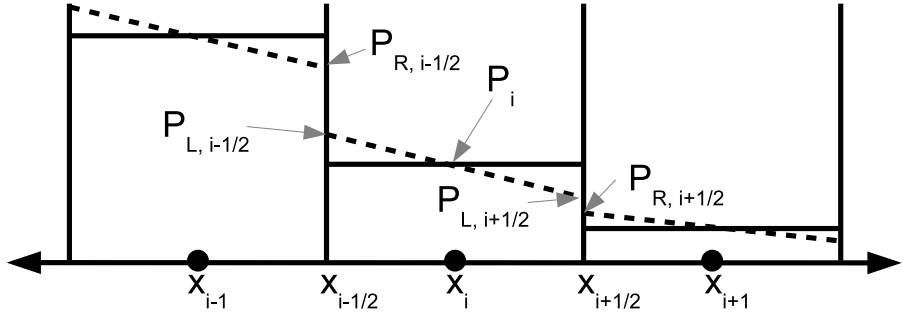}
\end{center}
\caption{Schematic of diagram of the volume average field of primitive
variables on a one dimensional grid (solid lines), the cell
reconstruction (dotted lines) and the location of the left and right
restricted states, $P_L$ and $P_R$.\label{f1}}
\end{figure}

We construct a spatially second order accurate integration procedure
via suitable reconstruction of the state in each computational cell
from the volume average state within that cell and its neighbors.  We
define a ``primitive variable'' operator, $L_p(\vec{Q}_i) = \vec{ P}_i
= \left[\rho, v_x, v_y, v_z, E - \rho \vec{v}^2/2 - \vec{B}^2/2, B_x,
  B_y, B_z \right]^T$, which converts the conserved state variables
into a form more suitable for interpolation.  Interpolation of the
primitive variables, rather than the conserved, has the advantage that
the reconstructed state at grid edges are guaranteed to have
non-negative pressure.  We write the reconstruction from the
volume-average state variables collated at grid centers to the grid
interface as:
\begin{eqnarray}
\vec{P}_{L,i\ph} = \vec{P}_{i} + \frac{1}{2} \vec{\phi}_{+,i} \nonumber \\
\vec{P}_{R,i\mh} = \vec{P}_{i} - \frac{1}{2} \vec{\phi}_{-,i}.
\label{p}
\end{eqnarray}
Figure \ref{f1} shows a cartoon schematic of the volume average field
of primitive variables on a one dimensional grid (solid lines), the
cell reconstruction (dotted lines) and the location of the left and
right restricted states.
\clearpage
\clearpage
The grid interface conservative fields are then computed as
\begin{eqnarray}
\vec{Q}_{L,i\mh} &= L_p^{-1}(\vec{P}_{L,i\mh}) \nonumber \\
\vec{Q}_{R,i\mh} &= L_p^{-1}(\vec{P}_{R,i\mh}).
\label{qedge}
\end{eqnarray}

The code implements a user selectable choice of three different second
order spatial reconstruction methods. The first is the MUSCL
reconstruction method of \cite{vanleer} using a slope limiter to
maintain monotonicity,
\begin{equation}
\vec{\phi}_{\pm,i} = \textrm{LIMITER}\left(\vec{P}_{i}-\vec{P}_{i-1},\vec{P}_{i+1}-\vec{P}_{i}\right).
\label{phi1}
\end{equation}
Many limiter functions may be constructed.  AstroBEAR implements a
choice of three limiters, which operate on the state vector parameters
in equation \ref{phi1} in a component-wise fashion.  In order of
decreasing levels of diffusion introduced into the scheme, these
limiters are the ``min-mod'' limiter,
\begin{equation}
\textrm{MINMOD}(x,y) = \left\{ \begin{array}{ll}
0 & \textrm{if~} xy<0 \\ 
\textrm{MIN}(\textrm{ABS}(x),\textrm{ABS}(y)) \textrm{SIGN}(x) & \textrm{otherwise}
\end{array} \right.,
\label{minmod}
\end{equation}
the limiter of vanLeer
\begin{equation}
\textrm{VL}(x,y) = \left\{ \begin{array}{ll}
0 & \textrm{if~} xy<0 \\ 
2\frac{xy}{x+y} & \textrm{otherwise}
\end{array} \right.,
\label{vl}
\end{equation}
and the ``monotonized-centered'' limiter
\begin{equation}
\textrm{MC}(x,y) = \left\{ \begin{array}{ll}
0 & \textrm{if~} xy<0 \\ 
\textrm{MIN}(2x,\frac{1}{2}(x+y),2y) & \textrm{otherwise}
\end{array} \right. .
\label{mc}
\end{equation}
The second reconstruction method is the local hyperbolic harmonic
variation of the piecewise hyperbolic method (PHM) of
\citet{marquina-phm}.  The PHM reconstruction prescribes the
components of $\vec{\phi}_{\pm,i}$ as
\begin{eqnarray}
\phi_{\pm,i} &=& \left\{ \begin{array}{ll}
0 & \textrm{if~} |\delta_L|<10^{-14} \textrm{~and~} |\delta_R|<10^{-14} \\
d~\Delta x~\eta_\pm & \textrm{otherwise}
\end{array} \right.
\\
\textrm{where} \nonumber \\
\eta_\pm &=& \left\{ \begin{array}{ll}
1 
& \textrm{if~} |\kappa|<10^{-5} \\
2 \kappa^{-2}\left(\log\left(\frac{2 \mp \kappa}{2 \pm \kappa}\right) \pm \frac{2\kappa}{2 \mp \kappa}\right) 
& \textrm{otherwise}
\end{array} \right. \nonumber
\\
\kappa &=& 2\left\{ \begin{array} {ll}
\sqrt{\frac{2}{1+\Delta x}}-1 
~ \textrm{if~} |\delta_L|<10^{-14} &\textrm{or~} (\delta_L\delta_R \geq 0 \textrm{~and~} \delta_R \leq \delta_L) \\
1-\sqrt{\frac{2}{1+\Delta x}} 
~ \textrm{if~} |\delta_R|<10^{-14} &\textrm{or~} (\delta_L\delta_R \geq 0 \textrm{~and~} \delta_L \leq \delta_R) \\
  \left\{ \begin{array}{ll} 
    \sqrt{\frac{d}{\delta_L}}-1 
    & \textrm{if~} \delta_L \leq \delta_R \\ 
    1-\sqrt{\frac{d}{\delta_R}}
    & \textrm{otherwise} 
  \end{array} \right\}
& \textrm{otherwise}
\end{array} \right. \nonumber
\\
d &=& 2\left\{ \begin{array} {lll}
\delta_R\left(\frac{\Delta x^2}{1+\Delta x^2}\right) 
& \textrm{if~} |\delta_L|<10^{-14} &\textrm{or~} (\delta_L\delta_R \geq 0 \textrm{~and~} \delta_R \leq \delta_L) \\
\delta_L\left(\frac{\Delta x^2}{1+\Delta x^2}\right) 
& \textrm{if~} |\delta_R|<10^{-14} &\textrm{or~} (\delta_L\delta_R \geq 0 \textrm{~and~} \delta_L \leq \delta_R) \\
\frac{\delta_R \delta_L}{\delta_R+\delta_L} 
& \textrm{otherwise,}
\end{array} \right. \nonumber
\\
\delta_L &=& \frac{P_{i}-P_{i-1}}{\Delta x} \nonumber
\\
\textrm{and} \nonumber
\\
\delta_R &=& \frac{P_{i+1}-P_{i}}{\Delta x} \nonumber
\end{eqnarray}
Note that $\eta$ has a removable singularity about $\kappa=0$ with
$\lim_{\kappa \to 0^+}(\eta) = \lim_{\kappa \to 0^-}(\eta) = 1$.  When
evaluated with 8-byte precision, the expression for $\eta$ begins to
diverge from its true solution for $|\kappa| < 10^{-5}$
and we set $\eta \to 1$ in this region using the piecewise expression
given above.  

The third method is the reconstruction procedure of the piecewise
parabolic method (PPM) of \cite{colella84}.  A detailed overview of
the PPM method is available from \cite{miller}, and \cite{mignone}.
Appendix B of \cite{mignone} already provides a concise description of
the PPM reconstruction procedure which we do not repeat here.  We note
that in the numerical examples using the PPM reconstruction presented
later in this paper, we have taken a different approach to numerical
oscillations than \cite{mignone}.  In particular, we maintain
monotonicity via the MINMOD limiter (equation \ref{minmod}), rather
than the more compressive Van Leer limiter and we do not included the
dissipation mechanisms of \S{B.1} of \cite{mignone}.

\subsection{Upwinded Numerical Flux} \label{upwind}
The numerical flux is computed by upwinding the waves associated with
the Riemann problem defined by $\vec{Q}_{L,i\mh}$ on the left and
$\vec{ Q}_{R,i\mh}$ on the right of each computational cell interface.
AstroBEAR implements three different methods for computing the upwinded
flux, the HLLD flux as described in as described in \cite{hlld} the
Roe flux, and the Marquina flux.

The Roe flux method \cite{roe-flux} calls for the decomposition of the
cell edge states into eigenmodes of a linearized system matrix.  In
the case of MHD we utilize the approximate linearized Riemann solver
of \citet{rj-rs}.  We write this decomposition in terms of the
eigenvalues $a_m(Q)$, right eigenvectors $\vec{R}_m(Q)$ and left
eigenvectors $\vec{L}_m(Q)$ given in \S2.2 of \cite{rj-rs} where the
subscript $m$ denotes the $m^{th}$ eigenmode.  Singularities which
arise, in certain limiting cases, in the normalization of the
eigensystem are avoided in the manner prescribed by
\cite{roe-balsara}.  The inter-cell flux across cell boundaries at
constant $x$ is computed as
\begin{eqnarray}
\vec{\tilde F}_{x,i\mh} &=& \frac{1}{2}\left(\vec{F}_x(\vec{Q}_{L,i\mh})+ \vec{F}_x(\vec{Q}_{R,i\mh})\right) - \nonumber \\
&& \frac{1}{2}\sum_{m=1}^8 \vec{L}_{m,i\mh} \left(a_{m,i} \left[\vec{Q}_{R,i\mh} - \vec{Q}_{L,i\mh}\right]\right) \vec{R}_{m,i\mh}.
\label{nf1}
\end{eqnarray}
where the eigendecomposition is carried out about a suitable average
of the states to the left and right of the cell interface which we
denote as $<\vec{Q}_{L,i\mh},\vec{Q}_{R,i\mh}>$.
\begin{eqnarray}
a_{m,i\mh} &= a_{m}(<\vec{Q}_{L,i-1/2},\vec{Q}_{R,i-1/2}>) \nonumber \\ 
\vec{L}_{m,i\mh} &= \vec{L}_{m}(<\vec{Q}_{L,i\mh},\vec{Q}_{R,i\mh}>) \nonumber \\ 
\vec{R}_{m,i\mh} &= \vec{R}_{m}(<\vec{Q}_{L,i\mh},\vec{Q}_{R,i\mh}>) \label{lin1}
\end{eqnarray}

The flux functions of \citet{marquina-flux}, first proposed by
\citet{shu-osher-eno} for scalar equations, take the form
\begin{eqnarray}
&\vec{\tilde F}_{x,i\mh} = \sum_{m=1}^8 \vec{\vec{\alpha}}_{i\mh}^L \vec{R}_{m,i\mh}^L + \vec{\vec{\alpha}}_{i\mh}^R \vec{R}_{m,i\mh}^R & \\
&\vec{\vec{\alpha}}_{i\mh}^L = \left\{ \begin{array}{ll}
\sum_{m=1}^8 \vec{L}_{m,i\mh}^L \vec{F}(\vec{Q}_{L,i\mh}) &\textrm{if~} a_{m,i\mh}^L a_{m,i\mh}^R > 0 \\ 
& \quad \textrm{ and } a_{m,i\mh}^L>0 \\
0 &\textrm{if } a_{m,i\mh}^L a_{m,i\mh}^R > 0 \\
& \quad \textrm{ and } a_{m,i\mh}^L \leq 0 \\
\frac{1}{2}\sum_{m=1}^8 \textrm{MAX}\left(|a_{m,i\mh}^L|,|a_{m,i\mh}^R|\right)\vec{L}_{m,i\mh}^L \vec{Q}_{L,i\mh} &\textrm{otherwise}
\end{array} \right. \noindent \\
&\vec{\vec{\alpha}}_{i\mh}^R = \left\{ \begin{array}{ll}
0 & \textrm{if } a_{m,i\mh}^L a_{m,i\mh}^R > 0 \\
& \quad \textrm{ and } a_{m,i\mh}^L > 0 \\
\sum_{m=1}^8 \vec{L}_{m,i\mh}^R \vec{F}(\vec{Q}_{R,i\mh}) &\textrm{if~} a_{m,i\mh}^L a_{m,i\mh}^R > 0 \\
& \quad \textrm{ and } a_{m,i\mh}^L \leq 0 \\
\frac{1}{2}\sum_{m=1}^8 \textrm{MAX}\left(|a_{m,i\mh}^L|,|a_{m,i\mh}^R|\right)\vec{L}_{m,i\mh}^R \vec{Q}_{L,i\mh} &\textrm{otherwise}
\end{array} \right. \noindent
\end{eqnarray}
The numerical flux across cell boundaries at constant $y$,
$\vec{\tilde F}_{y,j\mh}$, and $z$, $\vec{\tilde F}_{z,k\mh}$, are
computed in the analogous manner.  In this case a separate eight
decomposition of the system matrix on each side of the cell interface
is required.
\begin{eqnarray}
a_{m,i\mh}^L &= a_{m}(\vec{Q}_{L,i\mh}) \nonumber \\ 
a_{m,i\mh}^R &= a_{m}(\vec{Q}_{R,i\mh}) \nonumber \\ 
\vec{L}_{m,i\mh}^L &= \vec{L}_{m}(\vec{Q}_{L,i\mh}) \nonumber \\ 
\vec{L}_{m,i\mh}^R &= \vec{L}_{m}(\vec{Q}_{R,i\mh}) \nonumber \\ 
\vec{R}_{m,i\mh}^L &= \vec{R}_{m}(\vec{Q}_{L,i\mh}) \nonumber \\  
\vec{R}_{m,i\mh}^R &= \vec{R}_{m}(\vec{Q}_{R,i\mh}) \label{lin3}
\end{eqnarray}

In the coarse of carrying out simulations in the hydrodynamic limit,
we have found the Marquina flux to be advantageous for modeling
certain astrophysical phenomena (discussed below).  However, the
method fails to converge for some MHD shock tube test problems that
give rise to compound structures, that is, structures composed of a
shock and rarefaction of the same wave family moving together
\citep{brio}.  In particular, the method fails to converge for the
Riemann problem discussed in \S5 of \cite{rj-rs}.  We have overcome
this deficiency by constructing an adaptation of Marquina's flux which
introduces a small amount of numerical diffusion by utilizing the same
averaging procedure as the flux formula of \cite{roe-flux},
\begin{eqnarray}
a_{m,i\mh}^L &= a_{m}(\vec{Q}_{L,i\mh}) \nonumber \\ 
a_{m,i\mh}^R &= a_{m}(\vec{Q}_{R,i\mh}) \nonumber \\ 
\vec{L}_{m,i\mh}^L = \vec{L}_{m,i\mh}^R &= \vec{L}_{m}(<\vec{Q}_{L,i\mh},\vec{Q}_{R,i\mh}>) \nonumber \\ 
\vec{R}_{m,i\mh}^L = \vec{R}_{m,i\mh}^R &= \vec{R}_{m}(<\vec{Q}_{L,i\mh},\vec{Q}_{R,i\mh}>). \label{lin2}
\end{eqnarray}

For the methods calling for the decomposition of a linearized
approximation to the system matrix (the Roe flux and our adaptation of
the Marquina flux), we use the arithmetic average method of
\citet{rj-rs} where the system is linearized at the cell interface as
\begin{equation}
<\vec{P}^{RJ}_{i\mh}> = (\vec{P}^{RJ}_{L,~i\mh} + \vec{P}^{RJ}_{R,~i\mh})/2.
\end{equation}  
Numerical tests have shown that averaging the net of the magnetic and
thermal pressure rather than the thermal pressure alone provides more
accurate results.  Therefore we set
\begin{equation}
\vec{ P}^{RJ} = \left[\rho, v_x, v_y, v_z, P +
  \vec{B^2}/2, B_x, B_y, B_z \right]^T.
\end{equation}
In the puerly hydrodynamic limit the code uses the average state of
\citet{roe-rs}.  This averaging guarantees desirable property that
\begin{equation}
\frac{\partial \vec{F}(<\vec{Q}_{L,i\mh},\vec{Q}_{R,i\mh}>)}{\partial
\vec{Q}(\vec{Q}_L-\vec{Q}_R)}=\vec{F}_x(\vec{Q}_L)-\vec{F}_x(\vec{Q}_R)
\end{equation}
which is not true, in general, for the arithmetic average
linearization.  We therefore employ the arithmetic average
linearization of \cite{rj-rs} for MHD applications and revert to the
Roe average linearization only for purely hydrodynamic applications.

In general, the Roe flux provides the least diffusive formulation.
This is because the flux formulations based on the method of
\cite{marquina-flux} revert to the more diffusive local Lax-Freidrichs
upwinding for transonic eigenmodes.  This additional component of
numerical diffusion is is advantageous for simulating certain
astrophysical phenomena.  It is sufficient to prevent the development
of rarefaction shocks without the need to introduce an ``entropy fix''
\citep{harten-hyman}.  It is also sufficient to dampen the development
of carbuncles, even-odd decoupling and related numerical pathologies
\citep{sutherland} that are particularly problematic in grid-aligned
flows which exhibit strong radiative cooling.

\subsection{Temporal Reconstruction} \label{te}
The temporal update operation for the grid centered values is given by
equation \ref{update}.  Replacing $n \to t$ and utilizing any of the
methods for computing the numerical flux in the previous section
achieves an integration that is first order in time.  We implement
four different methods to obtain second order temporal accuracy by
performing this update using time-centered estimates of the numerical
flux.

The MUSCL-Hancock predictor-corrector temporal discretization
achieves second order accuracy by advancing the grid-face interpolated
states by a half-time-step using a one dimensional predictor step.
The predictor step is carried forward according to:
\begin{eqnarray}
\vec{Q}_{L,i\ph}^{t+\Delta t/2} &=& \vec{Q}_{L,i\ph}^{t} +
   \frac{\Delta t}{2 \Delta x} \left(\vec{F}_x(\vec{Q}_{R,i-1}^{t})-\vec{F}_x(\vec{Q}_{L,i+1}^{t})\right) + \frac{\Delta t}{2}\vec{S}(\vec{Q}_i) \nonumber \\
\vec{Q}_{R,i\mh}^{t+\Delta t/2} &=& \vec{Q}_{R,i\mh}^{t} +
   \frac{\Delta t}{2 \Delta x} \left(\vec{F}_x(\vec{Q}_{R,i-1}^{t})-\vec{F}_x(\vec{Q}_{L,i+1}^{t})\right)  + \frac{\Delta t}{2}\vec{S}(\vec{Q}_i). \label{predictor}
\end{eqnarray}
Note that the predictor step uses the cell centered, volume average
fluxes.  The Riemann problem at the cell interfaces is not solved and
the upwinded flux at the cell-faces are not needed for this step.

The MUSCL-Hancock corrector step calls for the construction of a
second order, time centered numerical flux which is computed by
applying any of the upwinding procedures of (\S \ref{upwind}) using
$\vec{Q}_{L,i\mh}^{t+\Delta t/2}$, and $\vec{Q}_{R,\mh}^{t+\Delta
t/2}$ as the left and right states for the Riemann problem at the
$i\mh$ cell interface.  We note the update to the time centered state
described here, use of this procedure in more than one dimension
requires the application of the operator split integrator to retain
the second order accuracy of the scheme for multidimensional flow.
The fully second order accurate update is therefore carried out via
application of equation \ref{splitupdate} with the time centered flux
with n=dt/2

The application of the predictor-corrector schemes like the
MUSCL-Hancock approach described above will in some cases necessitate
the application of a protection procedure to ensure pressure and
density positivity of the predictor interface states:
\begin{eqnarray}
\rho_{L,i\mh}^{t+\Delta t/2} &\leftarrow&
   \textrm{MAX}(\rho_{L,i\mh}^{t+\Delta t/2}, 10^{-2}\textrm{MIN}(\rho_{L,i\mh}^t+\rho_{R,i\mh}^t),10^{-14}) \nonumber \\
\rho_{R,i\mh}^{t+\Delta t/2} &\leftarrow&
   \textrm{MAX}(\rho_{R,i\mh}^{t+\Delta t/2}, 10^{-2}\textrm{MIN}(\rho_{L,i\mh}^t+\rho_{R,i\mh}^t),10^{-14}) \nonumber \\
P_{L,i\mh}^{t+\Delta t/2} &\leftarrow&
   \textrm{MAX}(P_{L,i\mh}^{t+\Delta t/2}, 10^{-4}\textrm{MIN}(P_{L,i\mh}^t+P_{R,i\mh}^t),10^{-14}) \nonumber \\
P_{R,i\mh}^{t+\Delta t/2} &\leftarrow&
   \textrm{MAX}(P_{R,i\mh}^{t+\Delta t/2}, 10^{-4}\textrm{MIN}(P_{L,i\mh}^t+P_{R,i\mh}^t),10^{-14}).
\end{eqnarray}

The second temporal integration option implemented in the code is the
two step Runge-Kutta temporal update operator of \citet{shu-tvdrk}.
In the first step $\vec{\tilde F}_{x,i\mh}^t$ and
$\vec{Q}_{x,i\mh}^{t+\Delta t}$ are computed via the application of a
first order update step given by equation \ref{update} with $n=\Delta
t$ and any operator split microphysical effects (e.g., source terms).
In the second step $\vec{\tilde F}(\vec{Q}_{x,i\mh}^{t+\Delta t})$ is
computed by a second application of the spatial interpolation and
upwinding procedure on the grid of $\vec{Q}_{x,i\mh}^{t+\Delta t}$
data.  The second order, time centered fluxes are then computed via
the interpolation formula,
\begin{equation}
\vec{\tilde F}_{x,i\mh}^{t+\Delta t/2} = \frac{1}{2}\left( \vec{\tilde F}(\vec{Q}_{x,i\mh}^{t}) + 
                                         \vec{\tilde F}(\vec{Q}_{x,i\mh}^{t+\Delta t}) \right).
\label{nft2}
\end{equation}
This flux may be used to carry forward a fully second order, unsplit
update via equation \ref{update}.  The unsplit nature of the
Runge-Kutta time stepping is a significant advantage of the method.
The Runge-Kutta scheme also has a comparative advantage in that it is
pressure positivity-preserving and therefore more robust.  However,
the method entails somewhat greater computational cost due to the
solution of twice the number of Riemann problems per grid cell per
time step as the MUSCL-Hancock approach.  In addition the scheme
suffers a somewhat restrictive time-step stability condition.  The
maximum numerically stable time-step $\Delta t$ is computed in terms of the
Courant condition as
\begin{equation}
\Delta t < MAX[(\frac{a_{m,i\mh,j,k}}{\Delta x})+
     MAX(\frac{a_{m,i,j\mh,k}}{\Delta y})+
     MAX(\frac{a_{m,i,j,k\mh}}{\Delta z})~\textrm{ for all }i,j,k].
\end{equation}
In practice, we estimate the next time-step increment from the maximum
wave speed encountered during the preceding integration sweeps as:
\begin{equation}
dt_{next} = CFL~MAX[(\frac{a_{m,i\mh,j,k}}{\Delta x},
                      \frac{a_{m,i,j\mh,k}}{\Delta y},
                      \frac{a_{m,i,j,k\mh}}{\Delta z})~\textrm{ for all }i,j,k].
\end{equation}
where $CFL$ is a user tunable parameter.  We typically choose, $CFL
\sim 0.8$ for one dimensional calculations and $CFL \sim 0.4$ for
multidimensional problems.  Future revisions of the code will include
the multidimensional corner transport upwind (CTU) reconstruction
method of \cite{colella} and enhancements to this method for the
integration of the MHD equations by \cite{gardiner}.  By explicitly
including the effect of transverse-propagating waves at each grid
interface, the CTU scheme retains numerical stability for larger
time-steps, $CFL<1$, while capturing greater accuracy.

\subsection{Constrained Transport Scheme} \label{ctsec}
While Faraday's law (equation \ref{faraday}) guarantees solenoidality
of $\partial B / \partial t$, and therefore the maintenance of
solenoidal magnetic field topologies, the Gudonov-based conservative
field update procedures described in the previous section do not
provide any such guarantee.  This is because no provision has been
made in the construction of the integration procedure that would
enforce the divergence constraint on the magnetic field.  Each of the
fluxes used in the conservative update procedure, are second order
approximations to the exact area averaged fluxes where
\begin{eqnarray}
\vec{F_x} &= \vec{\tilde F_x} + \vec{O}(\Delta x^3) \nonumber \\
\vec{F_y} &= \vec{\tilde F_y} + \vec{O}(\Delta y^3) \nonumber \\
\vec{F_z} &= \vec{\tilde F_z} + \vec{O}(\Delta z^3).
\end{eqnarray}
Therefore, the divergence of the magnetic field after a conservative
update will also contain high order truncation errors, with
$\vec{\nabla} \cdot \vec{B}= \vec{O}(\Delta x^3) + \vec{O}(\Delta y^3)
+ \vec{O}(\Delta z^3)$.  Local departures from $\vec{\nabla} \cdot
\vec{B} = 0$ usually grow rapidly, causing anomalous magnetic forces
and unphysical plasma transport which eventually destroys the correct
dynamics of the flow \citep{brackbill}.  Two strategies have emerged
for adapting Godunov-based MHD schemes so that the divergence-free
constraint is exactly maintained.  In the first approach a projection
operator is devised, usually by solving a Poisson equation which
removes numerical divergences from the grid
\citep{balsara98,jiang,kim,zachary,rjf}.  Solving the Poisson equation
is somewhat computationally expensive and particularly algorithmically
complex on AMR grid hierarchies.  The second approach utilizes a more
multidimensional approach toward the numerical quadrature of Faraday's
law by utilizing a conservative formulation of Stoke's theorem to
represent magnetic field components at staggered collocation points
\citep{balsara-spicer,dai,rj-ct}.  Following the nomenclature first
commissioned by \cite{evans}, this approach is commonly referred to as
``constrained transport'' (CT) in the literature.  The AstroBEAR code
utilizes the CT approach to maintain a divergence free field,
primarily due to the limitations of the first approach in AMR
applications.  Furthermore, \cite{balsara-comp} have demonstrated the
superiority of the staggered grid approach in the context of a
stringent astrophysically motivated test problem involving the
interplay of strong shocks with radiative cooling.

\begin{figure}[!h]
\begin{center}
\includegraphics[clip=true,width=0.5\textwidth]{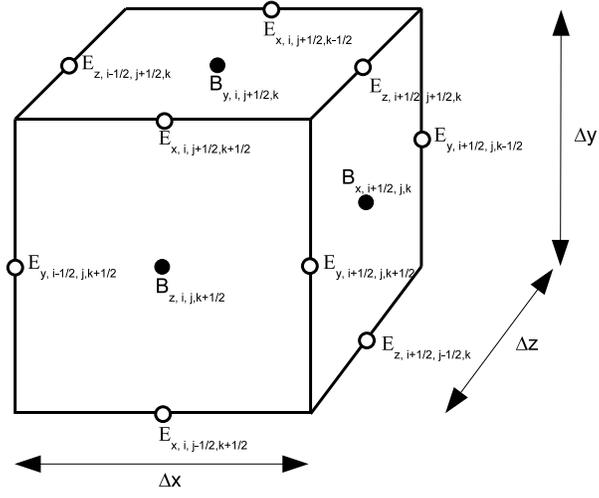}
\end{center}
\caption{ The location of the staggered electric and magnetic field
components on a computational cell centered at position $i, j,
k$. \label{f2}}
\end{figure}

The basis of the constrained transport approach is realized by
applying of Stoke's theorem to Faraday's law and integrating over
each face of a control volume.  This yields an expression for the
face-average normal component of $\partial B / \partial t$ at each
control volume interface.  Spatial discretization of the line integral
around the control volume interfaces calls for the average electric
field parallel to each edge of the control volume.  Temporal
discretization reveals an explicit update procedure for the normal
component of the magnetic field at each computational volume
interface.  The resulting discretized equations,
\begin{eqnarray}
B_{x,i\mh,j,k}^{t+\Delta t} &=& B_{x,i\mh,j,k}^t + \frac{\Delta t}{\Delta y \Delta z} \nonumber \\ 
 &&\left(
         \Delta y E_{y,i\mh,j,k\ph} - \Delta y E_{y,i\mh,j,k\mh} - \right. \nonumber \\
 &&\left. \Delta z E_{z,i\mh,j\ph,k} + \Delta z E_{z,i\mh,j\mh,k} \right) \nonumber \\
B_{y,i,j\mh,k}^{t+\Delta t} &=& B_{y,i,j\mh,k}^t + \frac{\Delta t}{\Delta x \Delta z} \nonumber \\ 
 &&\left(
         \Delta z E_{z,i\ph,j\mh,k} - \Delta z E_{z,i\mh,j\mh,k} - \right. \nonumber \\
 &&\left. \Delta x E_{x,i,j\mh,k\ph} + \Delta x E_{x,i,j\mh,k\mh} \right) \nonumber \\
B_{z,i,j,k\mh}^{t+\Delta t} &=& B_{z,i,j,k\mh}^t + \frac{\Delta t}{\Delta x \Delta y} \nonumber \\ 
 &&\left(
          \Delta x E_{x,i,j\ph,k\mh} - \Delta x E_{x,i,j\mh,k\mh} - \right. \nonumber \\
 && \left. \Delta y E_{y,i\ph,j,k\mh} + \Delta y E_{y,i\mh,j,k\mh} \right). \label{ct}
\end{eqnarray}
gives the desired CT update operator.  Note that the centered
difference discretization of the divergence of the magnetic field,
\begin{equation}
\left(\vec{\nabla \cdot B}\right)_{i,j,k} =  \frac{B_{x,i\ph,j,k} - B_{x,i\mh,j,k}}{\Delta x} + \frac{B_{y,i,j\ph,k} - B_{y,i,j\mh,k}}{\Delta y} + \frac{B_{z,i,j,k\ph} - B_{z,i,j,k\mh}}{\Delta z}
\end{equation}
is preserved 
\begin{equation}
\left(\vec{\nabla \cdot B}\right)_{i,j,k}^{t+dt}= \left(\vec{\nabla \cdot B}\right)_{i,j,k}^{t}
\end{equation}
and the CT update operator maintains a divergence free representation
of the magnetic field provided that the initial magnetic field is
divergence-free.  We emphasize that the CT update procedure requires
that the components of the magnetic field be collated at the center of
the zone faces to which they are orthogonal and that the component of
the electric field parallel to each computational cell interface be
known.  The spatial location of the desired electric and magnetic
field components are illustrated in figure \ref{f2}.
\clearpage
\clearpage

Specification of the CT update procedure is completed via a suitable
construction of the required electric field components at each grid
edge.  The fluxes of the MHD equation (equation \ref{fullmhd}) can be
expressed in terms of the electric field using Ohm's law (equation
\ref{ohm}).  The numerical inter-cell fluxes computed during the
conservative update step described in the preceding subsections,
provide a second order, shock capturing approximation to the
components of the electric field at the center of each computational
cell interface.  The CT update scheme, however, calls for the electric
field at the grid edges.  A reconstruction of the cell-face electric
fields to the cell edges that retains second order accuracy is given
by:
\begin{eqnarray}
E_{x,i,j\mh,k\mh}^{t+\Delta t/2} &=& k
  \left(\tilde f_{z~7,i,j,k\mh}^{t+\Delta t/2} + \tilde f_{z~7,i,j-1,k\mh}^{t+\Delta t/2} - 
  \tilde f_{y~8,i,j\mh,k}^{t+\Delta t/2} - \tilde f_{y~8,i,j\mh,k-1}^{t+\Delta t/2}\right)\nonumber \\
E_{y,i\mh,j,k\mh}^{t+\Delta t/2} &=& k
  \left(\tilde f_{x~8,i\mh,j,k}^{t+\Delta t/2} + \tilde f_{x~8,i\mh,j,k-1}^{t+\Delta t/2} - 
  \tilde f_{z~6,i,j,k\mh}^{t+\Delta t/2} - \tilde f_{z~6,i-1,j,k\mh}^{t+\Delta t/2}\right)\nonumber \\
E_{z,i\mh,j\mh,k}^{t+\Delta t/2} &=& k
  \left(\tilde f_{y~6,i,j\mh,k}^{t+\Delta t/2} + \tilde f_{y~6,i-1,j\mh,k}^{t+\Delta t/2} - 
  \tilde f_{x~7,i\mh,j,k}^{t+\Delta t/2} - \tilde f_{x~7,i\mh,j-1,k}^{t+\Delta t/2}\right), \label{E}
\end{eqnarray}
where the cell-face electric fields have been written in terms of cell
interface fluxes.  Setting $\tilde f_* = \tilde F_*$ and $k=1/4$
recovers the CT scheme of \citet{balsara-spicer}.  Many procedures for
averaging the cell interface flux components to construct the cell-edge
electric field components.  The CT scheme of \citet{rj-ct} can be
expressed in the form of equation \ref{E} by retaining only the
advective part of the inter-cell flux,
\begin{eqnarray}
\vec{f}_{x~6,7,8} = v_x \left[0,B_y,B_z \right]^T \nonumber \\
\vec{f}_{y~6,7,8} = v_y \left[B_x,0,B_z \right]^T \nonumber \\
\vec{f}_{z~6,7,8} = v_z \left[B_x,B_y,0 \right]^T
\end{eqnarray}
and setting $k=1/2$.  The upwinded, time centered, cell-face fluxes
$\tilde f^{t+\Delta t/2}_*$ are constructed from the cell centered
flux components $f^{t}_*$ according to the same procedure given in \S
\ref{upwind}.  This update procedure retains the upwinding abilities
of the conservative update scheme while maintaining a divergence free
solution to the magnetic field via the correspondence of the
components of the magnetic flux with the components of the electric
field called for by the CT update procedure.  Construction of the
electric field from the components of the intercell numerical flux
therefore has the advantage of introducing very little numerical
dissipation into the solution.  Of the schemes tested by \cite{toth},
those that utilize this approach to reconstructing the electric field
at grid edges produce the most accurate results.  One implementation
detail that should be noted is that the update of the components of
the magnetic field collated along the boundary of the computational
domain requires the flux components that are parallel to that boundary
extending into the first row of ghost cells along the boundary. The
conservative update procedure must therefore extend one row of
computational cells into the ghost cell region during integration
sweeps transverse to the boundary even though no conservative update
is applied to the boundary cells.  Extension into the ghost region
ensures that the components of the numerical flux required in the CT
update step are computed.

At the end of each CT update step, the volume centered cell solution
for the magnetic field, computed during the Gudonov update, is
discarded in favor of the solution provided by the CT update.  At this
stage in the algorithm, the cell centered the magnetic field is
recomputed according to the procedure:
\begin{eqnarray}
\tilde B_{x,i,j,k} &=& \frac{1}{2}\left(B_{x,i\ph,j,k}+B_{x,i\mh,j,k}\right) \nonumber \\ 
\tilde B_{y,i,j,k} &=& \frac{1}{2}\left(B_{y,i,j\ph,k}+B_{y,i,j\mh,k}\right) \nonumber \\
\tilde B_{z,i,j,k} &=& \frac{1}{2}\left(B_{z,i,j,k\ph}+B_{z,i,j,k\mh}\right).
\end{eqnarray}
Even though the solution of the magnetic field is advanced in time at
cell interfaces, we retain the grid of volume average magnetic field
components in order to compute the magnetic pressure at the cell
centers.  Also, the volume average magnetic field and total energy are
synchronized with the cell-face magnetic field after every CT update
step in order to preserve the volume averaged thermal energy.
\begin{equation}
\tilde \E_{i,j,k} \gets \E_{i,j,k} - \frac{\vec{B}_{i,j,k}^2}{2} + \frac{\vec{\tilde B}_{i,j,k}^2}{2} \label{efix}
\end{equation}
The total energy is thereby adjusted such that the CT update preserves
thermal energy. This optional step avoids numerical difficulty with
negative thermal pressure that arise in strongly magnetized problems
at the expense of energy conservation.  We view this as an acceptable
trade-off, particularly for astrophysical applications that are not
energy conserving due to radiative energy losses.

Because the normal component of the magnetic field is known at cell
faces, it is not necessary to interpolate these values during the
spatial reconstruction step of \S \ref{sr}.  Recalling that we have
defined the 6th, 7th and 8th components of the primitive field vector
as the x, y and z components of the magnetic field (equation \ref{p}),
we set
\begin{eqnarray}
P_{L,6,~i\mh} &=& P_{R,6,~i\mh} = b_{x,i\mh} \nonumber \\
P_{L,7,~j\mh} &=& P_{R,7,~j\mh} = b_{y,j\mh} \nonumber \\
P_{L,8,~k\mh} &=& P_{R,8,~k\mh} = b_{z,k\mh}
\end{eqnarray}
in lieu of the spatial reconstruction procedure when using an unsplit
scheme (equation \ref{update}).  With a direction split scheme
(equation \ref{splitupdate}) the partial update of the cell-centered
field from prior direction sweeps is interpolated back to the cell
edge and added to the cell edge state during the spatial
reconstruction step (\S sr) in order to retain second order accuracy
with multidimensional flow
\begin{eqnarray}
P_{L,6,~i\mh} &=& P_{R,6,~i\mh} = b_{x,i\mh} + \frac{1}{2} (B_{x,i}^{(\iota-1)}-B_{x,i}^{(0)}+B_{x,i-1}^{(\iota-1)}-B_{x,i-1}^{(0)}) \nonumber \\
P_{L,7,~j\mh} &=& P_{R,7,~j\mh} = b_{y,j\mh} + \frac{1}{2} (B_{y,j}^{(\iota-1)}-B_{y,j}^{(0)}+B_{y,j-1}^{(\iota-1)}-B_{y,j-1}^{(0)}) \nonumber \\
P_{L,8,~k\mh} &=& P_{R,8,~k\mh} = b_{z,k\mh} + \frac{1}{2} (B_{z,k}^{(\iota-1)}-B_{z,k}^{(0)}+B_{z,k-1}^{(\iota-1)}-B_{z,k-1}^{(0)})
\end{eqnarray}
where the superscript $(\iota-1)$ refers to the state after the
previous direction sweep.  While this procedure does maintain the second order
accuracy achieved via direction splitting it does suffer from the flaw
that the reconstructed states are not exactly divergence free.
Instead, the reconstructed states given by MUSCL-Hancock time stepping
procedure described in \S \ref{te} and any direction split scheme
suffers from the undesirable property that,
\begin{equation}
(b_{x,i\mh}^{t+dt/2}-b_{x,i\mh}^{t+dt/2})/dx +
(b_{y,i\mh}^{t+dt/2}-b_{y,i\mh}^{t+dt/2})/dy +
(b_{z,i\mh}^{t+dt/2}-b_{z,i\mh}^{t+dt/2})/dz = 0
\end{equation}
is not maintained exactly.  Also note that the CT update procedure is
applied after each stage of the multistage Runge-Kutta temporal update
procedure such described in \S\ref{te}.

We generally find the divergence error introduced via the split
reconstruction procedure is small and that because we apply a
divergence free CT update for the magnetic field, the effect of this
small error does not grow rapidly.  The effect of this error appears
as inexact evolution of the z-component of the magnetic field in the
magnetic field loop advection test of \cite{gardiner} as shown in \S
\ref{2D}.  \cite{gardiner} introduced a two dimensional unsplit scheme
and later a three dimensional extension thereof \cite{gardiner08}
where they devise an unsplit reconstruction that is not subject to
this error at the expense of considerably increased algorithmic and
computational complexity.  \cite{balsara-geom} has shown that unsplit
MHD schemes are less diffusive in some tests than the dimensionally
split counterpart.  Because of this we recommend the dimensionally
unsplit Runge-Kutta scheme which does provide for exactly divergence
free reconstructed states for AtroBEAR MHD applications.  Never the
less the direction split MUSCL-Hancock update has proven to be
reliable in earlier purely hydrodynamic works \citep{h1,h2,h3,h4}, and
for this reason we leave this option available for MHD applications.
We note that the MUSCL-Hancock scheme is efficient in terms of
computational cost, requiring only one Riemann solver per grid cell
per time step while retaining stability for $CFL < 1$ and that
exhaustive testing of this and other \citep{rj-ct} direction split MHD
schemes have shown good results.

\subsection{Summary or Numerical Methods}
The menu of options available in our code includes three spatial
reconstruction methods \S\ref{sr}: linear (MUSCL), piecewise
hyperbolic (PHM) and piecewise parabolic (PPM), two temporal
reconstruction methods \S\ref{te}: MUSCL-Hancock and Runge-Kutta, four
different Riemann solvers / upwinding procedures \S\ref{upwind}: the
HLLD flux, the Roe flux, the Marquina flux, and an adaptation of the
Marquina flux that is better suited for magnetized flow involving
compound wave structures and two different constrained transport
schemes for preserving the solenoidality constraint on magnetized
flows \S\ref{ctsec}: the method of \cite{balsara-spicer} and the
method of \cite{rjf}.  The user of the code may choose the method for
each of these operations which are summarized in table \ref{t1}.  The
advantage of this approach is that a given simulation may be performed
using several different methods.  The solution strategy which is
optimal for the physical regime of a given simulation may be readily
applied.

\clearpage
\begin{table}[!h] \caption{Numerical Method Options.\label{t1}}
 \begin{tabular}{l|l|l|l}
 Spatial Reconstruction & Temporal Reconstruction & Flux Function & CT Scheme \\
\tableline
 MUSCL & MUSCL-Hancock & Roe              & Ryu et al.       \\
 PHM   & Runge-Kutta   & Marquina         & Balsara \& Spicer \\
 PPH   &               & Adapted Marquina &                   \\
       &               & HLLD             &                   \\
 \end{tabular}
\end{table}
\clearpage
\section{Adaptive Mesh Refinement}\label{AMR}
\begin{figure}[!h]
\begin{center}
\includegraphics[clip=true,width=0.5\textwidth]{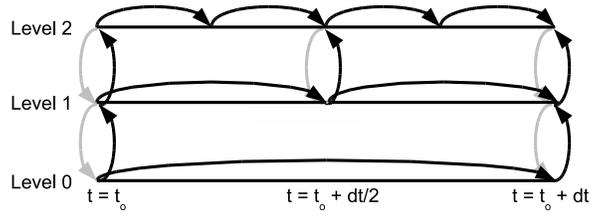}
\end{center}
\caption{A schematic of the update procedure for an AMR hierarchy of
three levels of refinement where each level has a refinement ratio of 2.
Curved horizontal arrows represent integration of all grids on a given
refinement level.  Gray vertical arrows represent restriction of
refined grids to their parent level.  Black vertical arrows represent
prolongation of the solution from the coarse level to its parent
level. \label{f3}}
\end{figure}

The central feature of the BEARCLAW framework on which AstroBEAR is
based is that it provides a framework for Adaptive Mesh refinement
(AMR).  Under AMR, regions of the flow that are susceptible to large
discretization errors are carried forward on a computational grid of
higher resolution while flow features not requiring high resolution
for adequate numerical convergence are carried forward on a
computational grid of lower resolution.  Two approaches to AMR have
emerged: (1) the block-based method of \citet{berger-oliger} and
\citet{berger-colella} which constructs a patchwork of refined grids
that optimally covers all of the cells on the next-coarsest level that
are heuristically identified for refinement (2) an alternative
approach where individual computational cells are refined or derefined
separately \citep{khokhlov}.  Our code employs the former approach.
In this section we give an overview of the AMR algorithm and stages of
the AMR algorithm that require special attention in handling grid face
magnetic field components.  For discussion we use the term ``parent
grid'' to designate an underlying block on the next coarser level,
``parent level'' to designate all of the grid blocks that are one
level coarser, ``child grids'' to designate those grid blocks that are
one level finer and ``child level'' to refer to grids that are
of higher resolution by one refinement ratio than the current grid.
Advancement of an AMR hierarchy of grids is carried out according to
the pseudo-code algorithm given in appendix \ref{amr}.  A schematic of
the update procedure is shown in figure \ref{f3} for an AMR hierarchy
of three levels of refinement where each level has a refinement ratio
of 2.  Curved horizontal arrows represent integration of all grids on
a given refinement level.  The algorithm is adaptive in time, with
each level advanced in time increments $dt_{level} = dt_{level-1}/r$
where $r$ is the refinement ratio of the level.  Gray vertical arrows
represent restriction and refluxing of the solution on refined grids
to their parent level.  Black vertical arrows represent prolongation
of the solution from the coarse level to its parent level.
\clearpage
\clearpage

\begin{figure}[!h]
\begin{center}
\includegraphics[clip=true,width=0.5\textwidth]{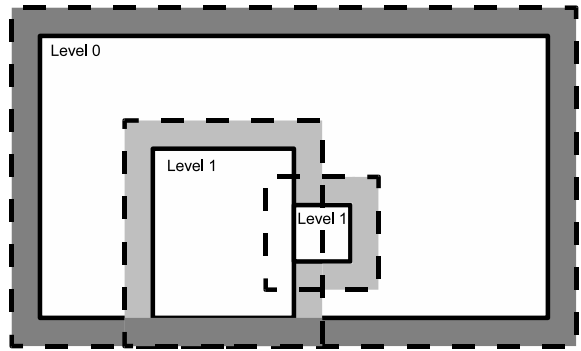}
\end{center}
\caption{An example of an AMR hierarchy containing one grid on the
root level (level=0), and one refined level (level=1).  The interior
of each grid is delineated by solid lines and the extended grid which
include the interior and ghost region of each grid are delineated by
dashed lines.\label{f4}}
\end{figure}

The first step of the update procedure, ``Set Ghost,'' calls for the
initialization of the ghost cells that are exterior to all grids on
the given level.  Figure \ref{f4} shows an example of an AMR hierarchy
containing one grid on the root level (level=0), and one refined level
(level=1).  The interior of each grid is delineated by solid lines and
the extended grid which include the interior and ghost region of each
grid are delineated by dashed lines.  The face-centered magnetic field
components that coincide with boundaries delineating the interior of
each grid are treated as interior cells.  We classify ghost cells into
three categories: 1) same-level ghost cells that coincide with the
interior of another grid on the same level appear white in the figure,
2) physical ghost cells that lie outside of the computational domain
appear dark gray in the figure, 3) child-level ghost cells that
coincide with the interior of grids on the parent level appear light
gray in the figure.  Same-level ghost zones are initialized to the
state of the interior of the coincident grid.  Physical ghost cells
are initialized according to user specified boundary conditions, the
code provides three physical boundary options; constant extrapolation,
reflecting or periodic.  The initialization of parent-level ghost
cells is carried out as a part of the ``Grid Adapt'' procedure which
will be discussed below.
\clearpage
\clearpage

\begin{figure}[!h]
\begin{center}
\includegraphics[clip=true,width=0.5\textwidth]{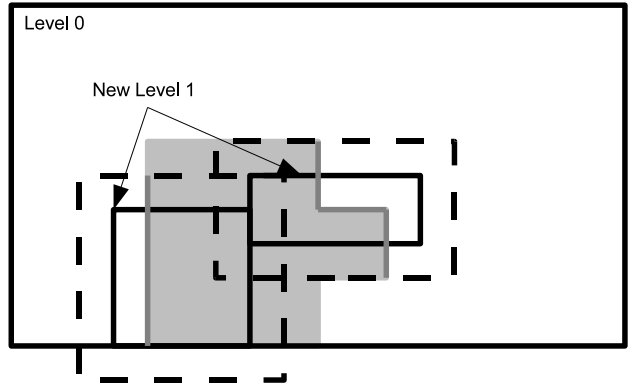}
\end{center}
\caption{The new arrangement of grid patches on the first refinement
level with interior boundaries and ghost boundaries delineated by
solid and dashed lines respectively.  The region coinciding with the
previous arrangement of AMR patches are shaded in gray. \label{f5}}
\end{figure}

The ``Grid Adapt'' procedure determines the arrangement of and
initialization of a new AMR hierarchy on a given level that tracks the
evolution of flow features in the solution.  A user-specified
truncation error estimation procedure is applied to each grid that is
one level coarser.  AstroBEAR uses either the maximum absolute value
of the primitive vector gradient or Richardson extrapolation
\citep{berger-oliger,berger-colella} as the options for error
estimation.  Zones on the parent grids with estimated error greater
than a user-specified tolerance are flagged for refinement.  We employ
the patch-wise clustering algorithm of \citet{berger-adapt} to
determine arrangement of refined grid patches that optimally overlays
all of the zones flagged for refinement.  For easier parallel
implementation, we require that each grid has only one parent.  In
figure \ref{f5}, the new arrangement of grid patches on the first
refinement level with interior boundaries and ghost boundaries
delineated by solid and dashed lines respectively.  The region
coinciding with the previous arrangement of AMR patches are shaded in
gray.  Interior regions of the new grid arrangement that coincide with
interior regions of the previous patchwork of grids on the same level
are initialized by copying the field values from the previous grids.
The previous patchwork of grids on this level are then released from
memory.  The ghost zones, and interior zones that do not coincide with
the interior of the previous grid patches are initialized from the
parent grid via a prolongation operator.  The code prolongs the
cell-centered conserved fields using interpolation from the parent
grid.  The cell-face grid of magnetic field components are initialized
using a divergence-preserving prolongation operator in order to
preserve the integrity of the CT update procedure.  We will discuss
this operator in detail in \S\ref{prolongation}.
\clearpage
\clearpage

\begin{figure}[!h]
\begin{center}
\includegraphics[clip=true,width=0.5\textwidth]{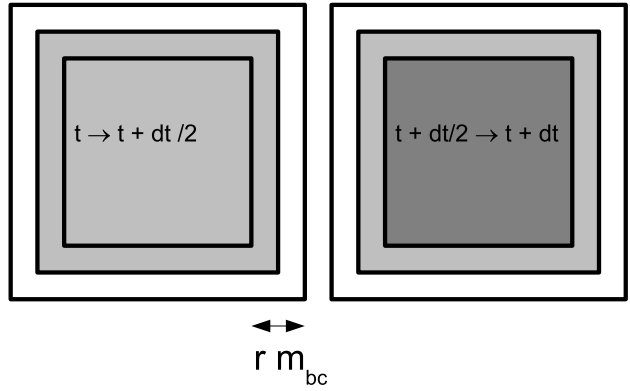}
\end{center}
\caption{A schematic of the arrangement of the extended ghost region
on a grid with a refinement ratio of 2.  The initial representation of
the field is at time $t$.  The first integration cycle, illustrated in
the left panel, carries all cells interior to the outer ring of
$m_{bc}$ ghost zones shaded in dark gray forward in time from $t$ to
$t+dt/2$.  On the second integration cycle, illustrated in the right
panel the interior cells shaded in dark gray are carried forward in
time from $t$ to $t+dt/2$. \label{f6}}
\end{figure}

The ``integrate'' step advances the solution on each grid on the given
level using one of the integration procedures discussed in
\S\ref{method} as specified by the user.  The time-adaptive nature of
the AMR engine imposes the difficulty that boundary information from
parent grids is not available during each step of child grid
integration cycle.  As shown in figure \ref{f3} for a refinement ratio
of two, child ghost zones are temporally synchronized with their
parent grids only every-other time cycle.  To accommodate this, we
incorporate an ``extended'' arrangement of ghost cells.  Each refined
grid carries a strip of ghost zones that extends the width $r \times
m_{bc}$ cells beyond the interior of the grid where $r$ is the
refinement ratio, and $m_{bc}$ is the number of ghost cells required
by the integration stencil.  The MUSCL-Hancock reconstruction operator
(\S\ref{sr}) without direction splitting requires two ghost cells,
yielding $m_{bc}=2$.  An extra ghost zone is required in
multidimensional problems when using direction splitting and
constrained transport in order to compute the flux components
necessary to compute the EMF at the grid corners, yielding $m_{bc}=3$.
The Runge-Kutta temporal integration method (\S\ref{te}) has been
implemented via additional rows of ghost cells to fully update the
interior region with $m_{bc}=4$.  The update of the extended region of
ghost cells is illustrated in figure \ref{f6} for a refinement ratio
$r=2$.  The initial representation of the field is at time $t$.  The
first integration cycle carries all cells interior to the first ring
of $m_{bc}$ ghost zones shaded in dark gray forward in time from $t$
to $t+dt/2$.  On the second integration cycle, the interior cells
shaded in dark gray are carried forward in time from $t$ to $t+dt/2$.
The second integration step can be carried out because the outermost
ring of $m_{bc}$ ghost zones are outside of the domain of influence on
the interior cells.
\clearpage
\clearpage

The ``Synchronize to Parent'' call carries forward the synchronization
of the solution on a given level to its parent grid.  The
synchronization procedure is composed of two steps: application of
coarse to fine level refluxing and restriction of the solution.  In the
refluxing step, a spatio-temporal average of the flux from child grid
interfaces along coarse / fine grid boundaries is compared to the
flux as computed during the integration of the coarse grid.  A
correction is applied to the parent grid so that the effective flux
across this boundary is equal to that as computed on the child grid.
The restriction step calls for the coarsening and injection of field
data from the interior of the child grids into their respective parent
grids.  We employ volume weighted average restriction of cell centered
conserved fields.  The restriction of face-centered magnetic field
components requires special attention to maintain the divergence-free
property of the magnetic field which will be discussed in the next
section.
\section{Divergence Preserving Restriction and Prolongation Operators} \label{prores}
In \S\ref{ctsec} we discussed the importance of satisfying the
$\vec{\nabla} \cdot \vec{B}=0$ in order to maintain the
integrity of the numerical solution to the MHD equations and
demonstrated an adaptation to Goduonov-based schemes that preserves
the divergence of the magnetic field throughout the calculation.  The
use of AMR imposes an additional challenge that must be overcome to
maintain the divergence of the magnetic field throughout the
calculation.  Specifically, the restriction step, which maintains the
consistency on coarse levels with the finer levels of refinement in
the grid, and the prolongation step, which initializes coarse
representations of the solution to finer grids in regions that have
been flagged for refinement must not introduce divergence errors into
the solution.  The volume average and bilinear interpolation
procedures utilized for the restriction and prolongation of the
cell-centered conserved fields cannot be adapted to operate on the
grid-edge magnetic fields in a divergence preserving manner.  Two
approaches to the divergence preserving prolongation have emerged.
\cite{balsara-amr} has generalized the divergence free reconstruction
procedure of \cite{balsara-spicer} to devise a piece-wise quadratic
interpolant that is divergence preserving.  \cite{li} present an
adaptation of Balsara's procedure that simplifies its implementation
for problems involving arbitrary refinement ratios.  \cite{toth-roe}
has devised a prolongation procedure by solving an algebraic system
which enforces the maintenance of the volume average curl and
divergence over a coarse grid cell.

In this subsections that follow, we present the restriction and
prolongation formulae employed in our code.  In particular, we present
the prolongation procedure of \citep{toth-roe} in a less compact form
than that of the original authors which are more readily transcribed
into computer code.  To simplify implementation our code only allows
refinement ratios of two between successive levels of refinement.

\subsection{Restriction}\label{restriction}
\begin{figure}[!h]
\begin{center}
\includegraphics[clip=true,width=0.5\textwidth]{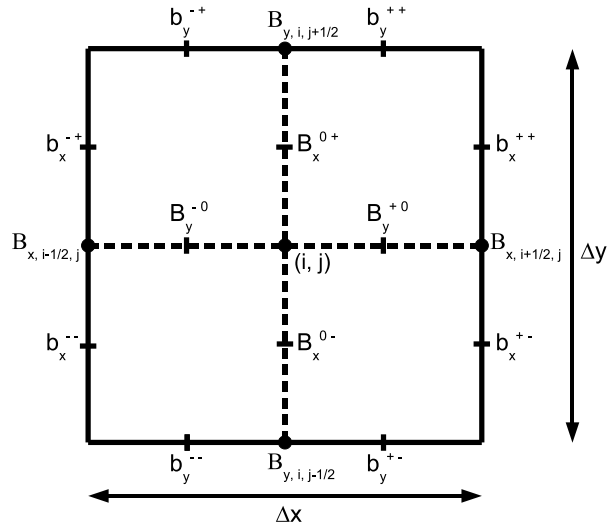}
\end{center}
\caption{Illustration of the locations on a two dimensional grid of
the staggered electric and magnetic field components with a refined
grid, delineated with dotted lines, that has been embedded within one
coarse grid cell, delineated by solid lines. \label{f7}}
\end{figure}

Figure \ref{f7} illustrates the locations of the magnetic field
components utilized by the constrained transport update procedure
where a refined grid (denoted with dotted lines) has been embedded
within one coarse grid cell (denoted with solid lines) for a two
dimensional calculation.  We denote the magnetic field components on
coarse grid faces as $\beta$, on refined cell faces that are
coincident with coarse grid faces (the exterior faces) as $b$, and on
refined cell faces that do not coincide with coarse grid faces (the
interior faces) as $B$.  We have adapted a short hand notation where
the first, second, and third character in the superscript to the
refined grid magnetic field components either represents the location
above (+), below (-) or aligned with (0) the center of the coarse cell
in each of the $x$, $y$, and $z$ directions.
\clearpage
\clearpage
We require that the fine-to-coarse grid synchronization step maintain
the cell interface magnetic field in an area-averaged sense,
\begin{eqnarray}
\beta_{x,i\pm1/2,j} &=& \frac{1}{2} \left(b_{x}^{\pm+} + b_{x}^{\pm-} \right) \nonumber \\
\beta_{y,i,j\pm1/2} &=& \frac{1}{2} \left(b_{y}^{+\pm} + b_{y}^{-\pm} \right) \label{res2}
\end{eqnarray}
in two dimensions, and 
\begin{eqnarray}
\beta_{x,i\pm1/2,j,k} &=& \frac{1}{4} \left(b_{x}^{\pm++} + b_{x}^{\pm-+} + b_{x}^{\pm+-} + b_{x}^{\pm--}\right) \nonumber \\
\beta_{y,i,j\pm1/2,k} &=& \frac{1}{4} \left(b_{y}^{+\pm+} + b_{y}^{-\pm+} + b_{y}^{+\pm-} + b_{y}^{-\pm-}\right) \nonumber \\
\beta_{z,i,j,k\pm1/2} &=& \frac{1}{4} \left(b_{z}^{++\pm} + b_{z}^{-+\pm} + b_{z}^{+-\pm} + b_{z}^{--\pm}\right) \label{res3}
\end{eqnarray}
in three dimensions.
Equations \ref{res2} \& \ref{res3} must be satisfied while preserving
the divergence of the magnetic field along coarse/fine grid
boundaries.  Simultaneous satisfaction of these properties along
coarse/fine grid interfaces and this restriction condition is
accomplished by applying a suitable restriction of the electric field
from fine to coarse grids before performing the CT update (equations
\ref{ct}) on the coarse level.  Manipulation of equations \ref{E}
subject to the constraint provided by equations \ref{res2}
\&\ref{res3} at all times yield the desired electric field restriction
operator as
\begin{eqnarray}
E_{z,i\mh,j\mh}^{t+dt/2} &= \frac{1}{2}(e_{z,i\mh,j\mh}^{t+dt/4} + e_{z,i\mh,j\mh}^{t+3dt/4}), \label{eres2}
\end{eqnarray}
in two dimensions and
\begin{eqnarray}
E_{x,i,j\mh,k\mh}^{t+dt/2} &= \frac{1}{4} (e_{x,i-1/4,j\mh,k\mh}^{t+dt/4} + e_{x,i-1/4,j\mh,k\mh}^{t+3dt/4} +
                                           e_{x,i+1/4,j\mh,k\mh}^{t+dt/4} + e_{x,i+1/4,j\mh,k\mh}^{t+3dt/4})  \nonumber \\
E_{y,i\mh,j,k\mh}^{t+dt/2} &= \frac{1}{4} (e_{y,i\mh,j-1/4,k\mh}^{t+dt/4} + e_{y,i\mh,j-1/4,k\mh}^{t+3dt/4} +
                                           e_{y,i\mh,j+1/4,k\mh}^{t+dt/4} + e_{y,i\mh,j+1/4,k\mh}^{t+3dt/4})  \nonumber \\
E_{z,i\mh,j\mh,k}^{t+dt/2} &= \frac{1}{4} (e_{z,i\mh,j\mh,k-1/4}^{t+dt/4} + e_{z,i\mh,j\mh,k-1/4}^{t+3dt/4} +
                                           e_{z,i\mh,j\mh,k+1/4}^{t+dt/4} + e_{z,i\mh,j\mh,k+1/4}^{t+3dt/4}),  \label{eres3}
\end{eqnarray}
in three dimensions where $\vec{E}$ is the electric field on the
coarse grid and $\vec{e}$ is electric field of the refined grid.  In
figure \ref{f3} note that level 0 is advanced by an increment from $t$
to $t+dt/2$ using the electric field computed at time $t+dt/2$ in one
step while the next child level, level 1, is integrated by the same
increment in two steps using the electric field at time $t+dt/4$ to
integrate from $t$ to $t+dt/2$ and electric field at time $t+3dt/4$ to
integrate from $t+dt/2$ to $t+dt$.  The temporal averaging of the
electric field is necessary due to the temporal refinement capability
of the code.  The time averaging in equations \ref{eres2} \&
\ref{eres3} ensures that evolution of the magnetic field across level
boundaries remains divergence free and consistent in the sense of
equations \ref{res2} \& \ref{res3}.

\subsection{Prolongation} \label{prolongation}
The prolongation step initializes a newly refined grid from its parent
grid that is coarser by one level of refinement.  The prolongation of
the face centered magnetic field is carried out in two steps.  In the
first stage, the exterior faces that coincide with the edges of an
already refined grid block are set by copying the field values from
the coincident face of the already refined grid.  The exterior faces
that do not coincide with the edges of an already refined region are
computed via a bilinear interpolation of the coarse representation of
the field given by
\begin{eqnarray}
b_x^{\pm+} &=& \beta_{x,i\pmh,j} + \frac{1}{2} \delta_y \beta_x \nonumber \\
b_x^{\pm-} &=& \beta_{x,i\pmh,j} - \frac{1}{2} \delta_y \beta_x \nonumber \\
b_y^{+\pm} &=& \beta_{y,i,j\pmh} + \frac{1}{2} \delta_x \beta_y \nonumber \\
b_y^{-\pm} &=& \beta_{y,i,j\pmh} - \frac{1}{2} \delta_x \beta_y,
\end{eqnarray}
in two dimensions and
\begin{eqnarray}
b_x^{\pm++} &=& \beta_{x,i\pmh,j,k} + \frac{1}{2} (~~\delta_y \beta_x + \delta_z \beta_x) \nonumber \\
b_x^{\pm+-} &=& \beta_{x,i\pmh,j,k} + \frac{1}{2} (~~\delta_y \beta_x - \delta_z \beta_x) \nonumber \\ 
b_x^{\pm-+} &=& \beta_{x,i\pmh,j,k} + \frac{1}{2} (-\delta_y \beta_x + \delta_z \beta_x) \nonumber \\
b_x^{\pm--} &=& \beta_{x,i\pmh,j,k} + \frac{1}{2} (-\delta_y \beta_x - \delta_z \beta_x) \nonumber \\
b_y^{+\pm+} &=& \beta_{y,i,j\pmh,k} + \frac{1}{2} (~~\delta_x \beta_y + \delta_z \beta_y) \nonumber \\
b_y^{+\pm-} &=& \beta_{y,i,j\pmh,k} + \frac{1}{2} (~~\delta_x \beta_y - \delta_z \beta_y) \nonumber \\ 
b_y^{-\pm+} &=& \beta_{y,i,j\pmh,k} + \frac{1}{2} (-\delta_x \beta_y - \delta_z \beta_y) \nonumber \\
b_y^{-\pm-} &=& \beta_{y,i,j\pmh,k} + \frac{1}{2} (-\delta_x \beta_y - \delta_z \beta_y) \nonumber \\
b_z^{++\pm} &=& \beta_{z,i,j,k\pmh} + \frac{1}{2} (~~\delta_x \beta_z + \delta_y \beta_z) \nonumber \\
b_z^{+-\pm} &=& \beta_{z,i,j,k\pmh} + \frac{1}{2} (~~\delta_x \beta_z - \delta_y \beta_z) \nonumber \\ 
b_z^{-+\pm} &=& \beta_{z,i,j,k\pmh} + \frac{1}{2} (-\delta_x \beta_z + \delta_y \beta_z) \nonumber \\
b_z^{--\pm} &=& \beta_{z,i,j,k\pmh} + \frac{1}{2} (-\delta_x \beta_z - \delta_y \beta_z),
\end{eqnarray}
in three dimensions, where we compute the spatial jumps using the same
slope limiter (equations \ref{minmod}-\ref{mc})as the base scheme:
\begin{eqnarray}
\delta_x \beta_{*} &=& \textrm{LIMITER}(\beta_{*,i+1,j\mh,k\mh} - \beta_{*,i,j\mh,k\mh},~\beta_{*,i,j\mh,k\mh} - \beta_{*,i-1,j\mh,k\mh}) \nonumber \\
\delta_y \beta_{*} &=& \textrm{LIMITER}(\beta_{*,i\mh,j+1,k\mh} - \beta_{*,i\mh,j,k\mh},~\beta_{*,i\mh,j,k\mh} - \beta_{*,i\mh,j-1,k\mh}) \nonumber \\
\delta_z \beta_{*} &=& \textrm{LIMITER}(\beta_{*,i\mh,j\mh,k+1} - \beta_{*,i\mh,j\mh,k},~\beta_{*,i\mh,j\mh,k} - \beta_{*,i\mh,j\mh,k-1}).
\end{eqnarray}

In the second stage of the prolongation procedure we interpolate from
the exterior faces of the refined grid cell to the interior faces.  In
three dimensions, the 12 refined inter-cell faces that are interior to
the coarse cell are constructed via an interpolation from the magnetic
field components collated at the 24 refined grid exterior face centers
that coincide with a coarse cell interface.  Following \cite{toth-roe}
we construct this interpolation to satisfy the constraint that it be
divergence and curl preserving.  For readability and conciseness, we
illustrate the steps required to derive such an interpolation
procedure in detail only for the two dimensional case.  We then
present the analogous solution in three dimensions.  The volume
average divergence, computed via the application of Gauss's law around
the perimeter of the exterior faces is:
\begin{equation}
d^{00}=\frac{1}{2} (\bar{b}_x^{++}+\bar{b}_x^{+-}-\bar{b}_x^{-+}-\bar{b}_x^{--}+\bar{b}_y^{++}+\bar{b}_y^{-+}-\bar{b}_y^{+-}-\bar{b}_y^{--}) \label{begin2d}
\end{equation}
where we have used the bar accent to denote division by the
discretization width in the normal direction on the coarse level (see figure \ref{f7}), i.e.
$\bar{b}_x=b_x/\Delta x$, and $\bar{b}_y=b_y/\Delta y$. The curl of the field is
constructed by bilinear interpolation of the requisite exterior field
components to the origin of the refined cell as:
\begin{equation}
c_z^{00}=\frac{\Delta y}{2 \Delta x}(\bar{b}_y^{++}-\bar{b}_y^{-+}+\bar{b}_y^{+-}-\bar{b}_y^{--})-\frac{\Delta x}{\Delta y}(\bar{b}_x^{++}-\bar{b}_x^{+-}+\bar{b}_x^{-+}-\bar{b}_x^{--}).
\end{equation}
The divergence centered at each of the four refined cell interiors is:
\begin{eqnarray}
D^{--}&=2 (\bar{B}_x^{0-}-\bar{b}_x^{--}+\bar{B}_y^{-0}-\bar{b}_y^{--}) \nonumber \\
D^{+-}&=2 (\bar{b}_x^{+-}-\bar{B}_x^{0-}+\bar{B}_y^{+0}-\bar{b}_y^{+-}) \nonumber \\
D^{-+}&=2 (\bar{B}_x^{0+}-\bar{b}_x^{-+}+\bar{b}_y^{-+}-\bar{B}_y^{-0}) \nonumber \\
D^{++}&=2 (\bar{b}_x^{++}-\bar{B}_x^{0+}+\bar{b}_y^{++}-\bar{B}_y^{+0}) \label{d2d}
\end{eqnarray}
and the curl at the origin of the refinement cells implied by the
interior field values is:
\begin{equation}
C_z^{00}=\frac{2 \Delta y}{\Delta x}(\bar{B}_y^{+0}-\bar{B}_y^{-0})-\frac{2 \Delta x}{\Delta y}(\bar{B}_x^{0+}-\bar{B}_x^{0-}). \label{end2d}
\end{equation}
The desired interpolation procedure is determined by imposing the
condition that each of the refined grid cells contributes equally to
the divergence of the refinement region,
$d^{00}=D^{--}=D^{-+}=D^{+-}=D^{++}$, and that the curl implied by the
interior faces equals the curl interpolated from the exterior faces,
$c_z^{00}=C_z^{00}$.  Because only three of four divergence conditions
are linearly independent, we have a system of four independent linear
equations for the four desired interior field values.  The solution to
the system, in matrix form, is:
\begin{eqnarray}
  \vec{\bar{b}_{ext}} &=& \left[ \begin{array}{cccccccc} \bar{b}_x^{--} & \bar{b}_x^{-+} & \bar{b}_x^{+-} 
	 & \bar{b}_x^{++} & \bar{b}_y^{--} & \bar{b}_y^{-+} & \bar{b}_y^{+-} & \bar{b}_y^{++} \end{array}\right]^T \nonumber \\
  \vec{\bar{B}_{int}} &=& \left[ \begin{array}{cccc} \bar{B}_x^{0-} & \bar{B}_x^{0+} & \bar{B}_y^{-0} & \bar{B}_y^{+0} \end{array}\right]^T \nonumber \\
  \vec{[A]} &=& \frac{1}{4}\left[ \begin{array}{cccccccc}
	 2 & 0 & 2 & 0 & 1 & -1 & -1 & 1 \\
	 0 & 2 & 0 & 2 & 1 & -1 & -1 & 1 \\
	 1 & -1 & -1 & 1 & 2 & 2 & 0 & 0 \\
	 1 & -1 & -1 & 1 & 0 & 0 & 2 & 2
    \end{array}\right] \nonumber \\
  \vec{\bar{B}_{int}} &=& \vec{[A]}\vec{\bar{b}_{ext}}. \label{pro}
\end{eqnarray}

In three dimensions, we follow the same procedure to derive the
solution for twelve interior refined cell face fields from 24 exterior
refined cell face fields.  In particular we have eight equations
(seven linearly independent) for the volume integrated divergence over
each of eight refined cell interiors,
$d^{000}=D^{---}=D^{--+}=D^{-+-}=D^{-++}=D^{+--}=D^{+-+}=D^{++-}=D^{+++}$,
and six equations (five linearly independent) by computing the three
components of the curl, each at two positions relative to the center
of the coarse cell, $c_x^{000}=C_x|_{-dx/4}=C_x|_{+dx/4}$,
$c_y^{000}=C_y|_{-dy/4}=C_y|_{+dy/4}$, and
$c_z^{000}=C_z|_{-dz/4}=C_z|_{+dz/4}$.  The solution of this system
for the desired interior field values yields the desired prolongation
procedure which can be written in the form of equation \ref{pro} as:
\begin{eqnarray}
  \vec{\bar{b}_{ext}} =& \left[ \begin{array}{cccccccc}
	 \bar{b}_x^{---} & \bar{b}_x^{--+} & \bar{b}_x^{-+-} & \bar{b}_x^{-++} & \bar{b}_x^{+--} & \bar{b}_x^{+-+} & \bar{b}_x^{++-} & \bar{b}_x^{+++}
	 \end{array} \right. \nonumber \\
      &\left. \begin{array}{cccccccc}
	 \bar{b}_y^{---} & \bar{b}_y^{--+} & \bar{b}_y^{-+-} & \bar{b}_y^{-++} & \bar{b}_y^{+--} & \bar{b}_y^{+-+} & \bar{b}_y^{++-} & \bar{b}_y^{+++}
	 \end{array} \right. \nonumber \\
      &\left. \begin{array}{cccccccc}
	 \bar{b}_z^{---} & \bar{b}_z^{--+} & \bar{b}_z^{-+-} & \bar{b}_z^{-++} & \bar{b}_z^{+--} & \bar{b}_z^{+-+} & \bar{b}_z^{++-} & \bar{b}_z^{+++}
    \end{array} \right]^T \nonumber \\
  \vec{\bar{B}_{int}} =& \left[ \begin{array}{cccccc} \bar{B}_x^{0--} & \bar{B}_x^{0-+} & \bar{B}_x^{0+-} & \bar{B}_x^{0++} & \bar{B}_y^{-0-} & \bar{B}_y^{-0+} 
    \end{array} \right. \nonumber \\
    & \left. \begin{array}{cccccc} 
	 \bar{B}_y^{+0-} & \bar{B}_y^{+0+} & \bar{B}_z^{--0} & \bar{B}_z^{-+0} & \bar{B}_z^{+-0} & \bar{B}_z^{++0} 
    \end{array}\right]^T \nonumber \\
%
%
%
%
  \vec{[A]} &= \frac{1}{16}\left[ \fontsize{8}{8}{\begin{array}{c@{\hspace{8pt}}c@{\hspace{8pt}}c@{\hspace{8pt}}c@{\hspace{8pt}}c@{\hspace{8pt}}c@{\hspace{8pt}}c@{\hspace{8pt}}c@{\hspace{8pt}}c@{\hspace{8pt}}c@{\hspace{8pt}}c@{\hspace{8pt}}c@{\hspace{8pt}}c@{\hspace{8pt}}c@{\hspace{8pt}}c@{\hspace{8pt}}c@{\hspace{8pt}}c@{\hspace{8pt}}c@{\hspace{8pt}}c@{\hspace{8pt}}c@{\hspace{8pt}}c@{\hspace{8pt}}c@{\hspace{8pt}}c@{\hspace{8pt}}c}
	   8 & 0 & 0 & 0 & 8 & 0 & 0 & 0 & 3 & 1 & -3 & -1 & -3 & -1 & 3 & 1 & 3 & -3 & 1 & -1 & -3 & 3 & -1 & 1 \\
	   0 & 8 & 0 & 0 & 0 & 8 & 0 & 0 & 1 & 3 & -1 & -3 & -1 & -3 & 1 & 3 & 3 & -3 & 1 & -1 & -3 & 3 & -1 & 1 \\
	   0 & 0 & 8 & 0 & 0 & 0 & 8 & 0 & 3 & 1 & -3 & -1 & -3 & -1 & 3 & 1 & 1 & -1 & 3 & -3 & -1 & 1 & -3 & 3 \\
	   0 & 0 & 0 & 8 & 0 & 0 & 0 & 8 & 1 & 3 & -1 & -3 & -1 & -3 & 1 & 3 & 1 & -1 & 3 & -3 & -1 & 1 & -3 & 3 \\
	   3 & 1 & -3 & -1 & -3 & -1 & 3 & 1 & 8 & 0 & 8 & 0 & 0 & 0 & 0 & 0 & 3 & -3 & -3 & 3 & 1 & -1 & -1 & 1 \\
	   1 & 3 & -1 & -3 & -1 & -3 & 1 & 3 & 0 & 8 & 0 & 8 & 0 & 0 & 0 & 0 & 3 & -3 & -3 & 3 & 1 & -1 & -1 & 1 \\
	   3 & 1 & -3 & -1 & -3 & -1 & 3 & 1 & 0 & 0 & 0 & 0 & 8 & 0 & 8 & 0 & 1 & -1 & -1 & 1 & 3 & -3 & -3 & 3 \\
	   1 & 3 & -1 & -3 & -1 & -3 & 1 & 3 & 0 & 0 & 0 & 0 & 0 & 8 & 0 & 8 & 1 & -1 & -1 & 1 & 3 & -3 & -3 & 3 \\
	   3 & -3 & 1 & -1 & -3 & 3 & -1 & 1 & 3 & -3 & -3 & 3 & 1 & -1 & -1 & 1 & 8 & 8 & 0 & 0 & 0 & 0 & 0 & 0 \\
	   1 & -1 & 3 & -3 & -1 & 1 & -3 & 3 & 3 & -3 & -3 & 3 & 1 & -1 & -1 & 1 & 0 & 0 & 8 & 8 & 0 & 0 & 0 & 0 \\
	   3 & -3 & 1 & -1 & -3 & 3 & -1 & 1 & 1 & -1 & -1 & 1 & 3 & -3 & -3 & 3 & 0 & 0 & 0 & 0 & 8 & 8 & 0 & 0 \\
	   1 & -1 & 3 & -3 & -1 & 1 & -3 & 3 & 1 & -1 & -1 & 1 & 3 & -3 & -3 & 3 & 0 & 0 & 0 & 0 & 0 & 0 & 8 & 8
    \end{array}}\right].
\end{eqnarray}

\section{Numerical Tests, Examples \& Discussion} \label{tests}
The AstroBEAR numerical schemes have been tested against a suite of
test problems, including the one dimensional tests of \cite{rj-rs} and
the two dimensional tests of \cite{rjf}.  Except for the failure of
the original Marquina flux formulation to converge for problems
involving compound MHD wave structures as discussed in \S\ref{upwind},
each of the methods recover results that are equivalent to those of
earlier researchers barring minor differences due to different levels
of numerical truncation error.  In the remainder of this section we
present the results of some of these tests.  The results of a few
popular hydrodynamic shocktubes are presented which illustrate the
differences between and limitations of each of the reconstruction and
upwinding methods.  In these cases we provide commentary which is
intended to provide guidance as to the optimal choice of methods given
the expected physical regime of a particular simulation.  We also
reproduce the results of several of the two dimensional tests of
\cite{rjf} using AMR.  These tests demonstrate the success of the
divergence preserving restriction and prolongation operators presented
in \S\ref{prores}.

\subsection{One Dimension}
\begin{figure}[!h]
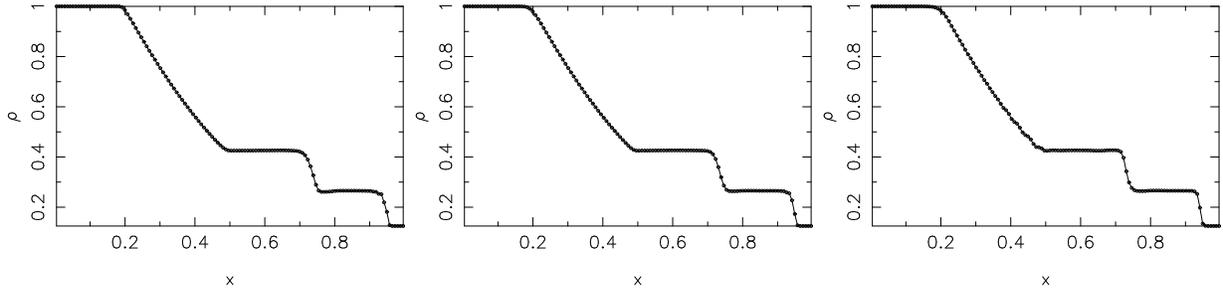

\begin{center}
    \includegraphics[angle=-90,clip=true,width=0.32\textwidth]{f8a.ps}
    \includegraphics[angle=-90,clip=true,width=0.32\textwidth]{f8b.ps}
    \includegraphics[angle=-90,clip=true,width=0.32\textwidth]{f8c.ps}
\end{center}
\caption{Result of the \cite{sod} shocktube (test problem 1 of
  \cite{toro}) using MUSCL/VL slope limiter (left), PHM (center) and
  PPM (right) spatial reconstruction.\label{f8}}
\end{figure}

Figure \ref{f8} shows the density resulting from the \cite{sod} shock
tube problem denoted as ``test problem 1'' in the book by \cite{toro}
using several spatial reconstruction methods.  The initial left and
right Riemann problem states, number of computational zones and final
time of the solution are presented in table \ref{t8}. In all three
cases the Runge-Kutta temporal integration and the flux formulation of
Roe was used.  We note that the result of this test is not very
sensitive to the choice of temporal reconstruction method or flux
formulations.  The shock tube consists from left to right of a
backward propagating rarefaction, a contact discontinuity, and a
forward propagating shock wave.  The numerical diffusion caused by the
truncation error of each of the spatial reconstruction methods is
readily apparent in the figure.  We characterize the diffusion of the
method by the width of the contact discontinuity.  The three methods
are presented in order of increasing diffusion.  The least diffusive
method, PPM resolves the contact across three zones, and the most
diffusive method, MUSCL resolves the contact across 8 zones.  We note
that the diffusion of the PPM and MUSCL methods may be improved by
choosing a more compressive slope limiter than the MINMOD limiter used
here at the expense of introducing oscillations near sharp
discontinuities in the flow.  Naturally, those methods which exhibit
the least numerical diffusion are the most accurate.  However, the PPM
and PHM methods tend to ``overshoot'' the solution near sharp
discontinuities, thereby introducing small oscillations into the
solution.  Such oscillations may drive numerical pathologies in
simulations of flow with very low pressure near discontinuities such
as persistently negative pressures.  We note that some researchers
have managed to reduce such oscillations by introducing additional
sources of diffusion to the base scheme (see, for example \S B1 of
\cite{mignone}).  Simulation of astrophysical phenomena involving
strong radiative cooling is particularly susceptible to this problem.
In such cases the optimal solution strategy is determined as a
trade-off between the desire for simultaneously low numerical
diffusion and low oscillation.
\clearpage
\clearpage
\begin{table}[!h] \caption{Sod Shocktube Parameters.\label{t8}}
 \begin{tabular}{l | l l}
 & left, $x<0.5$ & right, $x \ge 0.5$\\
 \tableline
$\rho$ & 1 & 0.125      \\
$v_x$  & 0 & 0          \\
$P$    & 1 & 0.1        \\
$\gamma$ & 1.4 & 1.4    \\
\tableline
grid zones & 128 &      \\
CFL & 0.9 (0.5 for PPM)&             \\
$t_{final}$ & 0.25 &    \\
 \tableline
 \end{tabular}
\end{table}
\clearpage

\begin{figure}[!h]
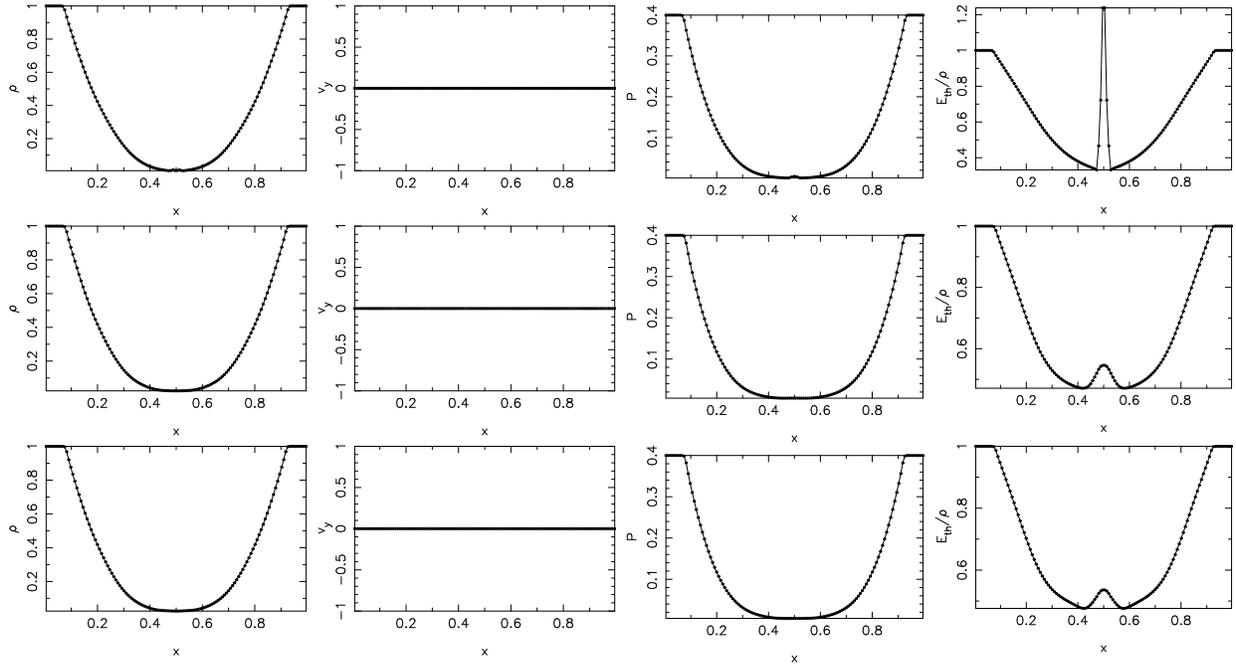

\begin{center}
   \includegraphics[angle=-90,clip=true,width=0.24\linewidth]{f9a.ps}
   \includegraphics[angle=-90,clip=true,width=0.24\linewidth]{f9b.ps}
   \includegraphics[angle=-90,clip=true,width=0.24\linewidth]{f9c.ps}
   \includegraphics[angle=-90,clip=true,width=0.24\linewidth]{f9d.ps} \\
   \includegraphics[angle=-90,clip=true,width=0.24\linewidth]{f9e.ps}
   \includegraphics[angle=-90,clip=true,width=0.24\linewidth]{f9f.ps}
   \includegraphics[angle=-90,clip=true,width=0.24\linewidth]{f9g.ps}
   \includegraphics[angle=-90,clip=true,width=0.24\linewidth]{f9h.ps} \\
   \includegraphics[angle=-90,clip=true,width=0.24\linewidth]{f9i.ps}
   \includegraphics[angle=-90,clip=true,width=0.24\linewidth]{f9j.ps}
   \includegraphics[angle=-90,clip=true,width=0.24\linewidth]{f9k.ps}
   \includegraphics[angle=-90,clip=true,width=0.24\linewidth]{f9l.ps}
\caption{Results of the \cite{einfeldt} ``1-2-0-3'' strong
rarefaction shocktube (test problem 2 of \cite{toro}) using the Roe
flux, the Marquina flux, and the adapted Marquina flux from top to
bottom.  \label{f9}}
\end{center}
\end{figure}

Figure \ref{f9} shows the result of the ``1-2-0-3'' strong expansion
shock tube of \cite{einfeldt}, denoted as ``test problem 2'' in the
book by \cite{toro}.  The initial left and right Riemann problem
states, number of computational zones and final time of the solution
is presented in table \ref{t9}.  This shock tube launches two
rarefaction waves, one propagating to the left and one propagating to
the right.  The upper three panels show the test results utilizing the
MUSCL - VL slope spatial and Ruge-Kutta temporal reconstruction
methods with different flux upwinding procedures.  This test problem was designed
to illustrate the failure mode common to linearized Riemann solvers
evident by the anomalous oscillation in velocity and anomalous rise
in specific thermal energy($E_{th}/\rho$) about the center of the grid
in panel 9a.  These anomalies are caused by the addition of small
amounts of thermal energy to the solution in regions where the
pressure becomes negative.  We note that specific thermal energy
anomalies such as these can result in a cascade of numerical
pathologies in multi-physics simulations involving temperature
dependent micro-physics.  \cite{einfeldt} showed that the more
diffusive HLLE Riemann solver, under certain conditions, guarantees
pressure positivity and therefore reduces such anomalies.
\cite{gardiner} have also demonstrated success applying the HLLE
solver only in those regions where the linearized solver produces
negative pressure or density.  In panel 9b we show that application of
the Marquina flux formulation also resolves the anomalous behavior.
In \S\ref{upwind} we presented an adaptation of the Marquina flux
formulation which is better suited to magnetized flow involving
compound wave structures.  The result of this calculation utilizing
the adapted flux formulation (panel 9c) shows that the adaptation
retains the desirable features of the Marquina flux for this test
problem.
\clearpage
\clearpage
\begin{table}[!h] \caption{Einfeldt Shocktube Parameters.\label{t9}}
 \begin{tabular}{l | l l}
 & left, $x<0.5$ & right, $x \ge 0.5$\\
 \tableline
$\rho$ & 1   & 1        \\
$v_x$  & -2  & 2        \\
$P$    & 0.4 & 0.4      \\
$\gamma$ & 1.4 & 1.4    \\
\tableline
grid zones & 128 &      \\
CFL & 0.8 &             \\
$t_{final}$ & 0.15 &    \\
 \tableline
 \end{tabular}
\end{table}
\clearpage

In figure \ref{f9} we reproduce the MHD shock tube test from figure 2a
of \cite{rj-rs}.  Following the notation of \cite{rj-rs}, we denote
the orientation angle of the magnetic field, $\Psi =
\tan^{-1}(B_z/B_y)$. The initial left and right Riemann problem
states, number of computational zones and final time of the solution
is presented in table \ref{t9}.  This Riemann problem demonstrates the
ability of the code to correctly capture a number of MHD
discontinuities.  The discontinuities in the flow from left to right
are, (1) fast shock, (2) rotational discontinuity, (3) slow shock, (4)
contact discontinuity, (5) slow shock, (6) rotational discontinuity,
and (7) fast shock.  We note that the PHM spatial reconstruction
employed for this test generally results in a level of truncation
error slightly lower than that of \cite{rj-rs} with a slightly more
diffusive result about of rotational discontinuities due to the lack
of a rotational discontinuity steepening procedure like that of
equations 2.96 - 2.98 of \cite{rj-rs}.  We note that when applied to
this particular problem, the MUSCL and PPM spatial reconstruction
methods result in lower and higher truncation error respectively, in a
manner that is similar to Sod problem discussed earlier.
\clearpage
\begin{figure}[!h]
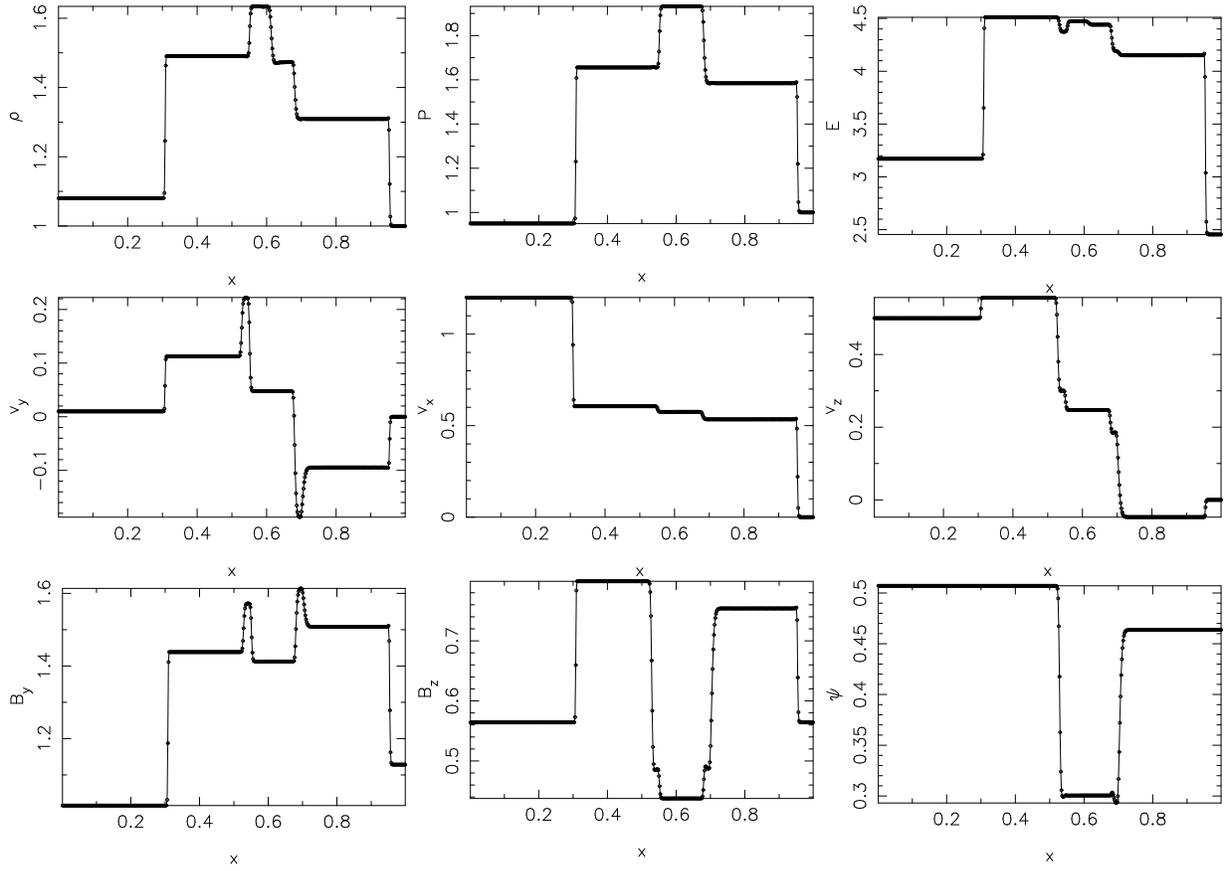

\begin{center}
   \includegraphics[angle=-90,clip=true,width=0.32\textwidth]{f10a.ps}
   \includegraphics[angle=-90,clip=true,width=0.32\textwidth]{f10b.ps}
   \includegraphics[angle=-90,clip=true,width=0.32\textwidth]{f10c.ps} \\
   \includegraphics[angle=-90,clip=true,width=0.32\textwidth]{f10d.ps}
   \includegraphics[angle=-90,clip=true,width=0.32\textwidth]{f10e.ps}
   \includegraphics[angle=-90,clip=true,width=0.32\textwidth]{f10f.ps} \\
   \includegraphics[angle=-90,clip=true,width=0.32\textwidth]{f10g.ps}
   \includegraphics[angle=-90,clip=true,width=0.32\textwidth]{f10h.ps}
   \includegraphics[angle=-90,clip=true,width=0.32\textwidth]{f10i.ps}
\end{center}
\caption{MHD shocktube result using PHM spatial reconstruction,
MUSCL-Hancock temporal reconstruction, and the Roe flux
function. \label{f10}}
\end{figure}
\clearpage
\begin{table}[!h] \caption{MHD Shocktube Parameters.\label{t10}}
 \begin{tabular}{l | l l}
 & left, $x<0.5$ & right, $x \ge 0.5$\\
 \tableline
$\rho$ & 1.08 & 1        \\
$v_x$  & 1.2  & 0        \\
$v_y$  & 0.01 & 0        \\
$v_z$  & 0.5  & 0        \\
$P$    & 0.95 & 1        \\
$B_x$  & $2/\sqrt{4\pi}$   & $2/\sqrt{4\pi}$ \\
$B_y$  & $3.6/\sqrt{4\pi}$ & $4/\sqrt{4\pi}$ \\
$B_z$  & $2/\sqrt{4\pi}$   & $2/\sqrt{4\pi}$ \\
$\gamma$ & 5/3 & 5/3    \\
\tableline
grid zones & 512 &      \\
CFL & 0.8 &             \\
$t_{final}$ & 0.2 &     \\
 \tableline
 \end{tabular}
\end{table}
\clearpage
\subsection{Two Dimensions} \label{2D}
In this section we present the results of several two dimensional
simulations to demonstrate the robustness of the divergence preserving
scheme in AMR applications.  The simulations presented in this section
which utilize the AMR functionality of the code apply an AMR hierarchy
of 3 levels using a refinement ratio of 2 between levels.

\begin{figure}[!h]
\begin{center}
   \includegraphics[angle=-90,clip=true,width=0.42\textwidth]{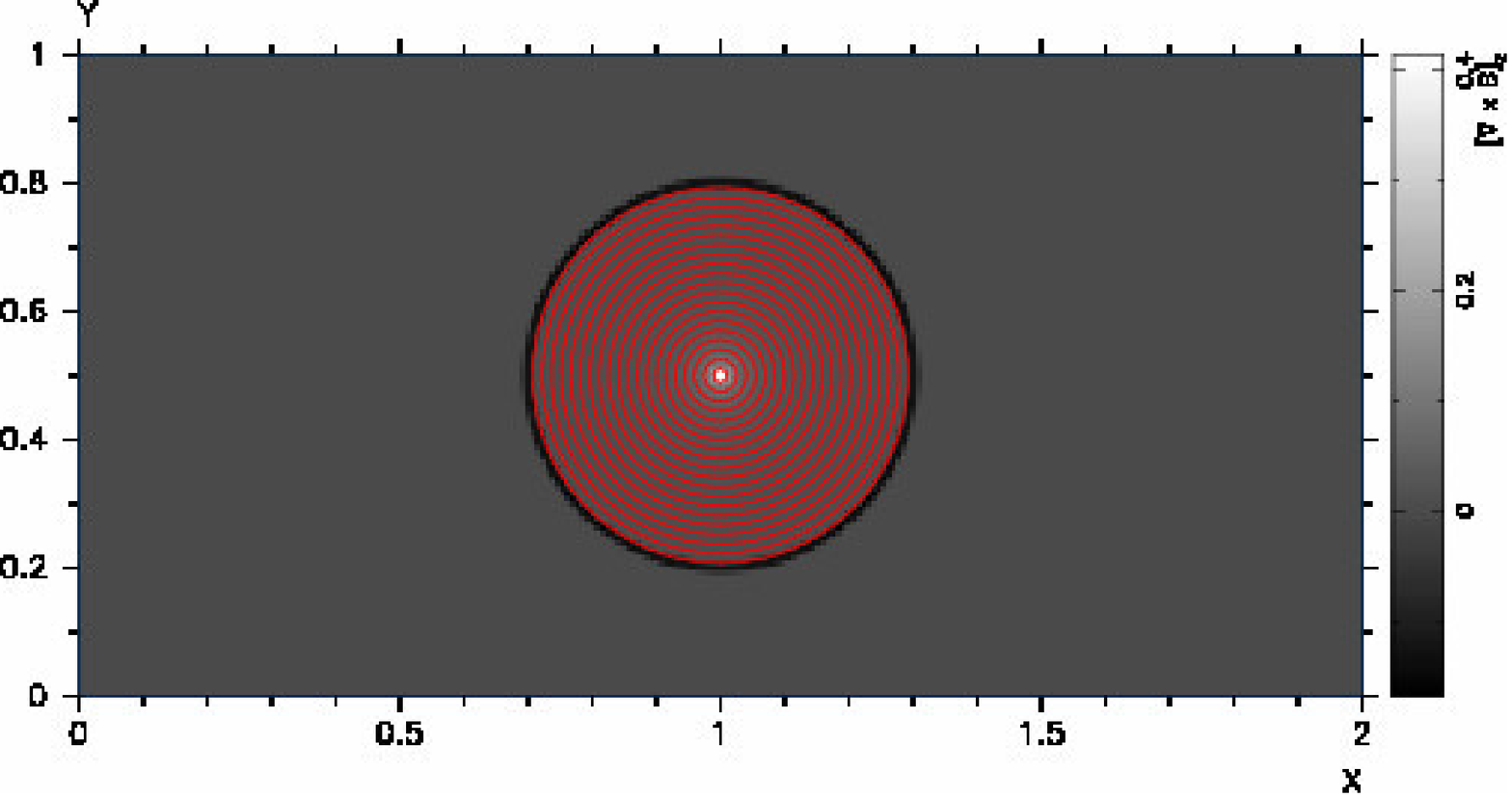}
   \includegraphics[angle=-90,clip=true,width=0.32\textwidth]{f11d.ps} \\
   \includegraphics[angle=-90,clip=true,width=0.42\textwidth]{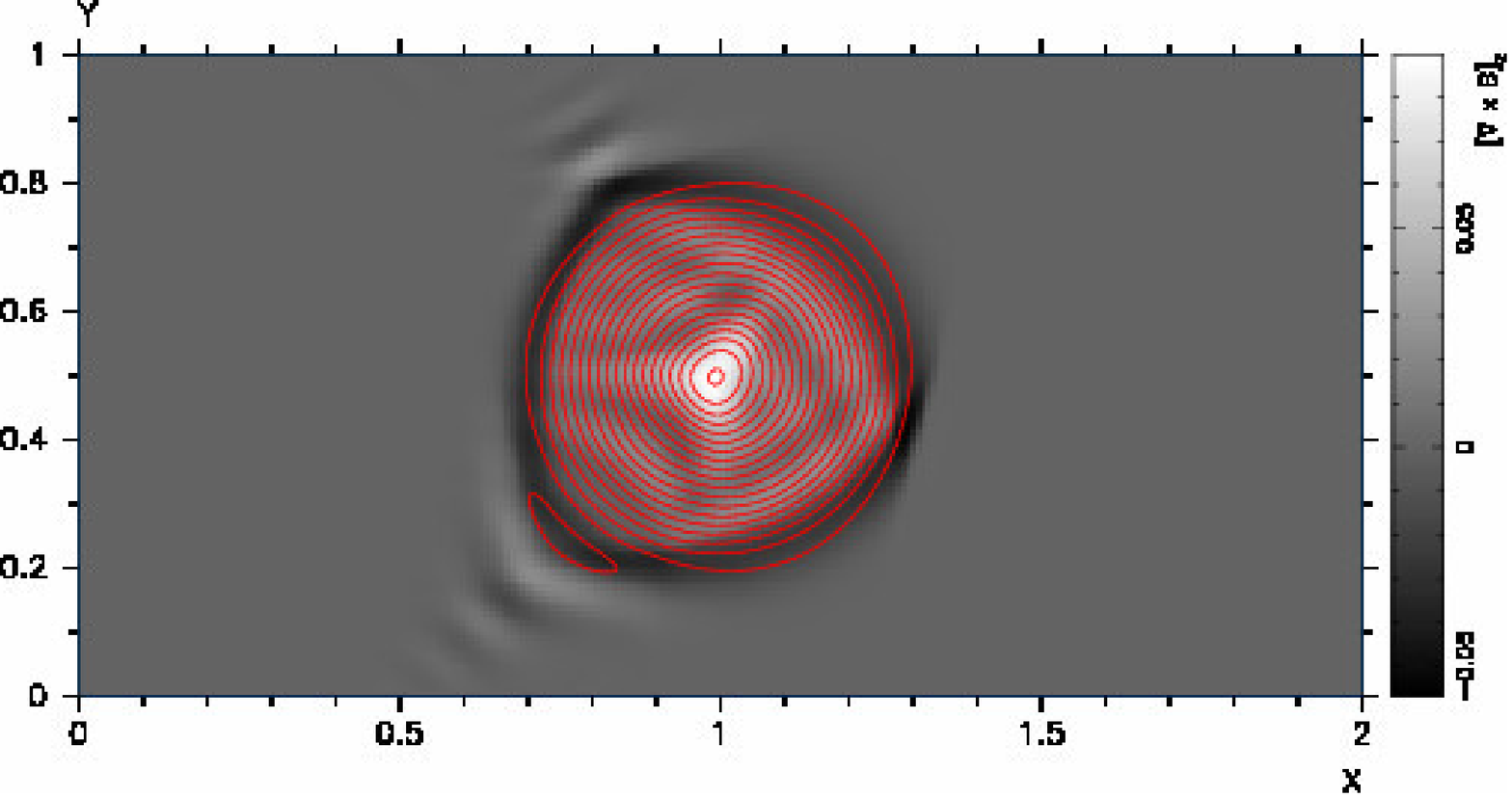}
   \includegraphics[angle=-90,clip=true,width=0.32\textwidth]{f11e.ps} \\
   \includegraphics[angle=-90,clip=true,width=0.42\textwidth]{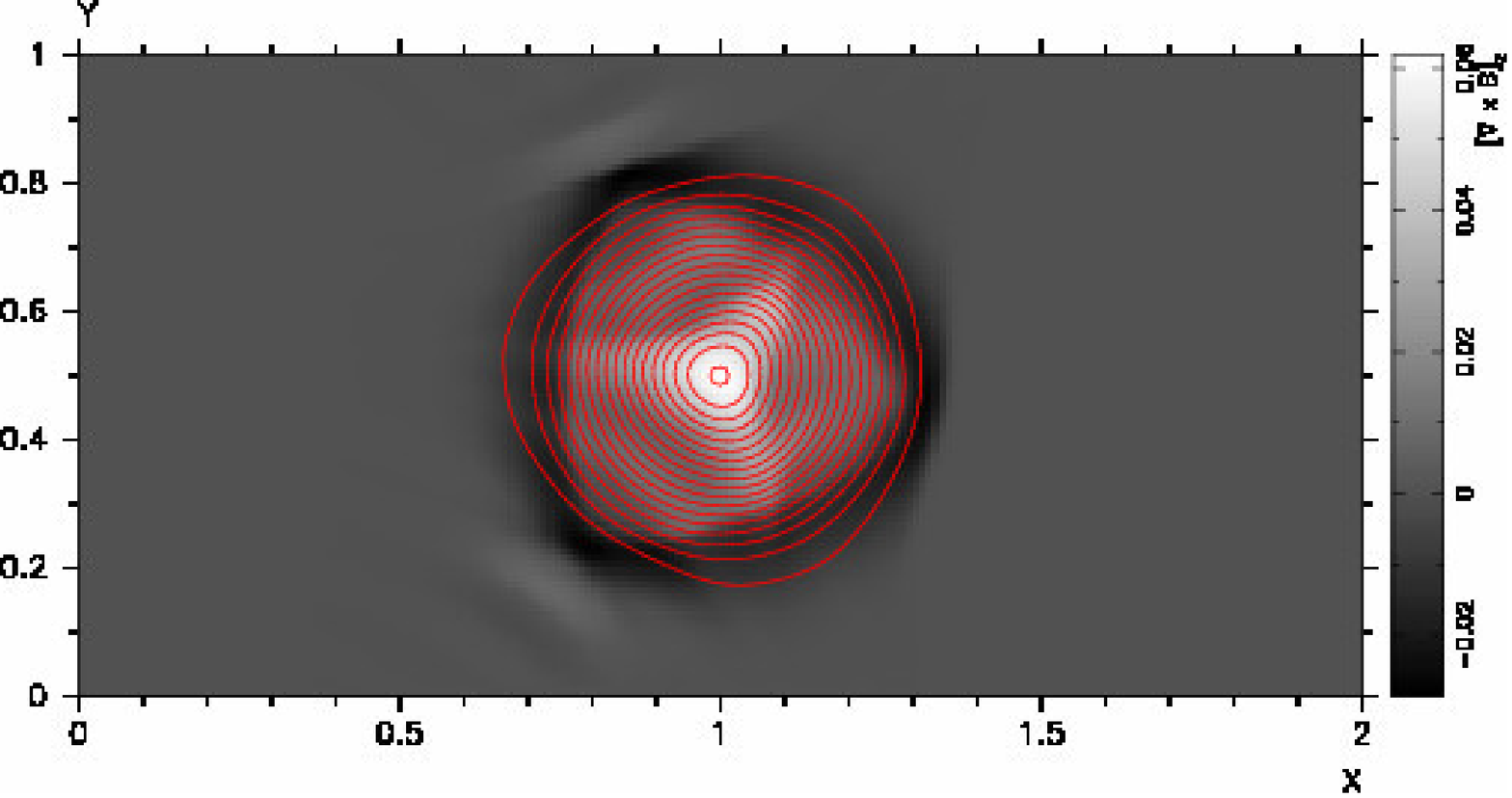}
   \includegraphics[angle=-90,clip=true,width=0.32\textwidth]{f11f.ps}
\end{center}
\caption{Left: Field loop advection magnetic field lines (red) over a
  greyscale representation of the current density $\vec{J} = \nabla
  \times \vec{B}$ for the initial condition (top) and after the
  advection of the loop through the periodic domain at $t=1$ for both
  the Runge-Kutta (center) and MUSCL-Hancock temporal integration
  (bottom).  Right: Evolution of the spatially averaged magnetic
  energy normalized to the initial analytic value $B_o$ for the
  MUSCL-Hancock (top) and Runge-Kutta (center) integration.  The grid
  resolutions shown correspond to $N=32,64~\textrm{and}~128$ for the
  dash-dot, dash and solid lines respectively.  Evolution of the
  spatial average of the normalized axial component of the magnetic
  field $|B_z|$ for the direction split MUSCL-Hancock integrator
  (bottom).  The plot shows decreasing error for the three resolutions
  $N=32,64~\textrm{and}~128$. \label{f11}}
\end{figure}

We begin with the problem of the advection of a weak magnetic field
loop following the prescription given in \cite{gardiner} and
\cite{gardiner08}.  The initial conditions given in table \ref{t11}.
The initial face-centered magnetic field is generated from the
analytic prescription of the vector potential ($\vec{B} = \nabla
\times \vec{A}$) at cell corners using a centered difference formula
on a grid extending from $0 \le L_x \le 2$, $0 \le L_y \le 1$ with
resolution $2N \times N$.  In the left column figure \ref{f11} we show
magnetic field lines (red) over a greyscale representation of the
current density $\vec{J} = \nabla \times \vec{B}$ for the initial
condition (top) and after the advection of the loop through the
periodic domain at $t=1$ for both the Runge-Kutta (center) and
MUSCL-Hancock temporal integration (bottom) with a resolution of
$N=128$.  In both cases monotized-center limited linear spatial
reconstruction with the flux function of Roe were used.  The problem
contains an initially singular current sheet along the surface and a
singular spike at the center which is very sensitive to error in the
evolution of the magnetic field.  The propagation of these singular
features through the grid serves as a very stringent test of the MHD
update algorithm.  We find that both methods maintain the correct
location of these singularities and maintain magnetic field contours
that are smooth and nearly symmetric about the current spike.

The top and center portions of the right column of figure \ref{f11}
show the evolution of the spatially averaged magnetic energy
normalized to the initial analytic value $B_o$ for the MUSCL-Hancock
and Runge-Kutta integration approaches respectively for different grid
resolutions.  We find that both methods show comparable accuracy and
the expected convergence properties for this quantity.  However, the
direction split MUSCL-Hancock scheme shows inexact evolution of the
axial component of the magnetic field $B_z$ and that this error
converges slowly as shown in the lower right panel of figure \ref{f11}
whereas the unsplit Runge-Kutta scheme maintains $B_z=0$ exactly to
machine precision.  This error is due to the failure of the direction
split scheme to produce the an exactly divergence free representation
of the magnetic field for the time-centered predictor state as
discussed in the last paragraph of \S \ref{ctsec}.  In particular, as
noted by \cite{gardiner08}, the evolution of the z-component of the
magnetic induction equation reduces to $\partial B_z/\partial t = v_z
(\partial B_x / \partial x + \partial B_y / \partial y)$ so that with
finite $v_z$ the exact evolution of $B_z$ will only be maintained for
unsplit schemes with exactly divergence free reconstruction.

\clearpage
\clearpage
\begin{table}[!h] \caption{MHD Loop Advection Parameters.\label{t11}}
 \begin{tabular}{l|ll}
\tableline
$\rho$ & 1 &      \\
$v_x$  & 2 &      \\
$v_y$  & 1 &      \\
$v_z$  & 1 &      \\
$P$    & 1 &    \\
$A_z$  &  $B_o (r-R)$ & $\textrm{if~} r \le R$ \\ 
       &  0           & $\textrm{otherwise}$ \\
$r$    &  $\sqrt{(x-1)^2+(y-1)^2}$ & \\
$B_o$  & $10^{-3}$ & \\
$R$    & 0.3 &    \\
$\gamma$ & 5/3 &  \\
\tableline
CFL & 0.9 (MUSCL-Hancock) \\
    & 0.45 (Runge Kutta)\\
 \tableline
 \end{tabular}
\end{table}
\clearpage

\begin{figure}[!h]
\begin{center}
   \includegraphics[angle=-90,clip=true,width=0.49\textwidth]{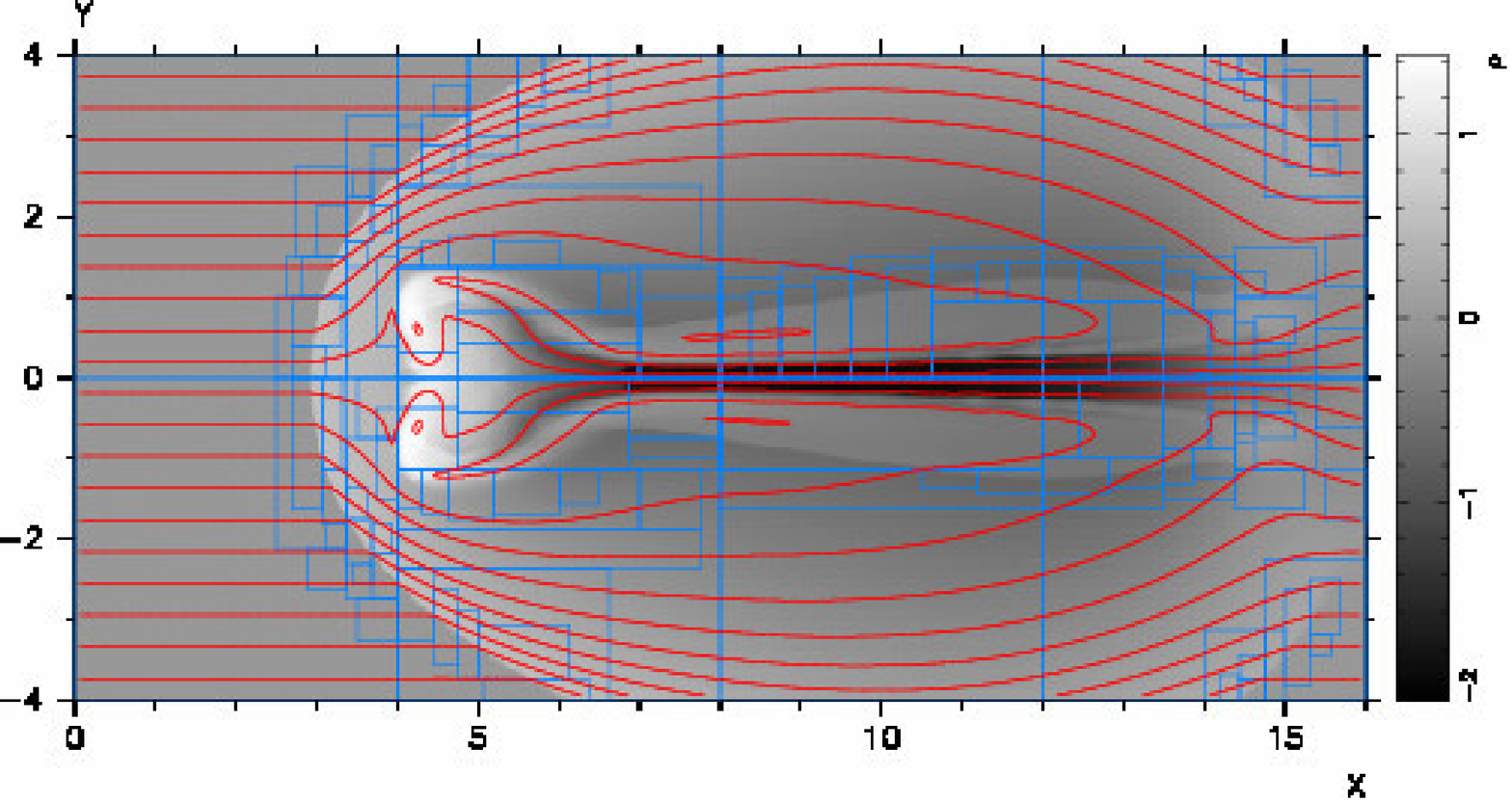}
   \includegraphics[angle=-90,clip=true,width=0.49\textwidth]{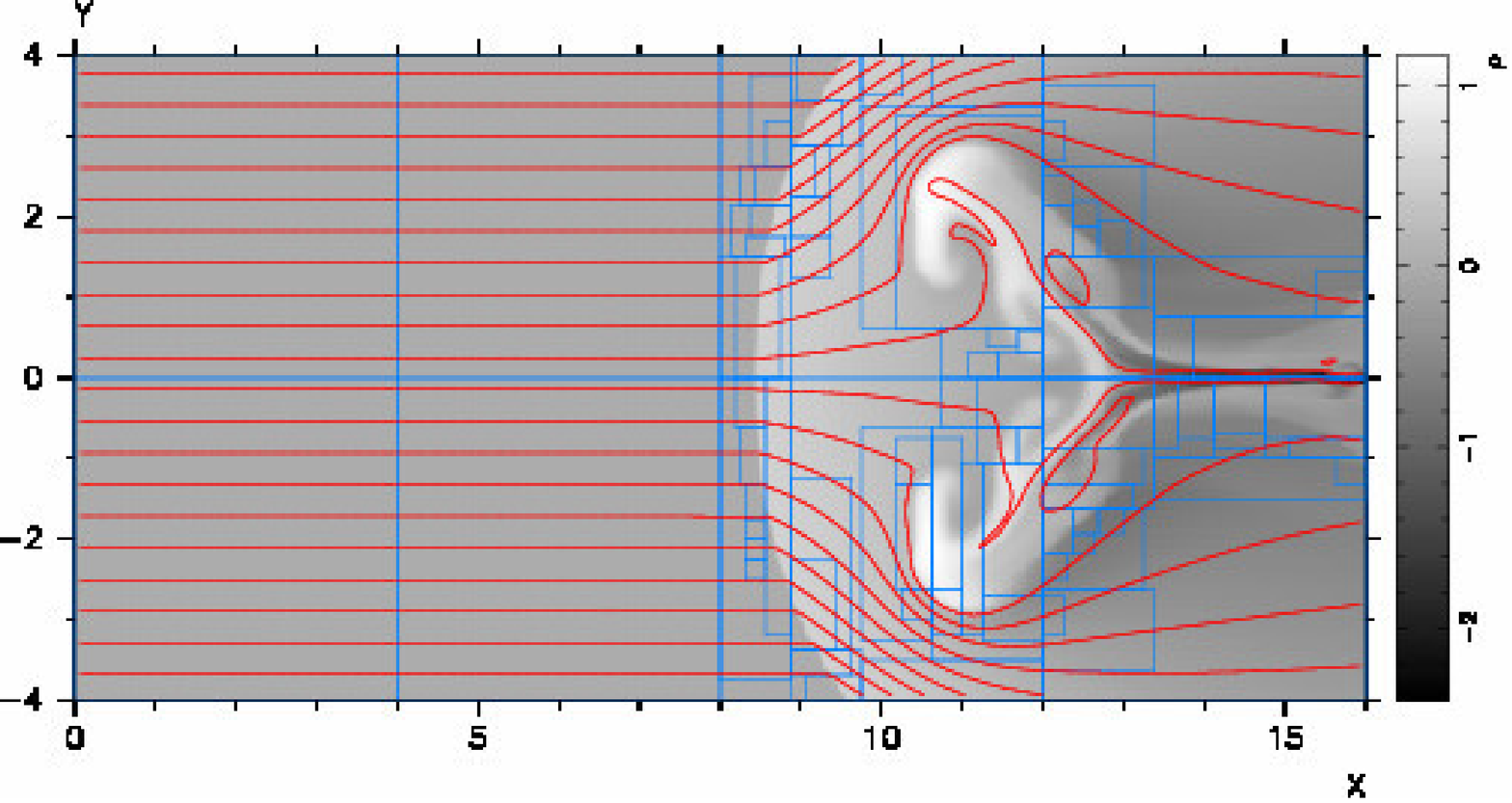} \\
   \includegraphics[angle=-90,clip=true,width=0.49\textwidth]{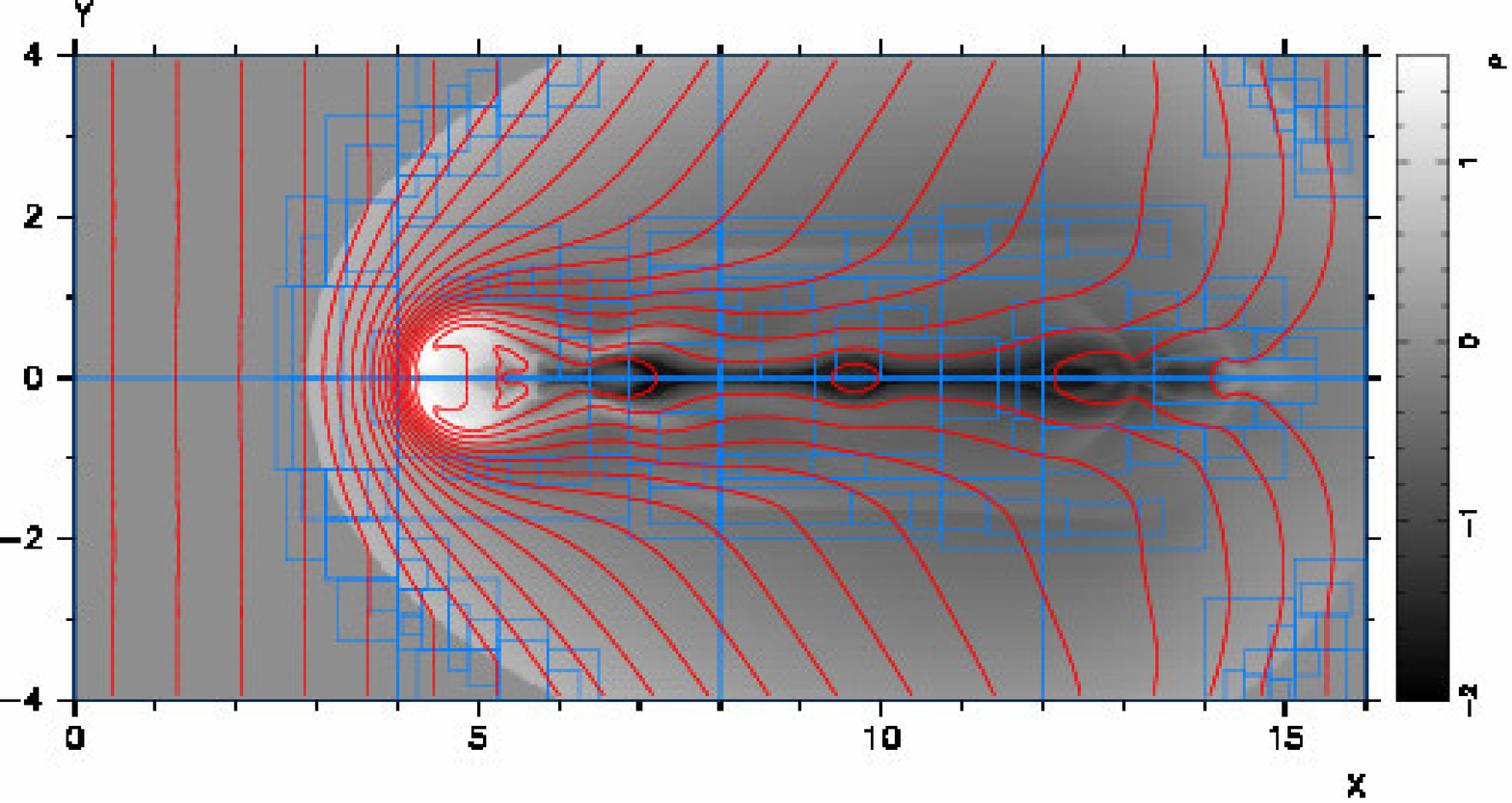}
   \includegraphics[angle=-90,clip=true,width=0.49\textwidth]{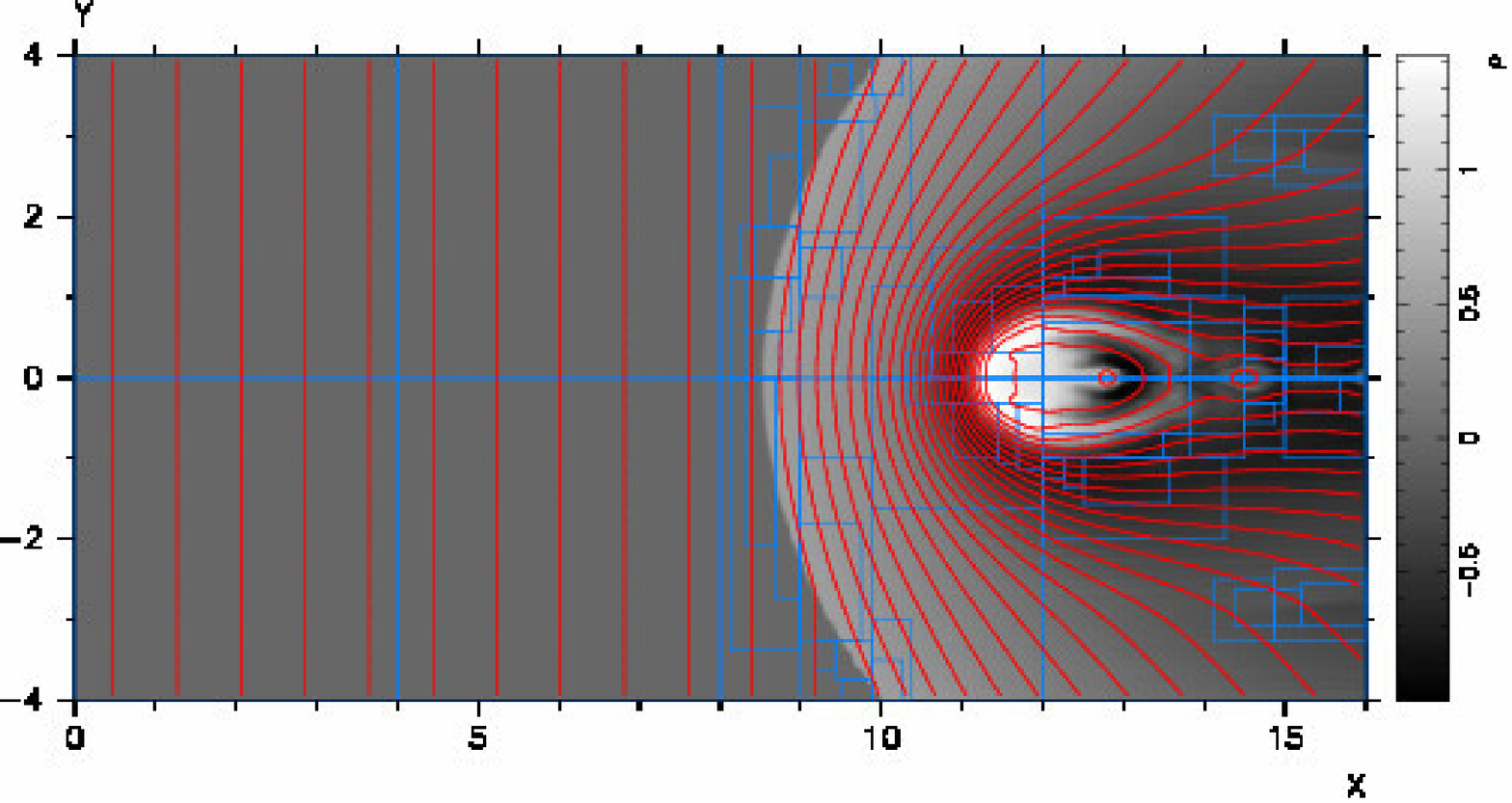} \\
   \includegraphics[angle=-90,clip=true,width=0.49\textwidth]{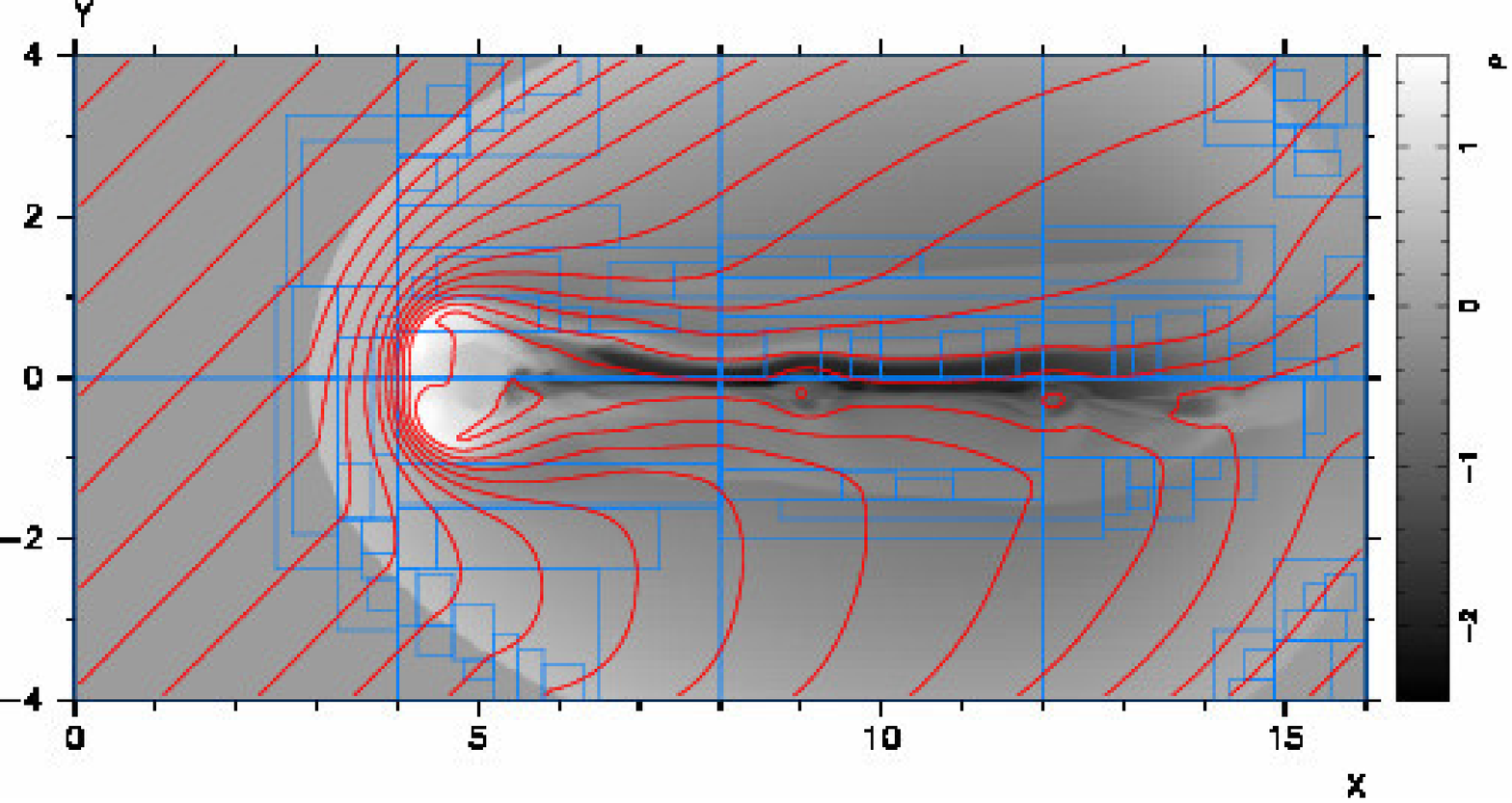}
   \includegraphics[angle=-90,clip=true,width=0.49\textwidth]{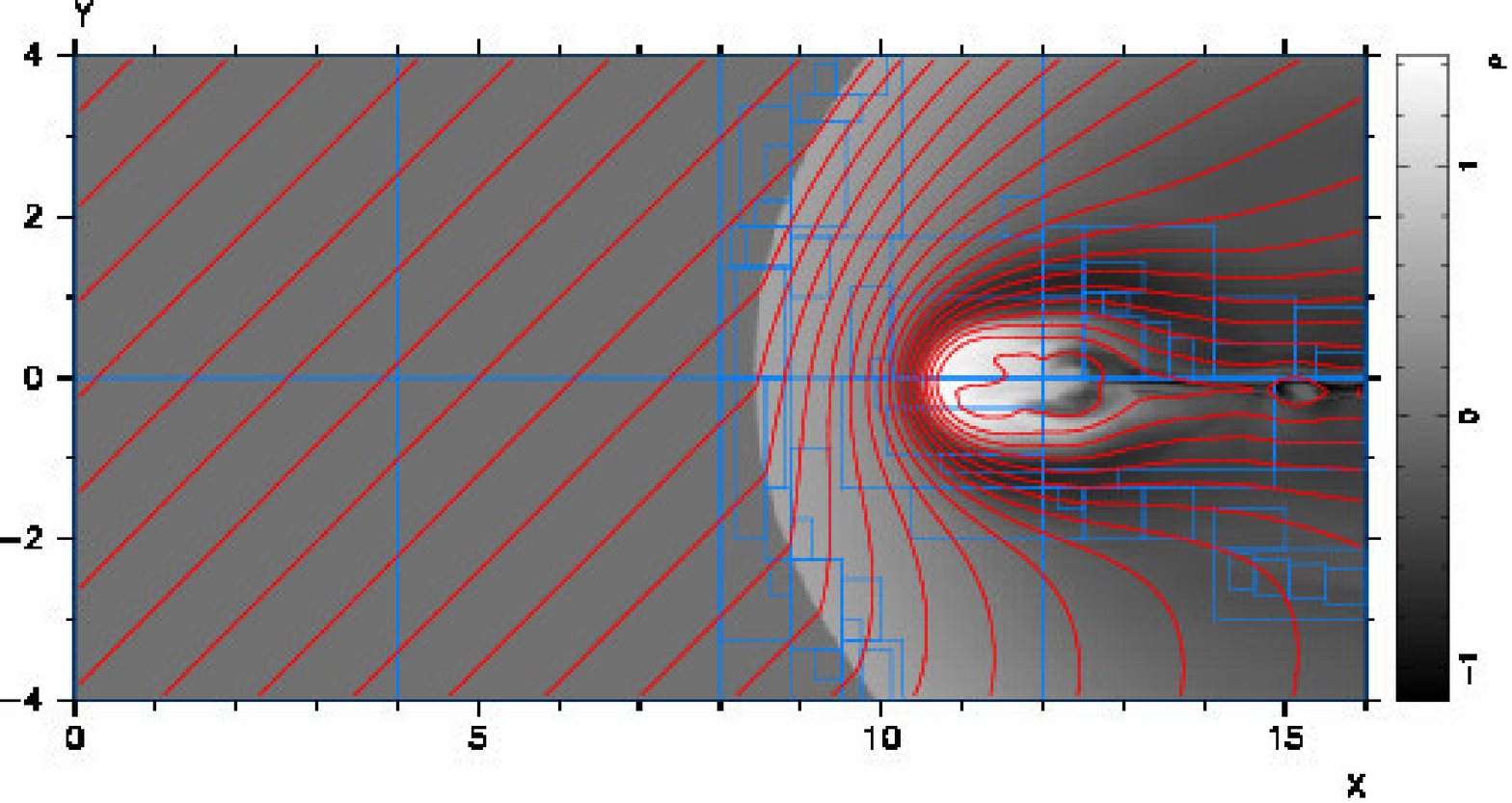}
\end{center}
\caption{Simulation results of a supersonic cylindrical cloud moving
through a magnetized medium with initial magnetic field oriented
parallel, perpendicular and diagonal to the propagation of the cloud
from top to bottom.  The logarithm of the density distribution is
presented in gray-scale, red lines delineate the magnetic field lines
and blue lines delineate regions of AMR-enhanced
resolution.\label{f12}}
\end{figure}

\begin{figure}[!h]
\begin{center}
   \includegraphics[angle=-90,clip=true,width=0.7\textwidth]{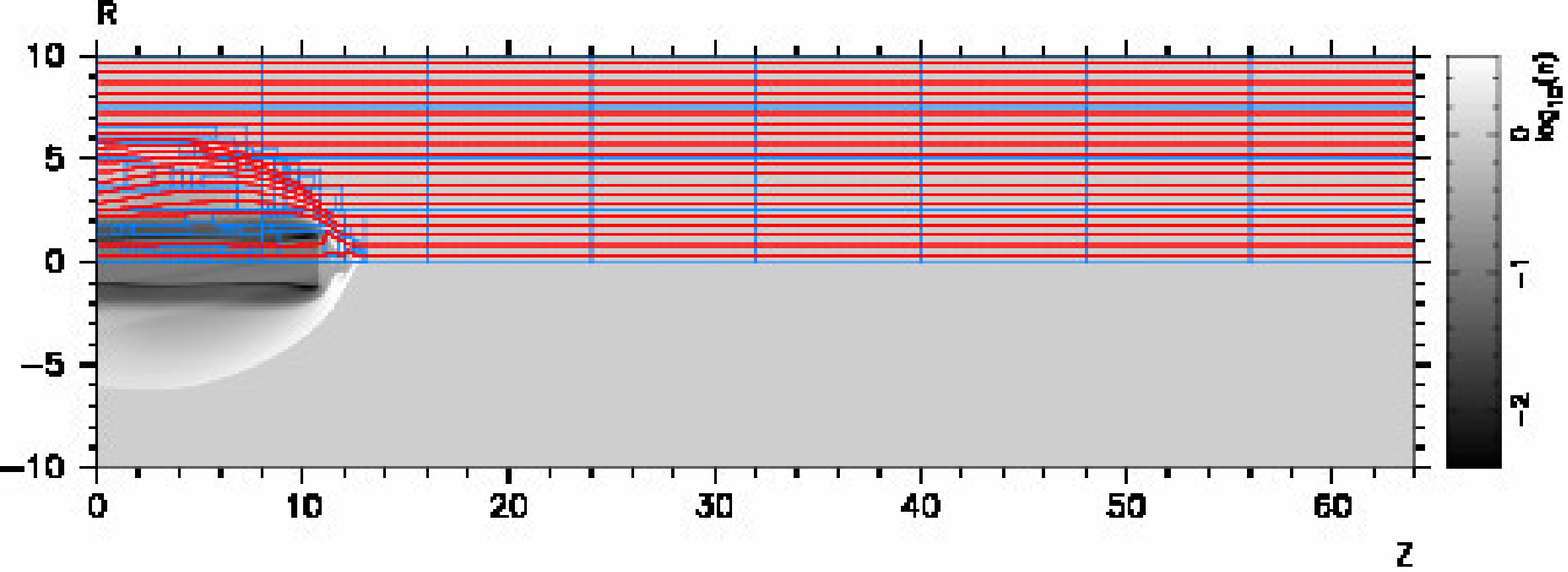}
   \includegraphics[angle=-90,clip=true,width=0.7\textwidth]{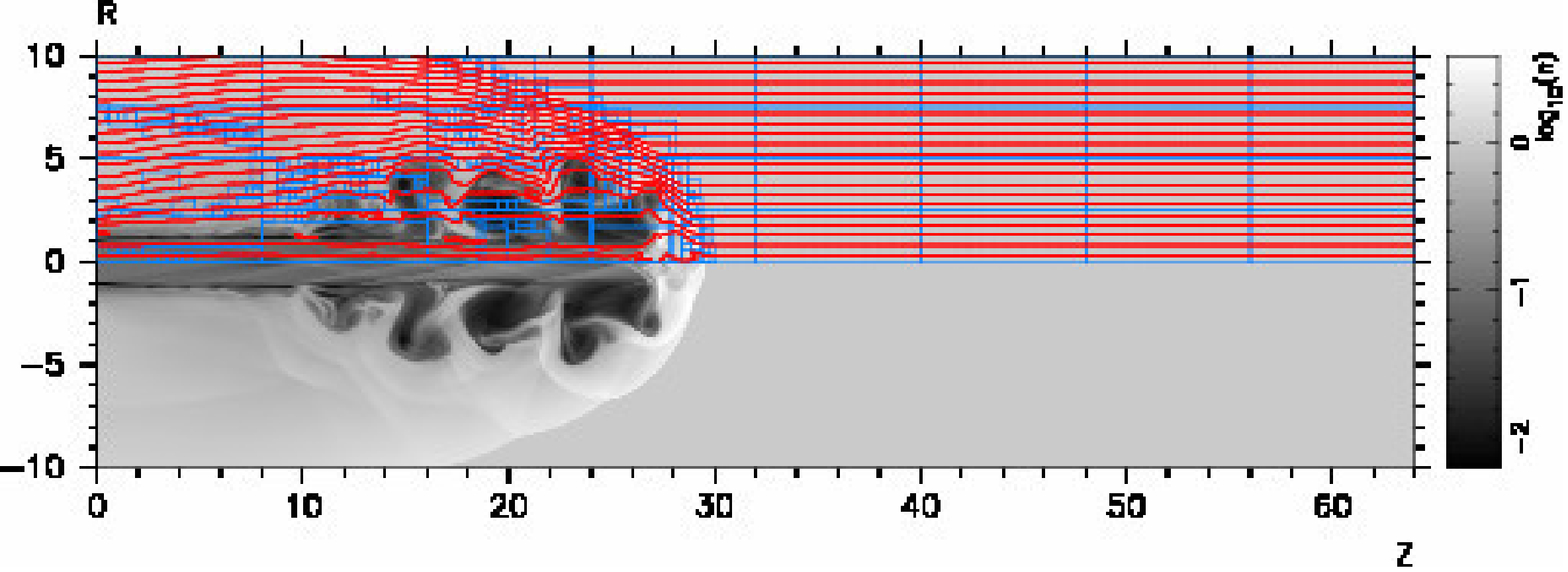}
   \includegraphics[angle=-90,clip=true,width=0.7\textwidth]{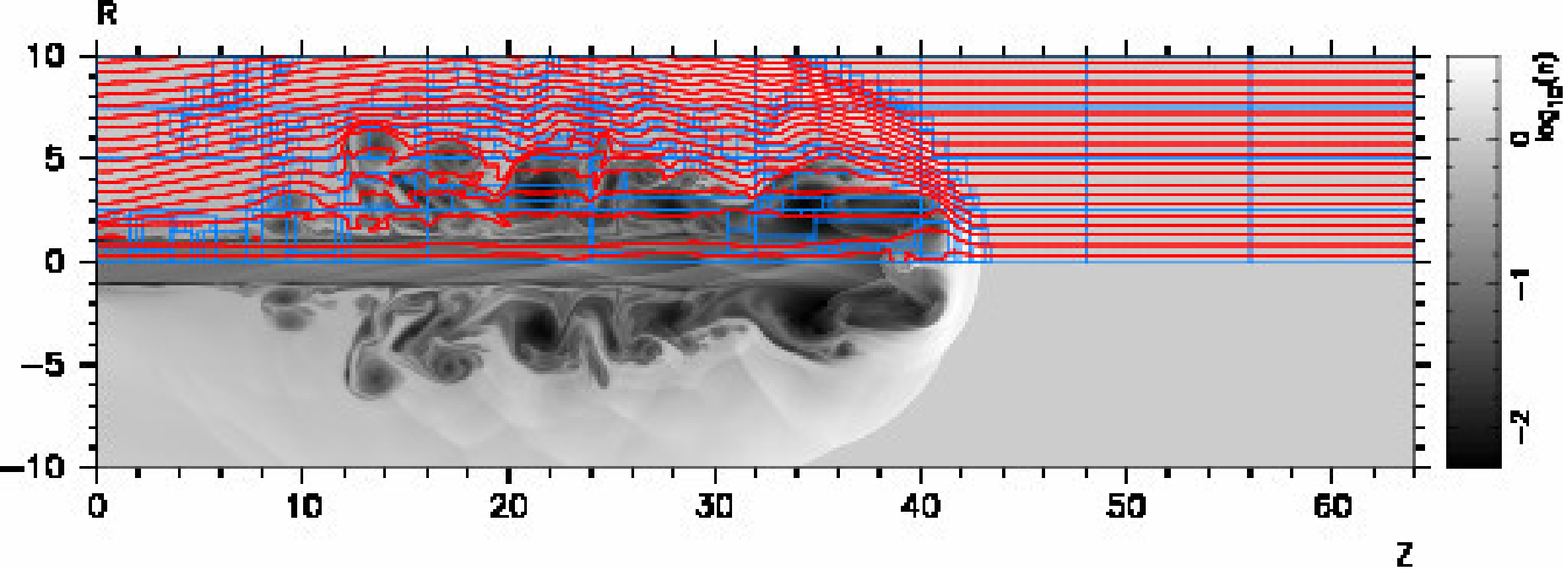}
   \includegraphics[angle=-90,clip=true,width=0.7\textwidth]{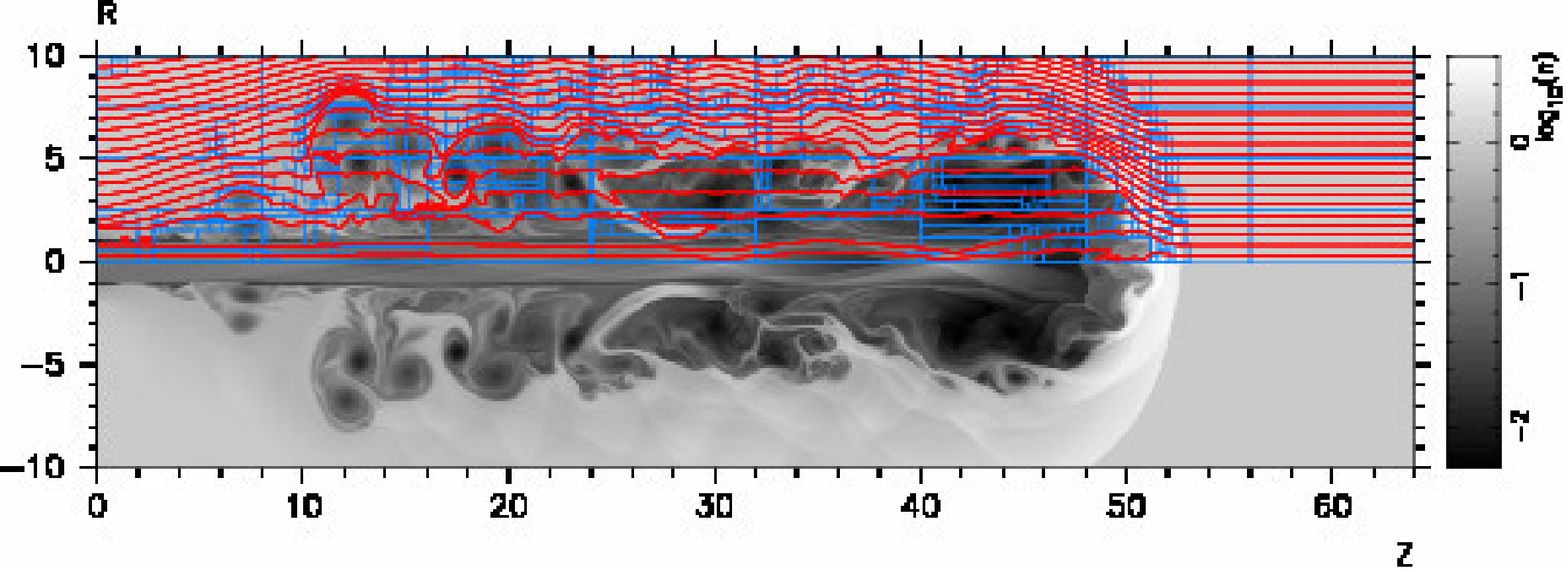}
   \includegraphics[angle=-90,clip=true,width=0.7\textwidth]{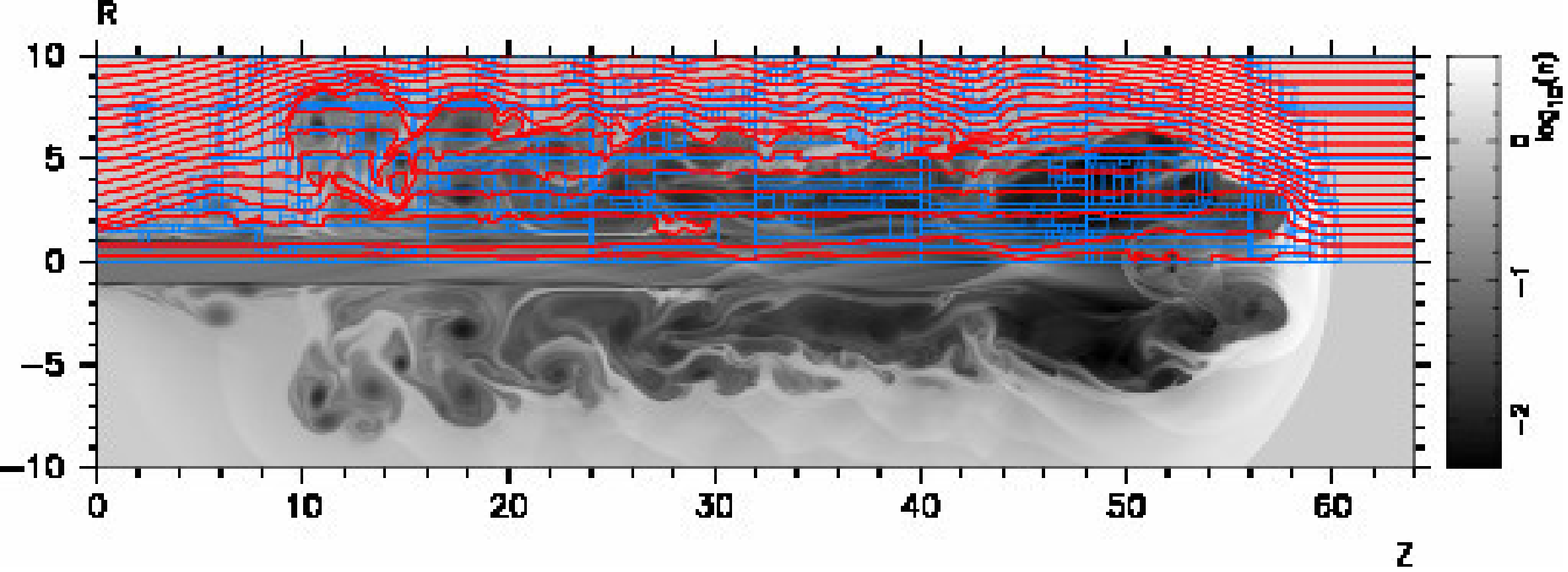}
\end{center}
\caption{Light MHD jet simulation.  The logarithm of the density
distribution is presented in gray-scale, red lines delineate the
magnetic field lines and blue lines delineate regions of AMR-enhanced
resolution at $t=2.43,~6.57,~10.62,~14.76,\textrm{ and }~18.00$ from
top to bottom. \label{f13}}
\end{figure}

The next set of simulations show the propagation of a cylindrical,
supersonic cloud through a magnetized medium where the magnetic field
is oriented parallel to the propagation of the cloud (figure
\ref{f12}, top), perpendicular to the propagation of the cloud (figure
\ref{f13}, middle), and $45^{\circ}$ to the propagation of the cloud
(figure \ref{f13}, bottom).  These simulations have been carried out
using PHM spatial reconstruction, MUSCL-Hancock temporal
reconstruction, the adapted Marquina flux and the constrained
transport evolution of the magnetic field of \cite{rjf}.  Each of
these simulations employ constant extrapolation conditions along each
boundary.  The AMR level with the finest resolution achieves an
effective resolution of 32 cells per cloud radius.  The initial cloud
density is smoothed linearly to that of the ambient environment and
the velocity is smoothed over the outer four computational cells of
the cloud.  The figures show the result of each simulation at two
evolutionary times, $t=2 t_{cr}$ in the left panel and $t=4 t_{cr}$ in
the right panel where $t_{cr}= 2 r_c \sqrt{\chi} / v_c$ is the cloud
crushing time where $r_c$ is the cloud radius, $v_c$ is the initial
cloud speed and $\chi$ is the density contrast of the cloud against
the ambient environment (see \cite{jones} for details).  The density
distribution is presented in gray-scale, red lines delineate the
magnetic field lines and blue lines delineate regions of AMR-enhanced
resolution.  We note that the turbulent wake behind the $45^{\circ}$
cloud and the associated early onset of tearing mode instability and
magnetic reconnection pose some degree of numerical difficulty which
requires the use of the more accurate and robust Runge-Kutta method
over the faster MUSCL-Hancock method.  The simulations use the same
initial physical parameters as the simulations presented in
\cite{rj-ct} and \cite{jones} which are presented in table \ref{t11}.

Because the shocked cloud simulations presented here do not utilize a
moving mesh, the final time ($t=4 t_{bc}$) of the simulations is
somewhat earlier than that of the earlier works ($t=6 t_{bc}$) of
\cite{rjf} and \cite{jones}.  The density distribution and magnetic
field lines at ($t=2 t_{bc}$) show agreement with results of the
earlier calculations at the same evolutionary time.  As pointed out in
\cite{rjf} and \cite{jones}, the nonlinear evolution of the cloud
depends sensitively on the exact initial perturbations which develop out
of the geometric mismatches between the cloud and the computational
grid.  The agreement of the AMR simulations presented here with the
earlier works demonstrates the robustness and accuracy of the
divergence preserving restriction and prolongation procedures
presented in \S\ref{prores}.

\clearpage
\clearpage
\begin{table}[!h] \caption{MHD Shocked Cloud Parameters.\label{t12}}
 \begin{tabular}{l | l l}
 & ambient & cloud     \\
 \tableline
$\rho$ & 1 & 10        \\
$v_x$  & 10 & 0       \\
$v_y$  & 0  & 0        \\
$v_z$  & 0  & 0        \\
$P$    & $1/\gamma$ & $1/\gamma$        \\
$\beta$  & 4  & 4 \\
$\gamma$ & 5/3 & 5/3    \\
\tableline
clump radius ($r_c$) & 1 &      \\
grid zones per cloud radius & 32 &      \\
initial cloud position & (2, 0) & \\
CFL & 0.4 &             \\
 \tableline
 \end{tabular}
\end{table}
\clearpage
%
In figure \ref{f13} we reproduce the light cylindrical MHD jet
simulation of \cite{rjf} and \cite{lind}.  A jet with a top-hat
profile is injected into a uniformly magnetized environment by
imposing the physical jet conditions presented in table \ref{t13}
inside of $r \le r_j$ along the $z=0$ boundary.  The symmetry of the
problem dictates the use of reflecting boundary conditions along the
$r=0$ boundary.  All other boundaries utilize constant extrapolation.
The simulation was carried forward on an AMR hierarchy with four
levels of refinement utilizing the PHM spatial integration method with
Runge-Kutta temporal integration, the Roe flux upwinding method, and
the constrained transport magnetic field evolution of \cite{rj-ct}.
The finest AMR level has a resolution of 32 cells per jet radius.  The
figures show the result of the simulation at 5 evolutionary times,
$t=2.43,~6.57,~10.62,~14.76,\textrm{ and }~18.00$.  Note these are the
same evolutionary times shown in the results of \cite{rj-ct} scaled to
the dimensionless units defined by the parameters listed in \ref{t13}.
The density distribution is presented in gray-scale, red lines
delineate the magnetic field lines and blue lines delineate regions of
AMR-enhanced resolution.  Because the simulations were carried out in
cylindrical symmetry, the lower half of each panel has been plotted by
reflecting the computational domain about the symmetry axis.  Barring
differences in the detail of the nonlinear evolution of the
Kelvin-Helmholtz shear flow behind the jet bow shock which arise from
the exact nature of the numerical perturbations present in the
simulation, the simulation results show excellent agreement with that
of \cite{rjf} at each evolutionary stage shown in the panels of figure
\ref{f13}.  The agreement of the AMR simulations presented here with
the earlier works demonstrates the robustness and accuracy of the
divergence preserving restriction and prolongation procedures
presented in \S\ref{prores} and the adaptations thereof for
cylindrical geometry presented in the appendix.

\clearpage
\clearpage
\begin{table}[!h] \caption{Light MHD Jet Parameters.\label{t13}}
 \begin{tabular}{l | l l}
 & ambient & jet \\
 \tableline
$\rho$ & 0.1 & 1        \\
$v_x$  & 0  & 20        \\
$v_y$  & 0  & 0        \\
$v_z$  & 0  & 0        \\
$P$    & $1/\gamma$ & $1/\gamma$        \\
$B_r$  & 0  &  0 \\
$B_{\theta}(r)$  & 0  &  $2 \sqrt{0.02/\gamma}~r/r_j$ \\
$B_z$  & $\sqrt{0.02/\gamma}$  &  $\sqrt{0.02/\gamma}$ \\
$\gamma$ & 5/3 & 5/3    \\
\tableline
jet radius $r_j$ & 1 &   \\
grid zones per jet radius & 32 &      \\
CFL & 0.4 &             \\
 \tableline
 \end{tabular}
\end{table}
\clearpage

\begin{figure}[!h]
\begin{center}
   \includegraphics[angle=-90,clip=true,width=0.55\textwidth]{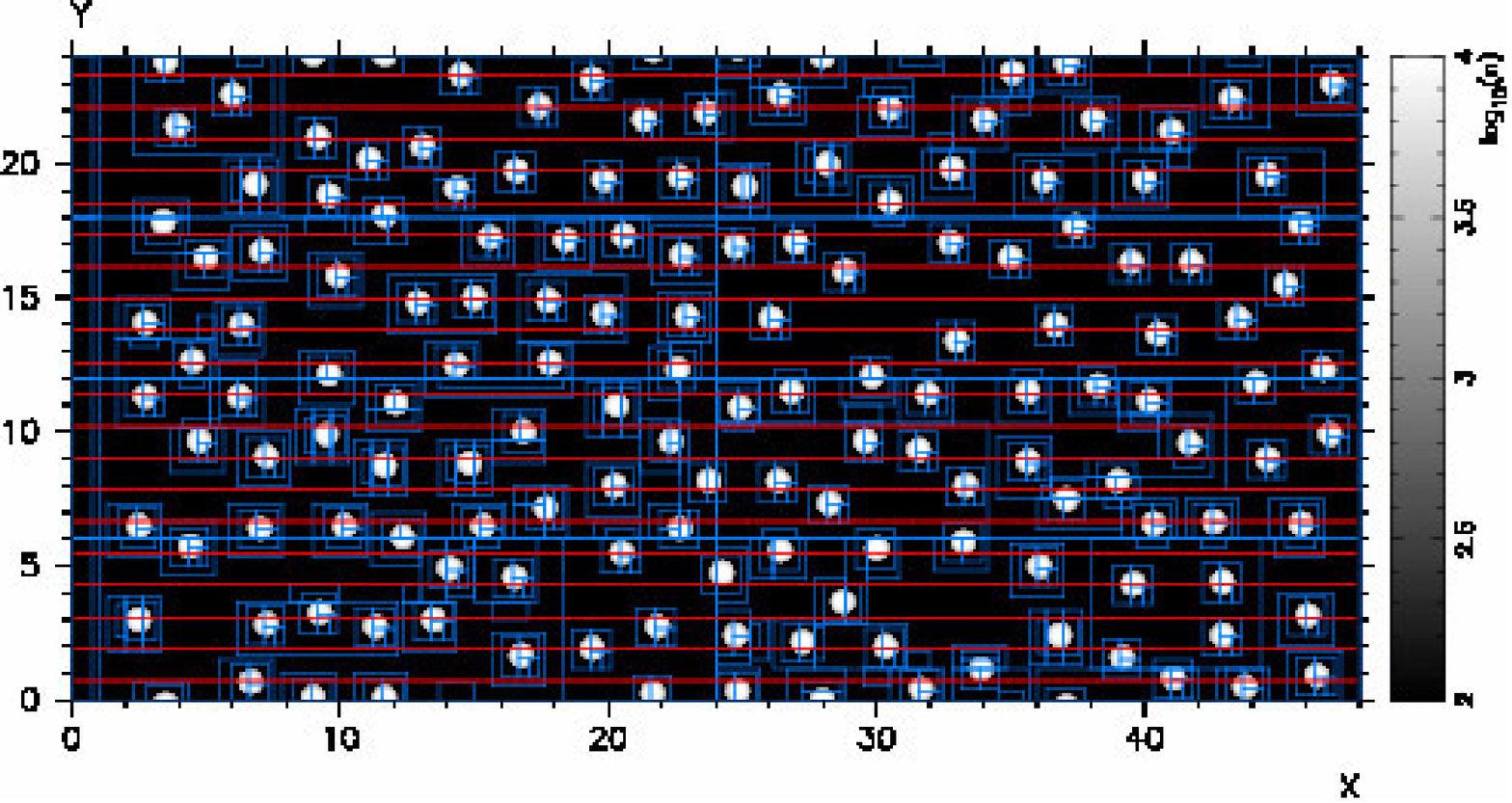} \\
   \includegraphics[angle=-90,clip=true,width=0.55\textwidth]{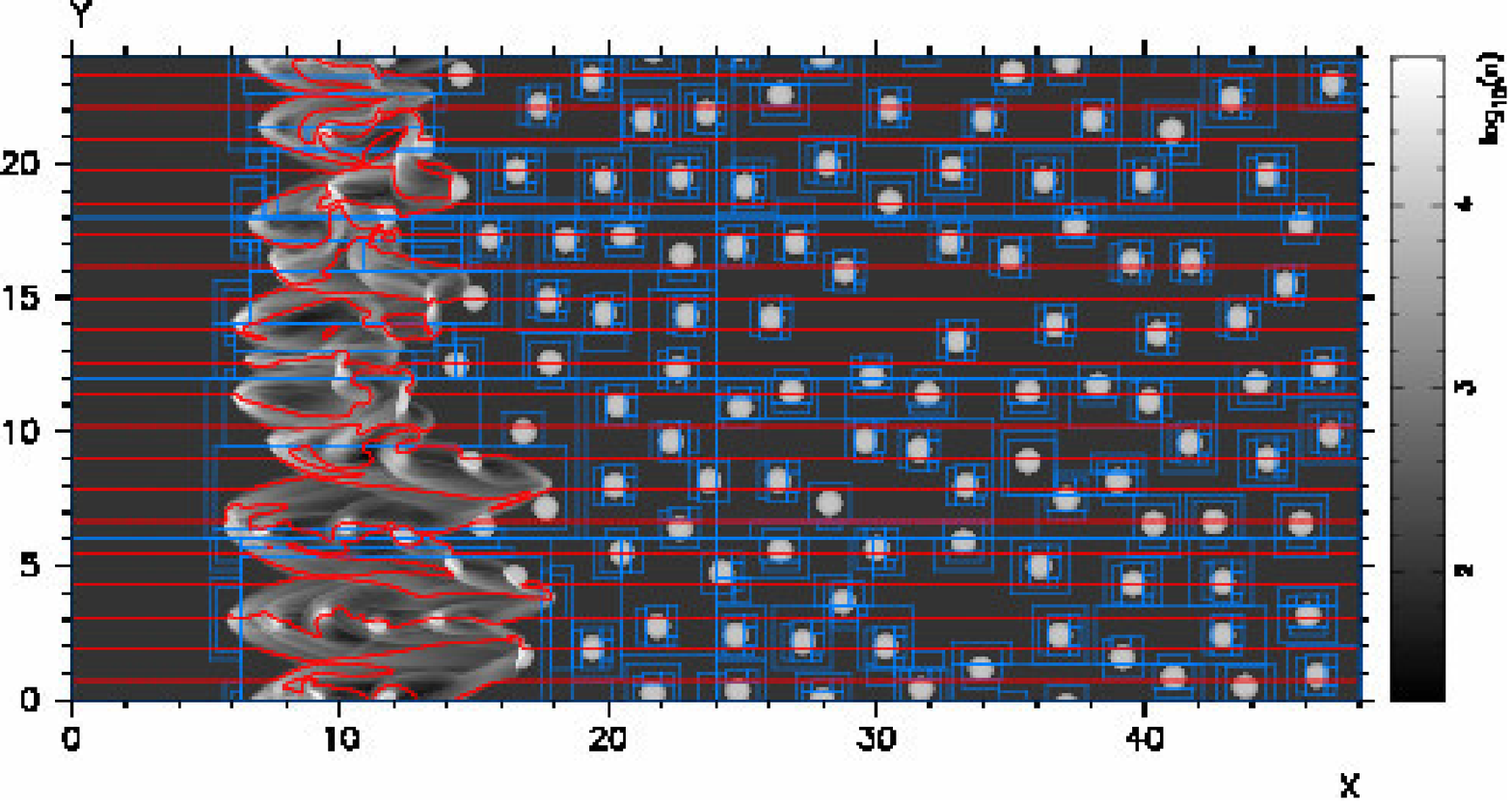} \\
   \includegraphics[angle=-90,clip=true,width=0.55\textwidth]{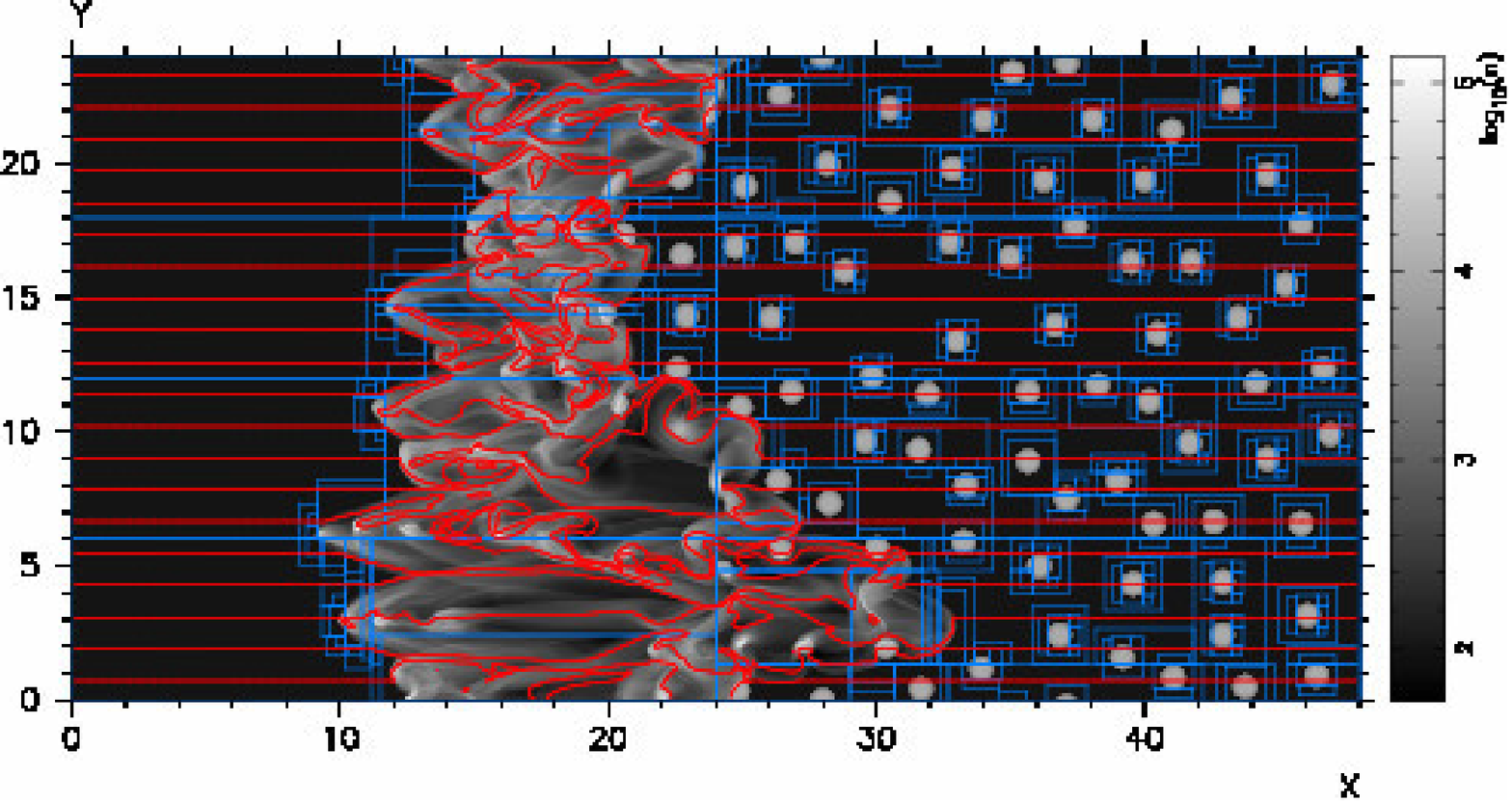} \\
   \includegraphics[angle=-90,clip=true,width=0.55\textwidth]{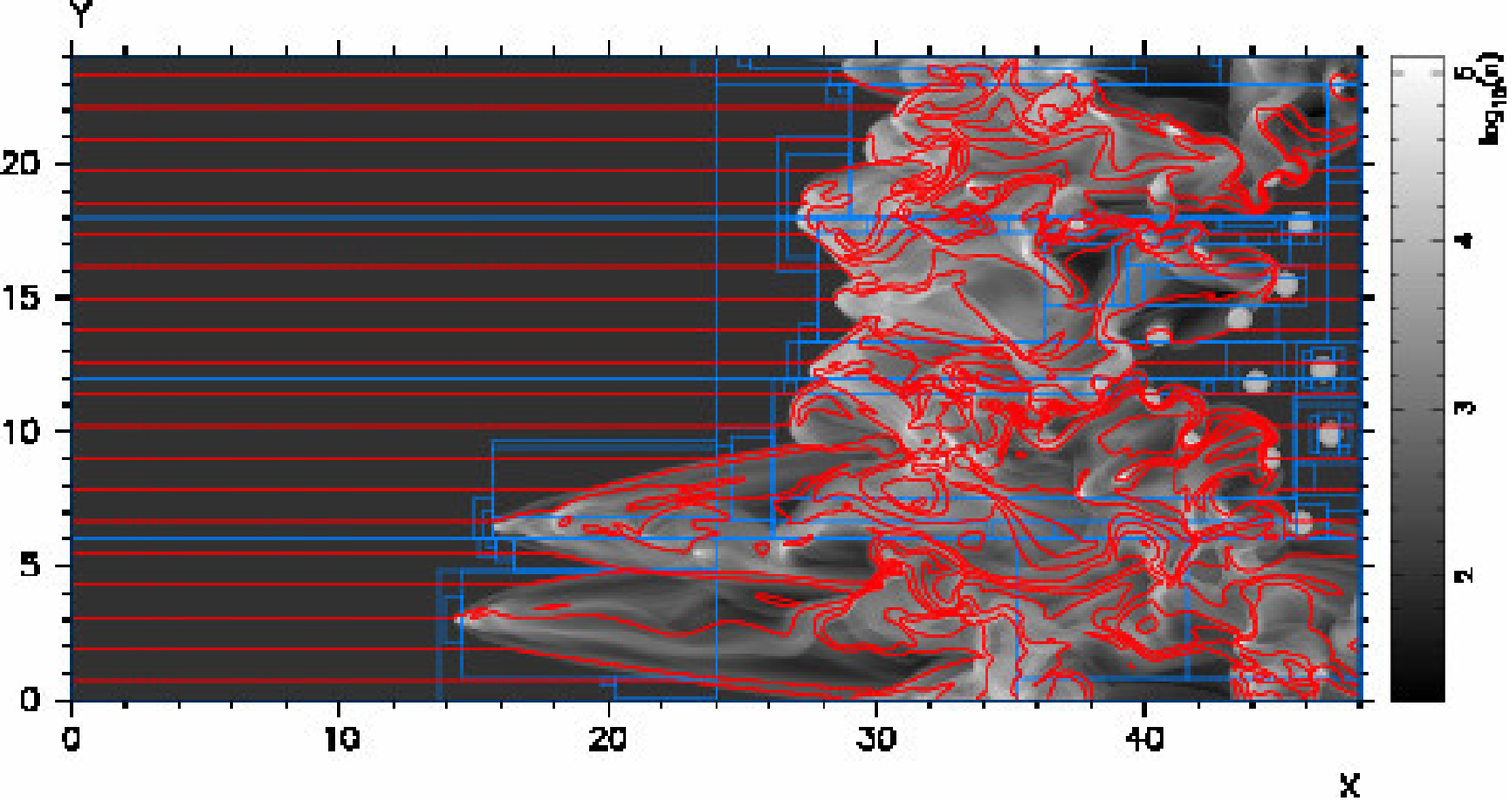}
\end{center}
\caption{ Shock propagation through multiple clouds with magnetic
field oriented parallel to the direction of shock propagation at
evolutionary time $t=0,~92,~184 \textrm{ and } 369~\textrm{yr}$.  The
logarithm of the density distribution in $cm^{-3}$ is presented in
gray-scale, red lines delineate the magnetic field lines, blue lines
delineate regions of AMR-enhanced resolution and the cloud diameter is
used as the length unit. \label{f14}}
\end{figure}

\begin{figure}[!h]
\begin{center}
   \includegraphics[angle=-90,clip=true,width=0.55\textwidth]{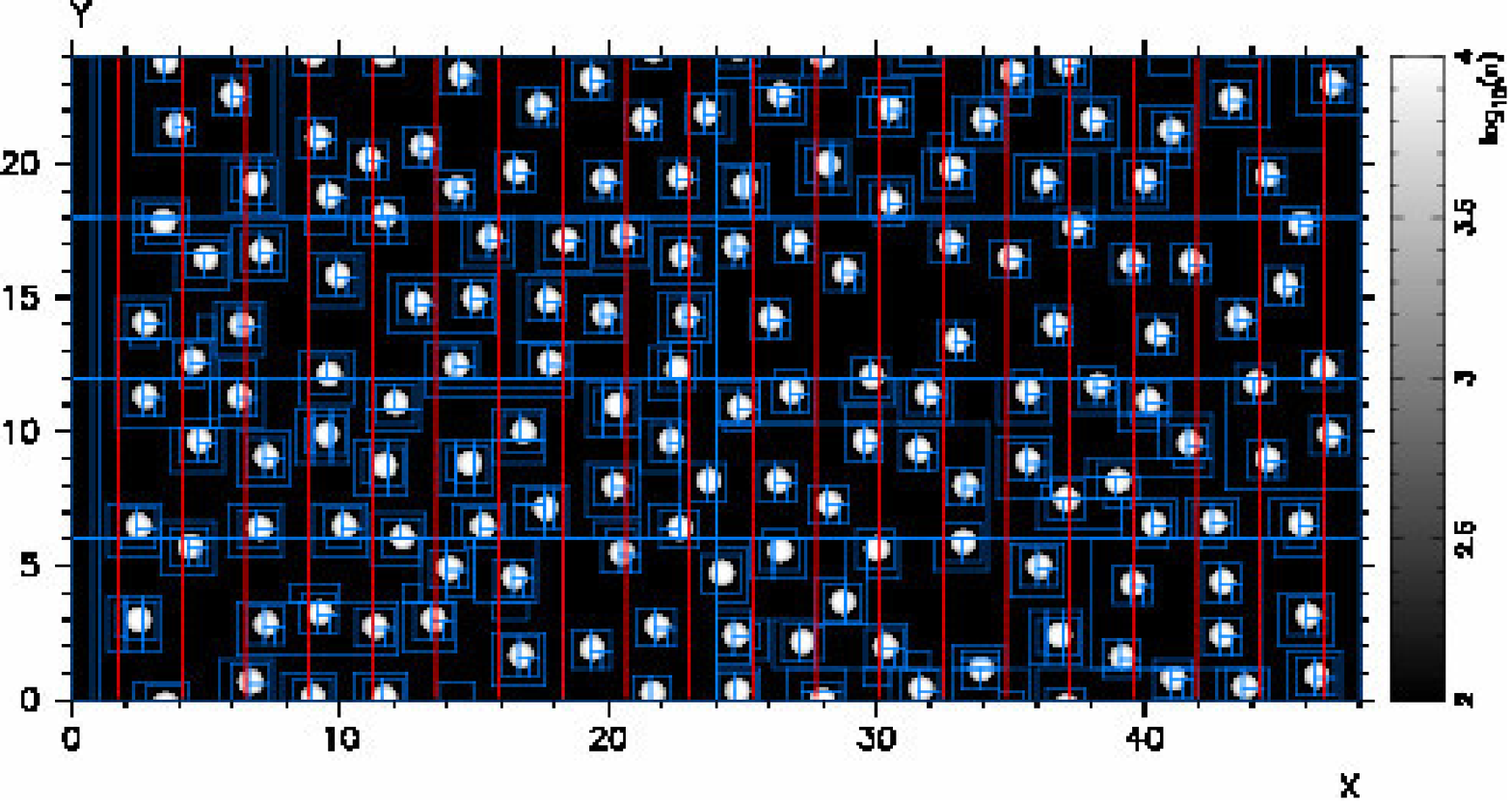} \\
   \includegraphics[angle=-90,clip=true,width=0.55\textwidth]{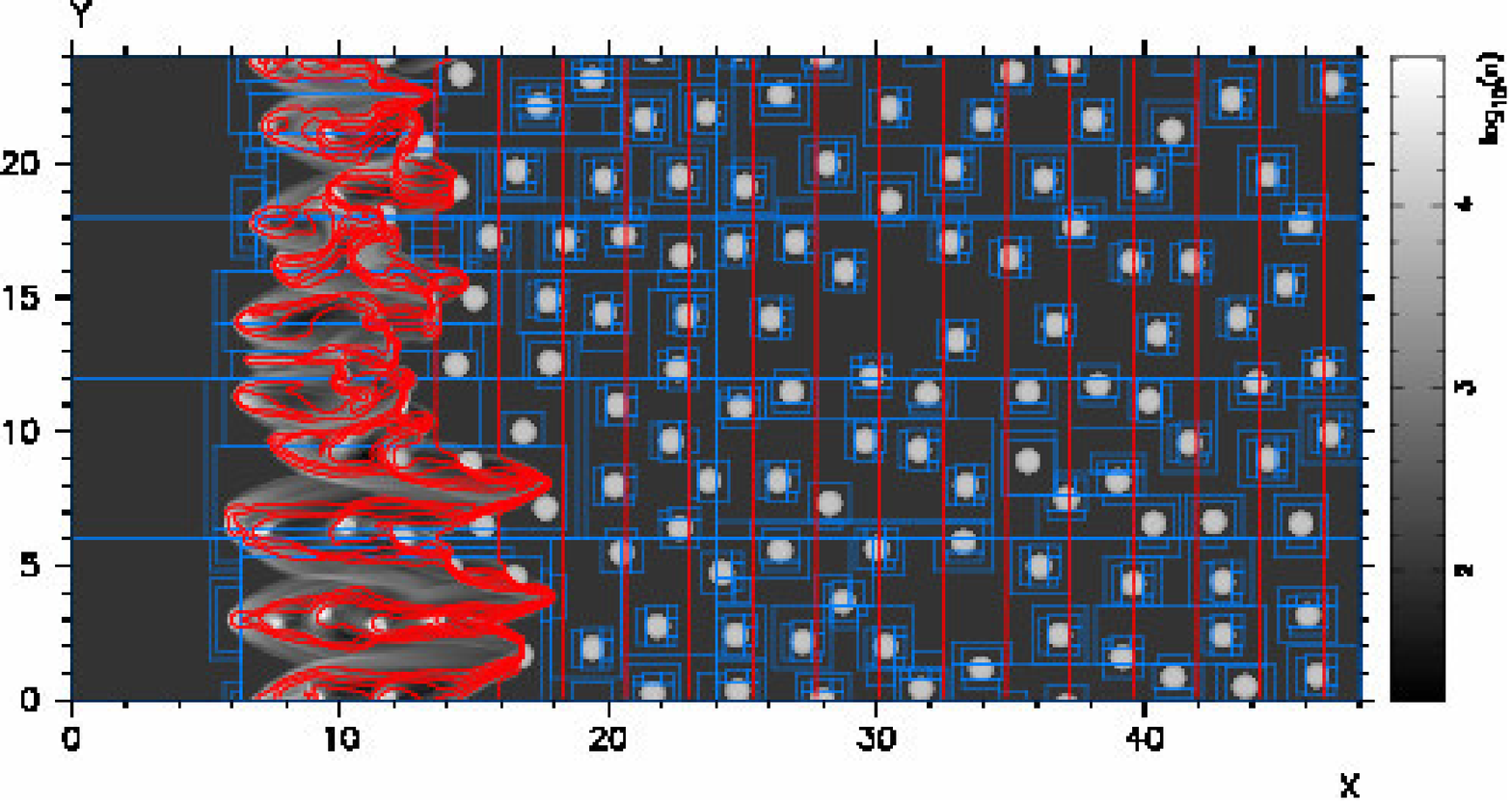} \\
   \includegraphics[angle=-90,clip=true,width=0.55\textwidth]{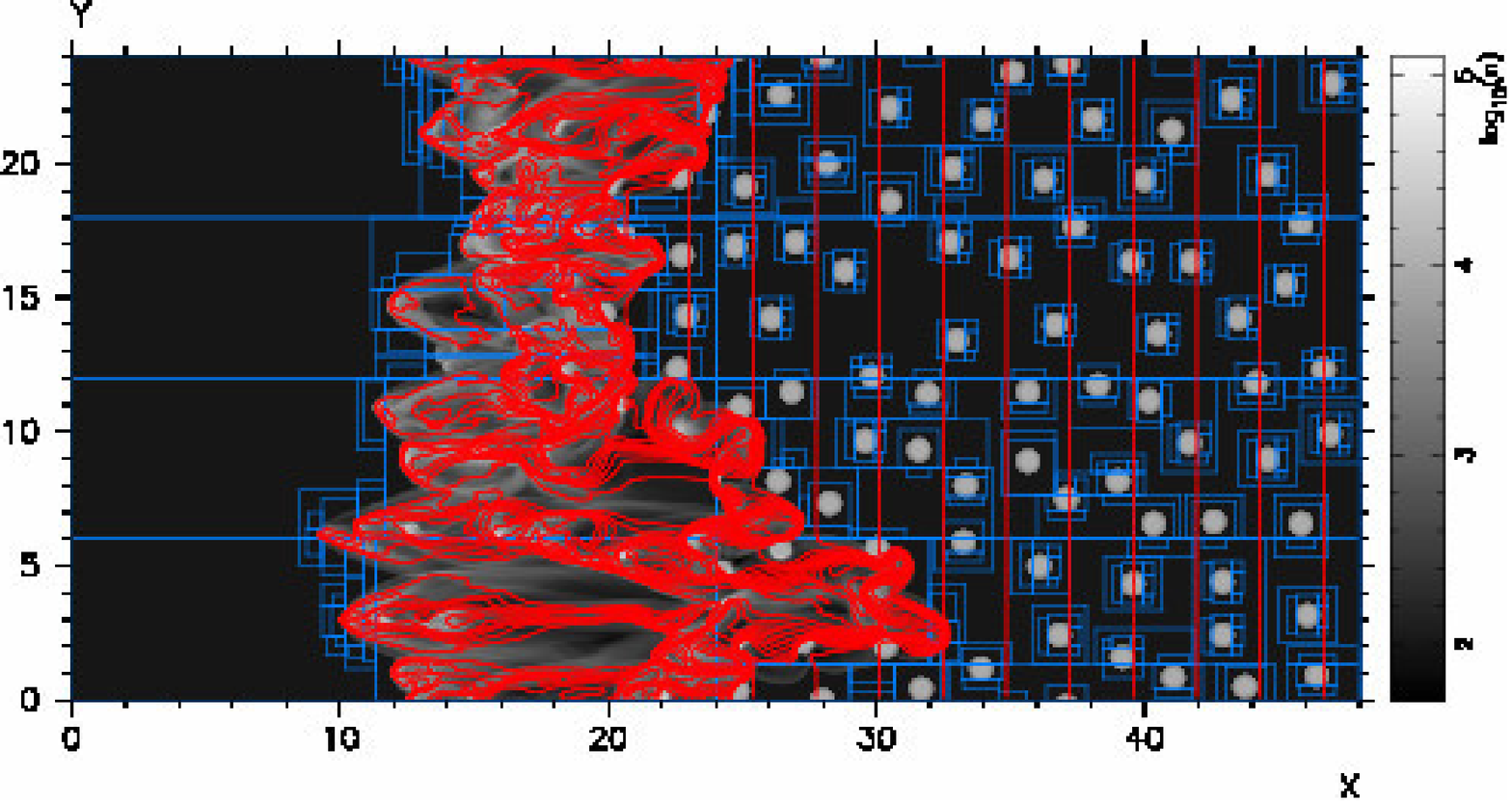} \\
   \includegraphics[angle=-90,clip=true,width=0.55\textwidth]{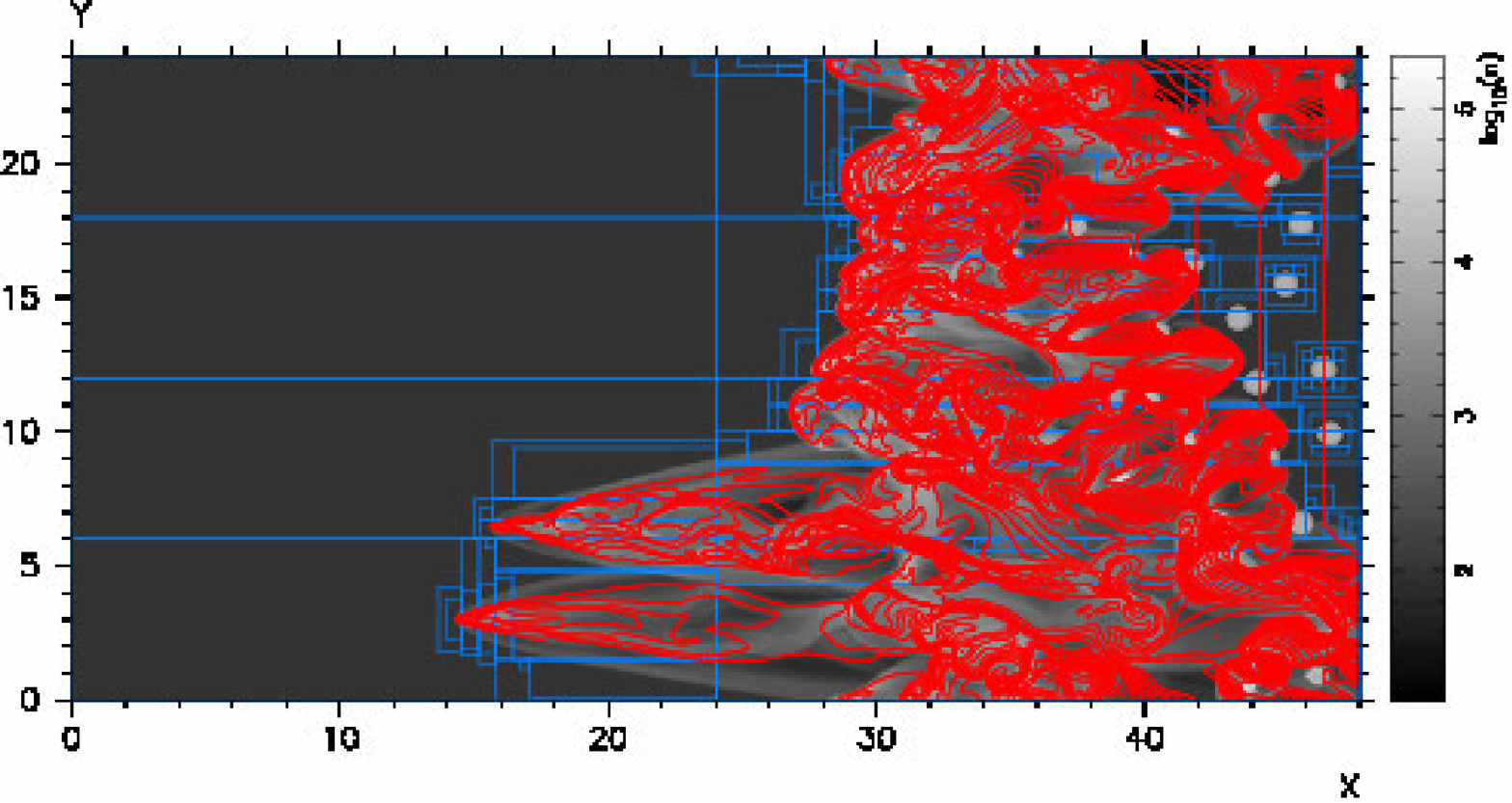}
\end{center}
\caption{Shock propagation through multiple clouds with magnetic field
oriented perpendicular to the direction of shock propagation at
evolutionary time $t=0,~92,~184 \textrm{ and } 369~\textrm{yr}$.  The
logarithm of the density distribution in $cm^{-3}$ is presented in
gray-scale, red lines delineate the magnetic field lines, blue lines
delineate regions of AMR-enhanced resolution and the cloud diameter is
used as the length unit. \label{f15}}
\end{figure}

To confirm the robustness of the code and to demonstrate its
application with radiative energy loss, we illustrate the propagation
of a strongly radiative shock through an inhomogeneous environment.
The inhomogeneities in these two dimensional calculations take the
form of a random distribution of 150 cylindrical clouds with 100 times
the density of the ambient environment at their center.  The density
of the clouds follow a hyperbolic tangent smoothing function to the
ambient density along the outer 50\% of the radius of each cloud.  The
inhomogeneities are initially in pressure balance with the ambient
environment.  The upper and lower boundaries of the computational
domain are periodic, the right boundary follows a constant
extrapolation and the wind state is held constant in the left boundary
throughout the simulation.  The physical state of the ambient
environment, the center of the clouds, and the ambient environment are
given in table \ref{t15}.  Figures \ref{f14} and \ref{f15} show the
results of the simulations at several evolutionary stages
($t=0,~92,~184 \textrm{ and } 369~\textrm{yr}$).  The simulations were
carried forward using MUSCL spatial reconstruction, Runge-Kutta
temporal integration, the adapted Marquina flux upwinding and the
constrained transport method of \cite{rjf}.  An operator split energy
sink source term is used to include the effects of radiative energy
loss via atomic line cooling using the cooling function of \cite{dm}.
The density distribution is presented in gray-scale, red lines
delineate the magnetic field lines and blue lines delineate regions of
AMR-enhanced resolution.  Figure \ref{f14} shows the case where the
magnetic field oriented parallel to the direction of wind and figure
\ref{f15} shows the case where the ambient environment is threaded
with a magnetic field that is oriented perpendicular to the direction
of an unmagnetized wind.  The turbulent nature of the flow pattern
that emerges in simulations, compounded sock compression ratios as high
as $\sim 12$ which are achieved via radiative losses results in the
development of persistently converging flow.  Such flows are
particularly problematic for codes that do not maintain the solenoidal
constraint on the magnetic field as local divergences tend to
accumulate in highly compressed regions of the flow
\citep{balsara-comp}.  These simulations, therefore, demonstrate the
robustness of the methods described in this paper for the simulation
of such flows.
\clearpage
\clearpage
%
\clearpage
\begin{table}[!h] \caption{Multiple Cloud Parameters.\label{t15}}
 \begin{tabular}{l | l l l}
 & ambient & cloud & wind \\ 
\tableline 
$\rho~(cm^{-3})$ & 100 & $10^4$ & 100 \\ 
$v_x~(km s^{-1})$ & 0 & 0 & 150 \\ 
$v_y$ & 0 & 0 & 0 \\ 
$v_z$ & 0 & 0 & 0 \\ 
$T~(k)$ & 200 & 0.2 & $10^4$ \\ 
$\gamma$ & 5/3 & 5/3 & 5/3 \\ 
\tableline 
cloud radius (AU) & 50 & & \\ 
grid zones per cloud radius & 12 & & \\ 
CFL & 0.4 & & \\ 
\tableline
 \end{tabular}
\end{table}
\clearpage
\subsection{Three Dimensions}
\begin{figure}[!h]
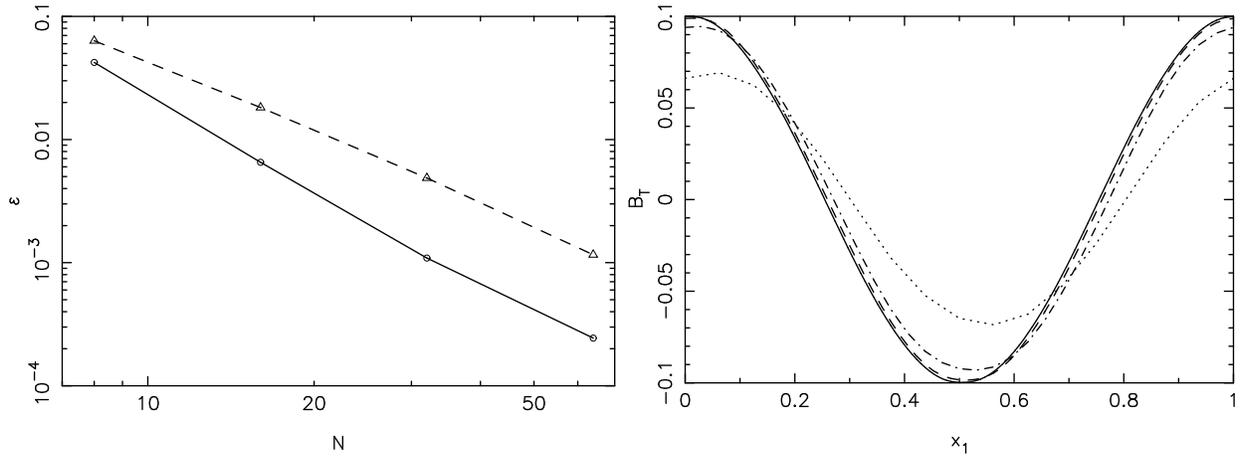

\begin{center}
   \includegraphics[angle=-90,clip=true,width=0.49\textwidth]{f16a.ps}
   \includegraphics[angle=-90,clip=true,width=0.49\textwidth]{f16b.ps}
\end{center}
\caption{Circularly polarized Alfven wave in three dimensions: Left:
  convergence of the $L_1$ error for the direction split MUSCL-Hancock
  temporal reconstruction (dotted) and the unsplit Runge Kutta
  temporal reconstruction (solid), Right: transverse component of the
  magnetic field (right) after one grid crossing $(t=1)$ for grid
  resolutions
  $N=8~\textrm{(dot)}~,16~\textrm{(dash-dot}~,32~\textrm{(dash)
    and}~64~\textrm{(solid)}~$.\label{f16}}
\end{figure}

Circularly polarized Alfven waves are an exact nonlinear solution to
the MHD equations.  We follow an approach similar to that of
\cite{gardiner08} and \cite{toth} by rotating the one
dimensional prescription of the problem (table \ref{t16}) onto a three
dimensional periodic grid of size $\sqrt{6}/2 \times \sqrt{6} \times
\sqrt{6}$ with resolution $N \times 2N \times 2N$ via the rotation:
\begin{eqnarray}
\left[ \begin{array}{c}
x \\ y \\ z
\end{array}\right] = 
\left[ \begin{array}{ccc}
\cos(\alpha) \cos(\beta) & -\sin(\alpha) & -\cos(\alpha) \sin(\beta) \\
\sin(\alpha) \cos(\beta) &  \cos(\alpha) & -\sin(\alpha) \sin(\beta) \\
\sin(\beta)              &  0            & \cos(\beta)
\end{array}\right]
\left[ \begin{array}{c}
x_1 \\ y_1 \\ z_1
\end{array}\right]
\end{eqnarray}
where $\alpha=\arcsin(1/\sqrt{5})$ and $\beta=\arcsin(1/\sqrt{6})$ so
that the initial state is periodic with the grid and the wave-vector
points along the diagonal of the computational domain.  Solutions have
been computed for one crossing time $(t=1)$ using monotonized-centered
limited linear spatial reconstruction, the Roe flux function and both
the unsplit Runge-Kutta and direction split MUSCL-Hancock temporal
integrators.  The left panel figure \ref{f16} shows the convergence of
the volume averaged norm of the $L_1$ error vector as measured with
respect to the initial state
\begin{equation}
\epsilon = \frac{1}{4 N^3} \sqrt{\sum_{nq} 
     \left( \sum_{i,j,k} \left| q_{i,j,k,nq}^{t=1} - q_{i,j,k,nq}^{t=0} \right| \right)^2}
\end{equation}
for grid resolutions of $N=8,16,32~\textrm{and}~64$.  Consistent with
the results of other authors on this test problem
\citep{toth,gardiner,gardiner08}, the solution error arises mainly
from the magnetic field components that are transverse to the wave
propagation direction.  The right panel of figure \ref{f16} shows the
steady convergence of the amplitude of the transverse components of
the magnetic $B_T = \sqrt{B_{y1}^2+B_{z1}^2}$ field after rotating
back into the unrotated space
\begin{eqnarray}
\left[ \begin{array}{c}
x_1 \\ y_1 \\ z_1
\end{array}\right] = 
\left[ \begin{array}{ccc}
\cos(\alpha) \cos(\beta) &  \sin(\alpha)\cos(\beta) & \sin(\beta) \\
-\sin(\alpha)             &  \cos(\alpha)            & 0           \\
-\cos(\alpha) \sin(\beta) & -\sin(\alpha)\sin(\beta) & \cos(\beta)
\end{array}\right]
\left[ \begin{array}{c}
x \\ y \\ z
\end{array}\right]
\end{eqnarray}
for the unsplit Runge-Kutta case.  Both integration techniques
converge in a manner consistent with the expected second order
accuracy with least squared power-law induces $\epsilon \propto
N^{-1.92}$ for the MUSCL-Hancock scheme and $\epsilon \propto
N^{-2.49}$ for the Runge-Kutta scheme.

\clearpage
\clearpage
\begin{table}[!h] \caption{Circularly Polarized Alfven Wave Parameters.\label{t16}}
 \begin{tabular}{l|l}
\tableline
$\rho$ & 1    \\
$v_1$  & 0    \\
$v_2$  & $a \sin(2 \pi x_1)$ \\
$v_3$  & $a \cos(2 \pi x_1)$ \\
$P$    & 1    \\
$B_1$  & 1    \\
$B_2$  & $a \sin(2 \pi x_1)$ \\
$B_3$  & $a \cos(2 \pi x_1)$ \\
$a$  & $0.1$  \\
$R$    & 0.3  \\
$\gamma$ & 5/3\\
\tableline
$\Delta t$ & $1/(3N)$ (MUSCL-Hancock) \\
           & $1/(6N)$ (Runge Kutta)   \\
 \tableline
 \end{tabular}
\end{table}
\clearpage
\section{Conclusion} \label{conclusion}
The staggered grid constrained transport schemes described in this
paper enable the application of high resolution shock capturing
methods to magnetized flow.  In this paper we have demonstrated that a
wide cross section of high resolution shock capturing schemes for
general conservation laws may be adapted for magnetized flow while
preserving the divergence free constraint on the magnetic field
topology exactly by conserving the surface integral of magnetic flux
over each computational cell in an upwind fashion.  The use of such
schemes on multi-resolution AMR grids is encumbered by the requirement
that the prolongation and restriction steps preserve the divergence
free topology of the magnetic fields. In this paper we have described
the application of prolongation and restriction operators which
maintain such topologies to machine precision.

The numerical schemes discussed here have been implemented and tested
in the AstroBEAR adaptive mesh refinement code.  The code utilizes a
modular design, enabling the user to choose from various methodologies
to tailor the numerical integration strategy to the requirements of
the application at hand.  The robustness of this approach to high
resolution, shock capturing MHD on AMR grid structures, and relative
advantages of the various numerical schemes implemented in the code
are demonstrated in the context of several numerical example problems.
The description of the numerical schemes presented in this paper
provides a concise recipe for their implementation which will enable
the reproduction of these outcomes by other researchers and the
interpretation of future works derived from the AstroBEAR code.

\acknowledgments We acknowledge support for this work from the Jet
Propulsion Laboratory Spitzer Space Telescope theory grant 051080-001,
Hubble Space Telescope grants 050292-001, HST-AR-11251.0,
HST-AR-11250.01, National Science Foundation grants AST-0507519,
AST-0406799, AST 00-98442 \& AST-0406823, DOE grant DE-F03-02NA00057,
the National Aeronautics and Space Administration grants JPL-1310438,
ATP04-0000-0016 \& NNG04GM12G issued through the Origins of Solar
Systems Program.  TWJ is supported in this work by NSF grant
AST0607674, NASA grant NNG05GF57G and by the University of Minnesota
Supercomputing Institute.  We also acknowledge the computational
resources provided by University of Rochester Information Technology
Services and the Laboratory for Laser Energetics.

\appendix
\section{Cylindrical Axisymmetry}
The conservative update procedure for integrating the ideal MHD
equations \S\ref{method} may be readily extended to the case
cylindrical axisymmetric flows via a change in the coordinate
variables and the addition of a source term to equation \ref{fullmhd}.
In particular, the coordinate subscripts transform as as $(x,y,z), \to
(r,\theta,z)$ and the source term becomes:
\begin{eqnarray}
\vec{S} = -\frac{1}{r} & [\rho v_r , \rho v_r^2-B_r^2+B_{\theta}-\rho v_{\theta}^2, 2 (\rho v_r v_{\theta} - B_{\theta}B_r), \rho v_r v_z - B_z B_r , \nonumber \\
  & v_r(E+P+\vec{B}^2/2)-B_r(\vec{B} \cdot \vec{v}) , 0 , 0 , v_r B_z - v_z B_r ]^T.
\end{eqnarray}
Due to axisymmetry, all differential terms in $\theta$ vanish.  We
handle the geometric source term, $\vec{S}$ separately from the
conservative update of the homogeneous part of the system using an
operator split approach.  The source term step $\frac{\partial
\vec{Q}}{\partial t}=\vec{S}$ is integrated via a fourth-order
Rosenbrock integration scheme for stiff systems of ordinary
differential equations \citep{numrec} using an adaptive time-step to
maintain the accuracy of the solution to a user specified level
(usually $1$ part in $10^4$).  Symmetry dictates that such simulations
be carried out in a half-meridional plane ($r \ge 0$) and that
reflecting boundary conditions be applied along the $r=0$ plane.

The constrained transport of \S\ref{ctsec} and prolongation /
restriction of \S\ref{prores} of grid-interface magnetic field
components, $B_r$ and $B_z$ may also be readily adapted for
cylindrical axisymmetric flow.  Readers interested in the extention to
more complicated geometries may refer to the work of
\cite{balsara-geom} which derives divegence free reconstruction
procedures in several three dimensional curvilinear geometries and on
tetrahedral meshes.  In cylindrical axisymmetry, the solenoidality
constraint on the magnetic field takes the form:
\begin{equation}
\vec{\nabla} \cdot \vec{B} = \frac{1}{r}\frac{\partial r B_r}{\partial r} + \frac{\partial B_z}{\partial z} = 0,
\end{equation}
and the component of the curl of the magnetic field orthogonal to the
symmetry plane takes the form:
\begin{equation}
\left[\vec{\nabla} \times \vec{B}\right]_\theta = \frac{\partial B_r}{\partial z} - \frac{\partial B_z}{\partial r}.
\end{equation}
These operators take the same form as the Cartesian case under the
change of variable $B_r \to r B_r$.  Therefore CT update formulae of
\S\ref{ctsec} and the prolongation and restriction forulae of
\S\ref{prores} are written for the case of axisymmetry by replacing
$(\tilde f_x, \tilde f_z)\to (\tilde f_r, \tilde f_z)$, $E_y \to
E_{\theta}$, $(B_x, B_z) \to (r B_r, B_z)$, and $(b_x, b_z) \to (r
b_r, b_z)$.  The CT integration procedure (equations \ref{ct}) can be
written for axisymmetric geometry by replacing $B_z \to r B_z$ and
$E_y \to r E_{\theta}$.

\section{Pseudo-code listing for the AMR engine.} \label{amr}
\begin{verbatim}
  SUBROUTINE AMR(level, dt)
    IF(level=0) 
      nsteps = 1
      Set Ghost(level)
    ELSE
      nsteps = r
    END IF
    DO n = 1, nsteps
      Distribute(level)
      IF(level < MaxLevel) 
         Grid Adapt(level + 1)
         Set Ghost(level+1)
      END IF
      Integrate(level,n)
      IF(level< MaxLevel)  AMR(level + 1, dt/r)
    END DO
    IF(level > 1) Syncronize to Parent(level)
  END SUBROUTINE AMR
\end{verbatim}

\end{document}